%% file: main.tex
\documentclass[12pt]{article}
\usepackage[utf8]{inputenc}
\usepackage{xcolor}
\usepackage{amsmath,amsthm}
\usepackage{amsfonts}
\usepackage{booktabs}
\usepackage{siunitx}
\usepackage{multirow}
\usepackage{graphicx}
\usepackage{hyperref}
\usepackage{longtable}
\usepackage{changepage}
\usepackage{mathrsfs}
\usepackage{subcaption}
\usepackage{biblatex}
\usepackage{algorithm}
\usepackage{algpseudocode}
\usepackage{standalone, tikz}
\usepackage{mathtools}
\usetikzlibrary{topaths, calc, trees, positioning, arrows, shapes, shapes.multipart, shadows, matrix, decorations.pathreplacing, decorations.pathmorphing,backgrounds}

\hypersetup{
    colorlinks=true,
    linkcolor=blue,
    filecolor=magenta,      
    urlcolor=cyan,
}

\def\I{\mathcal{I}}
\def\P{\mathcal{P}}

\def\Nat{\mathbb{N}}
\def\B{\mathcal{B}}
\def\H{\mathcal{H}}
\def\bfy{\mathbf{y}}
\def\bfY{\mathbf{Y}}

\def\fin{\text{fin}}
\usepackage{setspace}
\usepackage[margin=1in]{geometry}
\usepackage{xr}
\providecommand{\keywords}[1]
{
  \small	
  \textbf{\textit{Keywords---}} #1
}
\makeatletter
\newcommand*{\rom}[1]{\expandafter\@slowromancap\romannumeral #1@}
\makeatother

\newtheorem{thm}{Theorem}[section]

\newtheorem{prop}[thm]{Proposition}

\newtheorem{defn}[thm]{Definition}

\newtheorem{rmk}[thm]{Remark}%\endlocaldefs

\author{
  Zhang, Yuhua\\
  \texttt{zyuhua@umich.edu}
  \and
  Dempsey, Walter\\
  \texttt{wdem@umich.edu}
}
\title{Node-level community detection within edge exchangeable models for interaction processes}
\date{}

\begin{document}

\maketitle

\begin{abstract}

Scientists are increasingly interested in discovering community structure from modern relational data arising on large-scale social networks. 
While many methods have been proposed for learning community structure, few account for the fact that these modern networks arise from processes of interactions in the population.
% such as user post and comment exchanges.
We introduce block edge exchangeable models (BEEM) for the study of interaction networks with latent node-level community structure.  
The block vertex components model (B-VCM) is derived as a canonical example.  Several theoretical and practical advantages over traditional vertex-centric approaches are highlighted.  
In particular, BEEMs allow for sparse degree structure and power-law degree distributions within communities.  
Our theoretical analysis bounds the misspecification rate of block assignments, while supporting simulations show the properties of the network can be recovered. 
A computationally tractable Gibbs algorithm is derived.
We demonstrate the proposed model using post-comment interaction data from Talklife,  a large-scale online peer-to-peer support network, and contrast the learned communities from those using standard algorithms including spectral clustering and degree-correct stochastic block models.

\noindent \keywords{Large-scale Sparse Network, Edge Exchangeable Model, Community Detection, Interaction Process, Power-law Degree Distribution}

\end{abstract}

%\section{Abstract}
%Health scientists are increasingly interested in understanding the relationship between interaction process and  health  outcomes, 
%To serve the purpose of provide better supports to certain population, an important question in statistical network analysis is how to detect the community structures underlying these interactions. On the other hand, the properties of the network themselves have been largely ignored. For example, the sparsity and the power-law properties are properties that are frequently observed in real-life networks, but have not been emphasized much in the previous works.  In this paper, we propose a new model incorporating these two properties into detecting the community structures.  Our proposed model infers the cluster assignments of the nodes in the sparse network,  as well as estimates the power-law parameters within each cluster.  We show the misspecification rate of our model can be bounded and demonstrate it with simulated data.  Besides, the power-law properties of the network can be retrieved. The proposed method is  illustrated  using  Talklife  data,  which is mental  health-related  large-scale  online peer supporting network.\\

\section{Introduction}

%Para 1. Introduction of mobile health data. And why it is important to solve the problem.

The WorldWideWeb, social media, and technological innovation allow people from around the world to easily interact across geographical, cultural and economic boundaries.  Among its many benefits, increased connectivity enhances information flow, promotes community building, and facilitates the formation of support systems for individuals struggling with illness and addiction.
% Network data are gaining attentions these days. Such data can be collected in many different forms, aiming for different study purposes. 
In this paper, we focus on network data arising from sequences of interactions collected on social media platforms such as peer-to-peer support networks. 
%Scientists are primarily interested in improving peer support by understanding these digital interactions \cite{fortuna2020digital}.
Network data arising from such sequences naturally fit within models that treat the interaction as the statistical unit~\cite{crane2021sts} rather than the users.  Such networks often exhibit node-level community structure, i.e., an individual is more likely to comment on a post by someone from the same community.  

% Edge-exchangeable models were introduced as a natural framework for statistical analysis of interaction data~\cite{crane2016edge}.  While an attractive theoretical framework, the set of current edge exchangeable models do not allow for node-level latent communities.  
% In suicide research, for example, social support has been shown to be a preventative factor for future suicidal ideation \cite{kleiman2013social}.  
% To further und goals, this paper focuses on discovery of latent community structure.
Though there has been much progress on community detection, the current algorithms and models in the network literature do not account for the fact that many modern networks arise from processes of interactions. Edge exchangeable models~\cite{crane2016edge} are built specifically to analyze datasets containing these complex interactions.  Current models within the edge exchangeable framework can capture both global and local sparsity and power-law behavior~\cite{dempsey2021hierarchical}. While edge exchangeability is attractive as a theoretical framework, the set of current edge exchangeable models is inadequate to account for latent node-level communities which are common in modern network data.
% Meanwhile, it also incorporated the power-law structure and the sparsity of the network into the estimation of the network properties. 
% 

Motivated by the important fact that most common complex networks constructed from interaction data exhibit latent node-level community behavior, 
% The current set of edge exchangeable models focus mainly on the inference of the statistics of the network, but doesn't consider the community structure. 
we propose a new class of models in this paper, which extends the edge exchangeable framework to allow for community structure while permitting power-law degree and sparsity. Our main objective is the consistent identification of node-level latent communities as well as accurate estimates of the power-law parameters within each community.

% \walt{Need 3 paragraphs: p1: general $\to$ interaction data, p2: community detection is important, p3: this paper allows for community detection in interaction data}

\subsection{Relevant Prior Work}
%Para 2. The proposed methods that can be used to the analysis. What are the problems.
A popular model class for community detection is the class of stochastic block models (SBMs) \cite{holland1983stochastic}. The simplest version of the SBM assumes vertices within the same block have the same probability of forming an edge (i.e., interact) with other vertices, and within-block interactions are more likely than between-block interactions. Many associated methods have been proposed. These include but are not limited to spectral clustering based methods (\cite{rohe2011spectral}, \cite{chin2015stochastic}), Bayesian methods (\cite{van2018bayesian}, \cite{morup2012bayesian}), and pseudo-likelihood based methods (\cite{amini2013pseudo}, \cite{strauss1990pseudolikelihood}). The stochastic block model was later generalized to the degree-corrected stochastic block model (DC-SBM) \cite{karrer2011stochastic}, which allows for vertex degree heterogeneity. Theoretical guarantees of community recovery have been well established \cite{gao2018community, amini2013pseudo, zhao2012consistency}.

%Para 3. The contribution of this paper.
Prior literature has demonstrated empirical power law degree distribution in many communication and social networks~\cite{adamic2001search}. That is, such networks contain a few high degree nodes and many low degree nodes. This network feature may be exploited when performing community detection. Recent work has extended the SBM to account for power-law degree distributions within each community \cite{qiao2018adapting}. 
Though SBM and the DC-SBM have been successfully applied in many situations, these approaches do not account for the fact that many modern networks arise from processes of interactions and are therefore ill-suited for network data arising in these settings.  This is illustrated empirically via case study in Section~\ref{sec:Talklife}.

% \subsection{Motivation}

% Though much progress has been made, the SBM and its extensions do not account for the fact that many modern networks arise from processes of interactions. Edge exchangeable models~\cite{crane2016edge} are built specifically to analyze datasets containing these complex interactions.  Current models within the edge exchangeable framework can now capture both global and local sparsity and power-law~\cite{dempsey2021hierarchical}. While edge exchangeability is attractive as a theoretical framework, the set of current edge exchangeable models is inadequate to detect latent node-level communities which are common in modern network data.
% % Meanwhile, it also incorporated the power-law structure and the sparsity of the network into the estimation of the network properties. 
% % 

% Motivated by an important fact: most common complex networks constructed from interaction data exhibit latent community behavior, 
% % The current set of edge exchangeable models focus mainly on the inference of the statistics of the network, but doesn't consider the community structure. 
% we propose a new class of models in this paper, which extends the edge exchangeable framework to allow for community structure while permitting power-law degree and sparsity. Our main objective is the consistent identification of node-level latent communities as well as accurate estimates of the power-law parameters within each community.

%Part 4. The overview of the paper's structure.
\subsection{Outline and main contributions}

The main contributions of this paper are as follows: 
\begin{enumerate}
    \item We formally define network data with block structure and notation in Definition \ref{defn:blockintdata}. 
    \item We define, in Section~\ref{section:seqdesc}, the block vertex components models (B-VCM) - a subfamily of block edge exchangeable models that allow for node-level communities; we then prove a representation theorem for block edge exchangeable processes in Theorem~\ref{thm:repthm}.
    \item We establish basic statistical properties of the B-VCM in Section~\ref{sec:property}. In particular, we show the misspecification rate of block assignments can be bounded and the sparsity and power-law of the network can be guaranteed. 
    \item A computationally tractable Gibbs sampling inferential algorithm is derived in Section~\ref{sec:alg}.  Simulation results that support our models ability to capture power-law degree distributions and community structure are presented in Section~\ref{sec:simu}. 
    \item We then apply our method to a peer-to-peer mental health support network in Section~\ref{sec:Talklife}, contrasting the learned communities from those using standard algorithms including spectral clustering and degree-corrected stochastic block models.
%using the TalkLife data and the interpretation of the results. 
\end{enumerate}
We end the paper with a summary of the proposed methods and implications in Section~\ref{section:discussion}. Overall this paper presents a statistically rigorous, principled framework for incorporating node-level communities into the edge exchangeability framework.

\section{Block Edge Exchangeable Model}

\subsection{Motivating Example}

Our motivating example is TalkLife, a large-scale online peer support network for mental health. Talklife's users post short text snippets to which other users can react and/or comment. Each post consists of a poster and a set of commentators. Every post is flagged using machine-learning classifiers with at least one health-related topic, e..g, Anxiety/ Panic/ Fear suspected, Body Image/ Eating Disorders suspected and so on. The platform has collected millions of interactions among its users. 
A primary goal of TalkLife is to strengthen its peer support network and improve mental health outcomes for its users. 
% The network is of complex structures, and viewing the entire dataset as a whole network will lead to the loss of many features of the interactions among the users.
A natural question that arises as part of this work is node-level community detection from the large-scale interaction data generated by the platform.

\subsection{Notations and Data}
\label{sec:2.2}
%The overall structure of this section:

%Def 1. the interaction data, $\mathcal{P}$ and $fin(\mathcal{P})$.

%Def 2. the block structured interaction data. $B(\cdot): \mathcal{P}\rightarrow fin(P)$

%Def 3. The equivalence class: Let $\pho$ be a relabeling of the population, such that if $B(a)=B(b)$, then $B(\rho(a))=B(\rho(b))$.\\

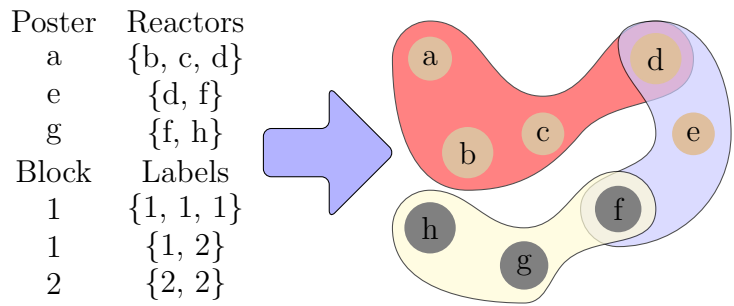
\begin{figure}[htp]
    \centering
    \input{z_concep_fig_h.tex} 
    \caption{Example of interaction data with block structure collected on TalkLife (left) and the corresponding network (right). 
    % In this example, there are 3 posts and 8 users involved. Post 1 is posted by User a and commented on by Users b, c, and d; Post 2 is posted by User e and commented on by Users d and f; Post 3 is posted by User g and commented on by Users f and h. The first 5 users are from block 1. The subsequent 3 users are from block 2.
    }
    % The network is constructed by connecting User 1 to User 3, 4, and 5; User 2 to User 6 and 7; User 10 to User 8 and 9. Note that Users 1, 2, 3, 4, 5, and 6 are from cluster 1, and Users 7,8,9, and 10 are from cluster 2.}
    % Therefore, 4 of the interactions are from cluster 1 to cluster 1, 1 of them is from cluster 1 to cluster 2, the rest are from cluster 2 to cluster 2.}
    \label{Fig.3.2}
\end{figure}

We start by defining block structured interaction data. Inspired by the previous work on edge exchangeability~\cite{crane2016edge}, we extend the definition of interaction data to the setting where there exists an underlying block structure. To start, we consider a simple example based on interaction data from Talklife.  Recall each TalkLife post consists of a poster $s$ and a set of commentators $\{ r_1, \ldots, r_k \}$.   Then a post on Talklife can be summarized by the ordered pair $\{ \{ s \}, \{ r_1, \ldots, r_k \} \} $. See Figure~\ref{Fig.3.2} for a visualization of Talklife's interaction data and its corresponding network construction. Here, posters and commentators are drawn from the same underlying population, which we denote as $\mathcal{P}$. Definition~\ref{defn:intdata} formalizes this example into a general definition of interaction data.  \\

\begin{defn}[Interaction data] \normalfont
\label{defn:intdata}
For a set $\P$, let $fin(\P)$ be the set of all finite multisets of $\P$. The interaction process of $\P$ is the correspondence $\I: \Nat \rightarrow fin(\P)$ between the natural numbers and finite multisets of $\P$. 
\end{defn}

Additional structure can be incorporated into Definition~\ref{defn:intdata}. In Talklife, for example, there is a single poster and multiple commentators.  Definition~\ref{defn:intdata} can capture this by replacing $fin (\P)$ with $fin_1(\P) \times fin(\P)$ where $fin_1(\P)$ are sets of size one from $\P$. This generalization can also be extended into other types of interactions. See~\cite{dempsey2021hierarchical} for additional examples of these extensions. 
% Note that the nodes are labeled in the order of appearance. If a new individual is chosen, that person is given the next label in the order of appearance. For example, if they are the $10$th person to be observed, they are labelled as individual $10$.

Using Figure~\ref{Fig.3.2} as an example, each interaction is represented by the user who posts and the users who comment. The users, regardless of being posters or commentators, are drawn from the population $\P \subseteq \Nat$. The interaction process is a correspondence $\I:\Nat \rightarrow fin_1(\P)\times fin(\P)$, i.e., $\I (1)=(\{a \} ,\{b,\ c,\ d\}), \I (2)=(\{e\},\{d,\ f\})$, and $\I(3)=(\{g\},\{f,\ h\})$.

Individuals who post and comment are likely to exhibit latent node-level community behavior.  That is, commentators and posters form a block structure.  In Figure~\ref{Fig.3.2}, for example, Users a, b, c, and d all belong to block 1, while Users f, g and h belong to block 2. Within-block interactions are more likely to be observed than between-block interactions. Note, however, that users from distinct blocks may appear within a single interaction, i.e., User f comments on a post by User e.  We formalize this idea in Definition~\ref{defn:blockintdata}. \\

%%%%%% BLOCK AND CLUSTER: CHOOSE 1 (OR CLARIFY)

\begin{defn}[Block structured interaction data] \normalfont 
\label{defn:blockintdata}

A \emph{K-block structure} is defined as the mapping $B: \mathcal{P}\rightarrow [K]$, i.e., for all $x\in\mathcal{P}$, $B(x)\in [K]$ is the block to which $x$ is assigned. The interaction process induced by $B$ is called the \emph{block interaction process}, which is the correspondence $\I_B:\Nat\rightarrow fin\textbf{}([K])$.

\end{defn}

In the running example shown in Figure~\ref{Fig.3.2}, $K=2$ and the interaction process induced by $B$ is $\I_B(1)=(\{1\},\{1,1,1\})$, $\I_B(2)=(\{1\},\{1,2\})$, and $\I_B(3)=(\{2\},\{2,2\})$. Note that for the pre-image $B^{-1}$, $B^{-1}(i)\cap B^{-1}(j)=\emptyset$ if $i\neq j$, and $\cup_{i=1}^K B^{-1}(i) = \P$. That is, $B$ partitions $\P$ into $K$ non-overlapping sets.

Above, interaction processes are distinct if they are defined on different populations or given distinct block structures.  Here, we formally address the equivalence of networks built from two different block structured interaction processes that can be linked via a set of bijections. Let $\rho: \mathcal{P}\rightarrow \mathcal{P}^{\prime}$ be the bijection which acts on $I$ via $\rho \mathcal{I}: \mathbb{N}\rightarrow fin(\mathcal{P}^{\prime})$. That is, for $( s_1, \ldots, s_m ) \in fin (\P) $:
$$\rho \circ (s_1, \ldots, s_m) = (\rho(s_1), \ldots, \rho(s_m)) \in fin(\P^\prime) $$
The bijection also acts on the block structure~$B$ defining the map $\rho B := B \circ \rho^{-1} : \P^\prime \rightarrow fin([K])$. Consider a second block structure,~$B': \P' \rightarrow fin([K])$. Then a second bijection~$\rho': [K] \to [K]$ acts on the block interaction process $I_{B'}$ via $\rho' \mathcal{I}_{B'}: \mathbb{N} \rightarrow fin([K])$.  In other words, the bijection~$\rho'$  preserves the block structure but changes the block labels. With this, the interaction-labeled network with block structure~$B$ is then the equivalence class constructed from the interaction network by quotienting out over pairs of bijections $(\rho, \rho')$ that preserve block structure:
$$\mathbf{Y}_{\mathcal{I}, B}=\bigcup_{\#\mathcal{P}^{\prime}=\#\mathcal{P}}
\left\{
\begin{array}{c c c} 
\mathcal{I^{\prime}}:\Nat\rightarrow fin(\mathcal{P}^{\prime}): &
\rho \mathcal{I}=\mathcal{I}^{\prime}, & 
\text{for }\rho: \mathcal{P}\rightarrow\mathcal{P}^{\prime}  \\
B' : \P^\prime \rightarrow [K]: &
\rho' \mathcal{I}_{\rho B} =\mathcal{I}_{B'} &
\text{for }
\rho': [K] \rightarrow [K]
\end{array}
\right\}
$$
where $\# P$ is the cardinality of the population.  Note we have only quotiented out labels of constituent elements and the blocks, so the network $\mathbf{Y}_{\mathcal{I}, B}$ still has uniquely labelled interactions with a unique node-level block structure.  To avoid cumbersome notation, we often write $\mathbf{Y}$ to denote the interaction-labeled network with block structure, leaving the associated interaction process $I$ and block structure~$B$ implicit.

For any $S \subset \mathbb{N}$, we define the \emph{restriction} of the interaction process $\mathcal{I}$ to $S$ by $I_{S}$.  This restriction induces a restriction of the interaction labeled network $\mathbf{Y}$ which we denote $\mathbf{Y}_{S}$.  For $S = [n] = \{1,\ldots, n\}$, we simply write $I_n$ to denote the restricted structured interaction process and $\bfY_n$ to denote the corresponding structured interaction network.   Moreover, we define the restriction of the interaction process to only the elements from a particular block~$b$ by $I_b$ and $\bfY_b$ to denote the corresponding structured interaction network. In our recurring example, for instance, $I_1 (2) = (\{e \}, \{ d \})$ where element $\{f\}$ is removed from~$I(2)$ as it belongs to block~$2$.  This restriction is useful in defining block-specific network properties.

\subsection{Exchangeable interaction-labeled networks with block structure}

For any finite permutation $\sigma : \Nat \to \Nat$, let $\mathcal{I}^\sigma$ denote the relabeled interaction process defined by $I^\sigma (i) = I (\sigma^{-1} (i))$ and the relabeled block interaction process defined by $I_B^\sigma (i) = I_B (\sigma^{-1} (i))$ for each $i \in \Nat$. Then $y^\sigma$ denotes the corresponding interaction labeled network constructed from $I^\sigma$. Note that $\sigma$ does not impact the block structure~$B$ but does impact the block structure process~$I_{B}^\sigma$. Note that the choice of representative from the equivalence class $(I,B)$ does not matter.  The interaction relabeling by $\sigma$ should not be confused with the bijections~$\rho$ and $\rho'$, which relabel constituent elements and block labels respectively.

In the remainder of the paper, we write~$\bfY$ to denote the random interaction labeled network with block structure~$B$.  Interaction exchangeability is characterized by the property that the labeling of the interactions (not the constituent elements or blocks) is arbitrary.  We now define block exchangeable interaction networks.

% $\mathcal{Y}$ to indicate the interaction network with edges labeled in , and $E_1, E_2, \ldots, E_m$ as the observed interactions.
% , where $E_j=(\{ S_j, C_j^{(s)}\}$, $\{ R_j, C_j^{(r)}\}),\ j\in [m]$, that is, the interaction $j$ is initiated from the sender $S_j$ of cluster $C_j^{(s)}$ and points to the receiver $R_j$ of cluster $C_j^{(r)}$. 

\begin{defn}[Block exchangeable interaction networks] \normalfont 
\label{defn:blockee}
A random interaction labeled network with block-structure~$\bfY$ is \emph{exchangeable} if $\bfY^\sigma =_{D} \bfY$ for all permutations $\sigma : \Nat \to \Nat$, where $=_D$ denotes \emph{equality in distribution}.
% network constructed from interactions within the same block $\mathcal{Y}_b\subset \mathcal{Y}_m$, $\in[K]$, is exchangeable if $\mathcal{Y}_b^{\sigma}=_{D}\mathcal{Y}_b$, for all permutations $\sigma:\ \mathcal{P}\rightarrow \mathcal{P}$ where $=_{D}$ denotes equality in distribution. We say an interaction process is block edge exchangeable if this definition holds for all $b\in [K]$.
\end{defn} 

We say an interaction process~$I$ and its associated block interaction process~$B$ are exchangeable if its corresponding interaction labeled network is exchangeable.

% Under our model assumption, given the nodes $S_j$ and $R_j$, and the corresponding cluster labels, the interaction $E_j|S_j,R_j$ is a Bernoulli random variable, with $\mathbb{E}(E_j|S_j,R_j)$ being specified by $C_j^{(s)}$ and $C_j^{(r)}$, denoted as $\mathbb{E}(E_j|S_j,R_j)=\B(C_j^{(s)},C_j^{(r)})$, where $\B$ is a $K\times K$ propensity matrix. In the remainder of this paper, $\B(c,)$ is used to indicate the $c$th row of $\B$, $c\in[K]$.

% Let $v(\mathcal{Y}_m)$ be the number of vertices in the network, and $m(\mathcal{Y}_m)$ be the total degree of the network. We subscript $c\in[K]$ to indicate the cluster specific statistics, e.g., $m(\mathcal{Y}_{m,c})$ is the total degree of cluster $c$ in the network~$\mathcal{Y}_m$. We now give the definition of the block edge exchangeable network.\\

\subsection{Sequential description of the B-VCM model}
\label{section:seqdesc}

\noindent Here we provide a sequential description of a particular family of block exchangeable interaction processes, termed the B-VCM, assuming the block structure~$B$ is observed; see Section~\ref{sec:property} for why this name was chosen.
%, termed the C-VCM
For ease of comprehension, the process is described in the context of TalkLife posts assuming a \emph{single commentator}.  Suppose $m$ posts $E_{[m]} := \{E_1,...,E_{m}\}$ have been observed along with the block assignments for each observed individual. That is, the post $j\in [m]$ is given by $E_j = (S_j, R_j)$ where $S_j \in \Nat$ is the sender and $R_j \in \Nat$ is the commentator with the associated blocks $(B (S_j), B(R_j) )$.  
% Let $\mathcal{H}_{n}^{(c)}$ denote the set of unique individuals in cluster $c \in [K]$ that have either posted or commented on a post in the first $m$ interactions. 
The post $E_{m+1}$ conditional on $E_{[m]}$ and the block structure~$B$ is constructed as follows.  First, the sender's block assignment is drawn according to
%$$P(C^{(s)}_{m+1}=c|\pi_1,...,\pi_K)\sim \text{Multinomial}(\pi_1,...,\pi_K)$$
\begin{equation}
\label{eq:clus_ass1}
P(B(S_{m+1}) = b \mid E_{[m]}; \omega_B)\propto
    \begin{cases}
    D_m (b) + \omega_B , & \text{if}\ b \in [K_m] \\
    (K - K_m) \omega_B, & \text{if}\ b \not \in [K_m]
    \end{cases}
\end{equation}
where~$D_m (b) = \sum_{j=1}^m I( B(S_j) = b)$ is the number of individuals from block~$b$ that have been observed in $E_{[m]}$, $K_m \leq K$ is the number of unique blocks observed in $E_{[m]}$, and $\omega_b > 0$ is a model parameter.
% Intuitively speaking, the probability of an interaction initiating from cluster $c$ is proportional to $\pi_c$, the pre-specified propensity, such that $\sum_{c=1}^{K}\pi_c=1$. 
% Eq~\eqref{eq:clus_ass} provides a sequential description of the same procedure by assuming the $\{\pi_1,..,\pi_K\}$ are generated from a symmetric Dirichlet distribution parameterized by $\eta$. Similar to picking a specific node, the clusters are ranked by the order of appearance.

Let $D_{m} (i)$ be the number of times individual $i \in \Nat$ has been observed (either as a poster or commentator) in the first $m$ interactions~$E_{[m]}$.  
Similar to the Chinese Restaurant Process \cite{pitman2002combinatorial}, given parameters $0<\alpha_b<1$ and $\theta_b > -\alpha_b$ for each~$b \in [K]$, poster $S_{m+1}$ can be either chosen from one of the previously observed nodes, or an unobserved node in block $b$ with probability:
  \begin{equation}
  \label{eq:select_prob}
    P(S_{m+1} = s \mid E_{[m]}, B(S_{m+1}) = b )\propto
    \begin{cases}
      D_{m} (s) -\alpha_b, & \text{if}\ s \in V_m \cap B^{-1} (b) \\
      \theta_b+\alpha_b \sum_{i\in V_m \cap B^{-1} (b)} D_{m} (i), & \text{if}\ s \not \in V_m, s \in B^{-1} (b)
    \end{cases}
  \end{equation}
where $V_m$ is the set of the observed senders in the first $m$ interactions.
Given the sender, a similar process is used to choose the $(m+1)st$ commentator. The block assignment for $R_{m+1}$ is chosen according to:

%$$P(C_{m^*+1}^{(r)}=c'|C_{m^*+1}^{(s)}=c,\B)\sim \text{Multinomial}(\B(c,1),...,\B(c,K))$$

\begin{equation}
\label{eq:clus_ass2}
P(B(R_{m+1}) = b' | B(S_{m+1})=b, E_{[m]}) \propto
    \begin{cases}
    D_m (b,b')  + \zeta_{B} & \text{if}\ b' \in [K_{m,b}] \\
    (K-K_{m,b'}) \zeta_{B}, & \text{if}\ b' \not \in [K_{n,b}]
    \end{cases}
\end{equation}
where~$D_m (b, b') = \sum_{j=1}^m [I( B(S_j) = b, B(R_j) = b')]$ is the number of interactions from block $b$ to block~$b'$ that have been observed in $E_{[m]}$, where $K_{m,b} \leq K$ are the number of unique blocks observed as a commentator from this block, and $\zeta_{B} > 0$ is a model parameter.
%  I(C_{i}^{(s)}=c,C_{i}^{(r)}=k)$ is the count of the interactions between cluster $c$ and $k$ among $m$ interactions. Similar to the logic as described before, we assume the propensity matrix $\B(c,)$ is generated from a symmetric Dirichlet distribution parameterized by $\nu$.
% Let the set of previously observed nodes in cluster $c'$ in the first $m$ interactions be $\H_{m}^{(c')}$. Note that if $c'=c$, $\H_{m}^{(c')}=\H_{m}^{(c')}\cup \{s\}$, and $D_{m}(s,c')=D_{m}(s,c')+1$. 
Finally, the receiver $R_{m+1}$ is selected with probability whose expression takes the same explicit form as that of Eq~\eqref{eq:select_prob}. Note that in the above sequential description, blocks and individuals within blocks are labelled in order of appearance. 
 
% Let $\mathbf{C}_{m}$ denote the cluster assignments of all observed nodes in the first $m$ interactions. 
Let~$\bfy_m$ denote the observed interaction-labelled network with block structure~$B$ after $m$ interactions,~$v(\bfy_b)$ denote the number of non-isolated vertices in $\bfy_m$ belonging to block~$b$, and~$m(\bfy_b)$ is the total degree of block~$b$. Then, based on the above sequential description, the probability of observing~$\bfy_m$ takes the explicit form:
%\begin{equation}
%\label{eq:jointform1}
%    \begin{aligned}
%P(\mathcal{Y}_m=\mathbf{y}_m|\{\alpha_c ,\theta_c\}, \mathbf{C}_m)=&\prod_{c=1}^{K}\pi_c^{L_c}\times \prod_{j=1}^{m}P(S_j=s|{\alpha_c,\theta_c},\H_{j}^{(c)},C_{j}^{(s)}=c) \\
%&\times P(R_j=r|\alpha_{c'},\theta_{c'},\{S_j\},\H_{j}^{(c')},C_{j}^{(r)}=c')\times\prod_{c'=1}^{K} \B(c,c')^{W_m(c,c')}
%\end{aligned}
%\end{equation}
\begin{align}
\label{eq:jointform1}
\begin{split}
\underbrace{\frac{[K-1]_{-1}^{K_m}}{[K\omega_B]_1^{m}} \prod_{b \in [K_m]} [\omega_B]_1^{D_m(b)}}_{(\rom{1})} 
&\times 
\underbrace{\prod_{b \in [K_m]} \frac{[\alpha_{b}+\theta_{b}]_{\alpha_b}^{v(\mathbf{y}_{b})-1}}{[\theta_{b}+1]_1^{m(\mathbf{y}_b)-1}}
\prod_{j\in v(\bfy_b)}
[1-\alpha_{b}]_1^{D_m(j)-1}}_{(\rom{2})} \\
\times
&\underbrace{\prod_{b \in [K_m]}
\frac{[K-1]_{-1}^{K_{m,b}}}{[K\zeta_B]_1^{m_b}} \prod_{b' \in [K_{m,b}]} [\zeta_B]_1^{D_m(b,b')}}_{(\rom{3})}
\end{split}
\end{align}
where for real number $x$ and $a$, and non-negative integer N, the operation on x, $[x]^{N}_a=x(x+a)...(x+(N-1)a)$. Component $(\rom{1})$ of~\eqref{eq:jointform1} is the probability of an interaction initiating from block $b$. Component $(\rom{2})$ of~\eqref{eq:jointform1} is the joint distribution of all the nodes within the each block $b$. Component $(\rom{3})$ of~\eqref{eq:jointform1} is the propensity of connection between block $b$ and $b'$.  Proposition~\ref{prop:blocke2} proves the sequential description yields a block exchangeable interaction labelled network as desired.

% where $L_c$ is the number of observed interactions that are initiated from cluster $c$; $W_m(c,c')$ is the number of observed interactions from cluster $c$ to $c'$. Part $(\rom{1})$ of Eq~\eqref{eq:llk} is the probability of an interaction initiating from cluster $c$. Part $(\rom{2})$ of Eq~\eqref{eq:llk} is the joint distribution of all the nodes within the same cluster $\H_{m}^{(c)}$ which can be expressed explicitly as \cite{pitman2002combinatorial}: 

% \begin{equation}
% \label{eq:pochhammertime}
% P(\H_{m}^{(c)}|\alpha_{c},\theta_{c},\{C_m\})=\frac{[\alpha_{c}+\theta_{c}]_{\alpha_c}^{v(\mathbf{y}_{c})-1}}{[\theta_{c}+1]_1^{m(\mathbf{y}_c)-1}}\prod_{j\in \H_{m}^{(c)}}[1-\alpha_{c}]_1^{D_m(j,c)-1} 
% \end{equation}
% where for real number $x$ and $a$, and non-negative integer N, the operation on x, $[x]^{N}_a=x(x+a)...(x+(N-1)a)$. Part $(\rom{3})$ of Eq~\eqref{eq:llk} is the propensity of connection between cluster $c$ and $c'$. By replacing Part(~\rom{2}) with Eq~\eqref{eq:pochhammertime}, we have:

% \begin{equation}
% \label{eq:llk}
% P(\mathcal{Y}_m=\mathbf{y}_m|\{\alpha_c, \theta_c\}, \mathbf{C}_m)=\prod_{c=1}^{K}\pi_c^{L_c}\times 
% \frac{[\alpha_{c}+\theta_{c}]_{\alpha_c}^{v(\mathbf{y}_{c})-1}}{[\theta_{c}+1]_1^{m(\mathbf{y}_c)-1}}\prod_{j\in \H_{m}^{(c)}}[1-\alpha_{c}]_1^{D_m(j,c)-1}
% \times\prod_{c'=1}^{K} \B(c,c')^{W_m(c,c')}
% \end{equation}

\begin{prop}
\label{prop:blocke2}
The B-VCM with parameters $\Psi = (\omega_{B}, \{ \theta_c, \alpha_c \}, \zeta_B)$ determines a block exchangeable
interaction-labeled network for all $\Psi$ in the parameter space.
\end{prop}

\subsection{Statistical Properties}
\label{sec:property}

The B-VCM is an example of block exchangeable interaction-labeled networks.  A question is whether B-VCMs represent all block exchangeable interaction labeled networks. Here, we show that this is not the case by providing a representation theorem. We focus on the setting where each interaction $(s,r)$ is either never observed or observed infinitely often, i.e.,~the ``blip-free'' setting~\cite{crane2016edge}.  This setting is most relevant to the statistical applications we consider. We then leverage this representation theorem to demonstrate connections between the proposed B-VCM and the degree-corrected stochastic block model. 

% Under the exchangeable assumption, we provide a formulation of the interaction process in Section~\ref{section:seqdesc} (C-VCM). It is noticeable that the formation is not unique. We show the representation theorem for block structured interaction processes. Inspired by the edge exchangeable model \cite{crane2016edge}, we focus on the "blip-free" setting of each cluster. That is, each interaction within the same cluster is either never observed or observed infinitely often. 
We first define the $\fin_2([K])$-simplex, denoted $\mathcal{F}_K$, and the restricted $\fin_2(\Nat)$-simplex, denoted $\mathcal{F}_{\bar b}$ where $\bar b \in \fin ([K])$:
\begin{equation*}
\begin{split}
\mathcal{F}_K:=\{(f_s)_{s\in \fin_2([K])}: f_s\ge 0\text{ and }\sum_{s\in \fin_2([K])} f_s =1\} \\
\mathcal{F}_{\bar b}:=\{(f^\prime_s)_{s\in \fin_2(\Nat)}: B(s) = \bar b, f^\prime_s \ge 0\text{ and }\sum_{s \in \fin_2(\Nat), B(s) = \bar b} f_s^\prime =1\}
\end{split}
\end{equation*}
Let $\phi_K$ be a probability measure on $\mathcal{F}_K$ and define $f_K \sim \phi_K$ be the random variables drawn from the measure.  For each $\bar b \sim f_K$, we draw~$f_{\bar b} \sim \phi_{\bar b}$ a random variable drawn from a probability measure~$\phi_{\bar b}$ on $\mathcal{F}_{\bar b}$. 
Each~${\bf f} := (f_K, \{ f_{\bar b} \})$ determines a distribution on finite multisets of $\mathbb{N}$ relative to the block structure~$B$ by
$$
\text{pr} ( E = (s, r) | {\bf f}; B) = f_K (b, b') \times f_{(b,b')} (s,r)
$$
where~$\bar b = (b,b')$, $B(s) = b$, and $B(r) = b'$. This determines a block interaction exchangeable network.  
% Letting~$X_1, X_2, \ldots, $
% The Eq~\eqref{eq:pochhammertime} as described in the previous section can be one explicit form of $f_c$. Given $\{f_c\}$ and sets $\{S_c\}\in \mathcal{P}_c$ and $\{R_c\}\in\mathcal{P}_c$, we have the following statement.\\
%let the sequence of interactions $E_1$, $E_2$,... be generated by:
%$$P(E_j=\{S_j,R_j\}|f_c)=f_c$$

\begin{thm}[Representation Theorem]
\label{thm:repthm}
Let $\bfY$ be the interaction exchangeable network relative to block structure~$B$ that is blip-free. Then there exists probability measures $\phi_K$ on $\mathcal{F}_K$ and $\phi_{\bar b}$ on $\mathcal{F}_{\bar b}$ such that $\mathcal{Y} \sim \epsilon_{\phi}$ where $\phi = (\phi_K, \{ \phi_{\bar b} \})$ and
$$
\epsilon_{\phi}(\cdot)=
\int_{\mathcal{F}_K \times \{ \mathcal{F}_{\bar b}\} } \epsilon_{(f_K, \{ f_{\bar b} \}) }(\cdot) \phi(df)
$$
\end{thm}

\begin{rmk}
If $K=1$, then $f_1 (1,1) = 1$, and so we only draw $\phi_{(1,1)}$ on $\mathcal{F}_{(1,1)}$.  This recovers the representation theorem for edge exchangeable networks without block structure~\cite{crane2016edge}. \end{rmk}

Theorem~\ref{thm:repthm} presents a general framework for generating interaction exchangeable networks relative to block structure~$B$. Proof can be found in Section A.4 of the supplementary materials. Using the representation theorem, the B-VCM corresponds to particular choices of distributions~$\phi_{F}$ and $\phi_{\bar b}$. Specifically, the B-VCM corresponds to the following choices: (1) the propensity of block~$b$ initiating an interaction corresponds to $\pi = (\pi_1, \ldots, \pi_K)$ chosen from the symmetric Dirichlet distribution with parameter~$(\omega_B, \ldots, \omega_B)$ on the $(K-1)$-simplex; (2) the propensity of block~$b'$ commentating on an interaction initiated by block~$b$ corresponds to $\B(b,\cdot) = (\B(b,1), \ldots, \B(b,K))$  each chosen from the symmetric Dirichlet distribution with parameter~$(\zeta, \ldots, \zeta_B)$ on the $(K-1)$-simplex; and (3) propensities per block~$b$ the propensities per block $f_b = (f_s^{(b)}))_{s \geq 1: B(s) = b}$ are chosen from the Griffiths–Engen–McCloskey (GEM) distribution with parameter $(\alpha_b, \theta_b)$ for each~$b \in [K]$.  Then:
\begin{equation}
    \label{eq:repthme2}
P( E =  (s, r) \mid \pi_b, f, \B) = 
\underbrace{\left[ \pi_b \times \mathcal{B}(b,b^\prime) \right]}_{(f_K)} \times \underbrace{\left[ f_{s}^{(b)} \times f_{r}^{(b^\prime)} \right]}_{(f_{\bar b})}
\end{equation}

\begin{rmk}[Connection to Degree-corrected SBMs] \label{rmk:connectionsbm}
\normalfont 
% The $f_s^{(c)}$ and $f_{r}^{(c')}$ can be interpreted as the degree parameter under the e2 framework, the value of which is determined by the power-law property. $\mathcal{B}$ determine the propensity of observing the interaction from cluster c to c'. The extra parameter in the model $\pi_c$ is the prior on the power law structure of cluster c. Note that though these two models are similar in many ways, they still have different properties. For example, the DC-SBM can be approximated through Poisson distribution. But the same approximation is not feasible in our model.

From~\eqref{eq:repthme2}, we can relate the proposed B-VCM to the degree-corrected stochastic block model (DC-SBM). The DC-SBM models adjacency matrices $A=[A_{i,j}]$ of a random graph~$G = (V,E)$. The probability of an edge between node $i$ and $j$ according to the DC-SBM is $\mathbb{E}(A_{i,j})=\theta_{i}\theta_{j} P (B(i),B(j))$.  In~\eqref{eq:repthme2}, node-level parameters~$\theta_i$ are replaced by block-specific vertex propensities~$f_s^{(b)}$ and the likelihood of link between clusters~$P(b, b')$ is replaced by the likelihood of initiating and then interacting with that other cluster $\pi_b B(b, b')$.
Note $\sum_{b'=1}^K P(b,b')$ need not sum to one but~$\sum_{b'=1}^K B(b,b')=1$. 
\end{rmk}

\subsection{Statistical properties of B-VCM}

We next highlight two key statistical properties, the first being network properties of the B-VCM and the second being an inferential consistency property under an approximate local updating rule. 

\subsubsection{Network Properties} 

For an interaction-labeled network~$\bfY$, let $v(\bfY)$ denote the number of non-isolated receivers; $e(\bfY)$ be the number of interactions; $M_k (\bfY)$ is the number of interactions with $k$ receivers; $N_k (\bfY)$ is the number of receivers that appear exactly $k$ times; and $d(\bfY) = (d_k(\bfY))_{k \geq 1}$ is the degree distribution where $d_k(\bfY) = N_k (\bfY) / v(\bfY)$.  These are global statistics that do not depend on the interaction labels.  We define block-specific versions by superscripting each statistic by $b \in [K]$ to denote the statistic on the network~$\bfY_b$, i.e., each interaction restricted to those elements that belong to block~$b$.  For instance, $v^{(b)} (\bfY)$ is the number of non-isolated receivers from block~$b$.  The statistics~$e^{(b)} (\bfY)$, $M^{(b)}_k (\bfY)$, $N^{(b)}_k (\bfY)$, $d^{(b)} (\bfY)$, and $d^{(b)}_k (\bfY)$ are defined similarly.

\begin{defn}[Global and block-level sparsity] 
Let $(\bfY_m)_{m \geq 1}$ be a sequence of interaction-labeled networks for which $e(\bfY_m) \to \infty$ as $m \to \infty$.  The sequence~$(\bfY_m)_{m \geq 1}$ is \emph{sparse} if
$$
\lim \sup_{m \to \infty} \frac{e(\bfY_m)}{v(\bfY_m)^{m_{\bullet} (\bfY_m)}} = 0,
$$
where $m_\bullet (\bfY_m) = e(\bfY_m)^{-1} \sum_{k \geq 1} k M_k (\bfY_m)$ is the average arity (i.e., number of elements) of the interactions in the restricted network~$\bfY_b$.  A non-sparse network is \emph{dense}. We say the sequence $(\bfY_m)_{m \geq 0}$ is $b$-locally sparse if
$$
\lim \sup_{m \to \infty} \frac{e^{(b)}(\bfY_m)}{v^{(b)} (\bfY_m)^{m_{\bullet}^{(b)} (\bfY_m)}} = 0,
$$
where $m_\bullet^{(b)} (\bfY_m) = e^{(b)}(\bfY_m)^{-1} \sum_{k \geq 1} k M^{(b)}_k (\bfY_m)$ is the average arity (i.e., number of elements from block $b$) of the interactions in~$\bfY_m$.  A network that is not $b$-locally sparse is \emph{$b$-locally dense}.
\end{defn}

For~$(X_n)_{n \geq 1}$ a sequence of positive random variables and $(y_n)_{n \geq 1}$ a sequence of positive non-random variables, let $X_n \simeq y_n$ indicate $\lim_{n \to \infty} X_n / y_n$ exists almost surely and equals a finite positive random variable.  
% Given the empirical importance of network sparsity, a critical question is whether network sparsity holds for both the block-specific sub-networks as well as the entire network. 
% In many real world networks the number of the edges in the network grows linearly with respect to the number of the nodes. %One of the examples is the TalkLife network, as shown in Figure~\ref{fig:test}. In each of these small sub-networks, which are subsets of the large network, the degree distribution shows the power-law property. This property is also observed in the large network. 
Theorem~\ref{eq:sparsity} shows the canonical B-VCM model may be either globally and locally sparse and/or dense. 

\begin{thm}%[Sparsity]
\label{eq:sparsity}
Let~$(\bfY_m)_{m \geq 1}$ be a sequence of random exchangeable interaction-labeled networks drawn from the B-VCM. Then, for all~$b \in [K]$, the expected number of non-isolated vertices in $\bfY_m$ belonging to block $b$ satisfies 
$$
E[ v^{(b)} (\bfY_m) ] \simeq \frac{\Gamma(\theta_b + 1)}{\alpha_b \Gamma (\theta_b + \alpha_b)} (\mu_b \cdot m)^{\alpha_b}
$$
where~$\mu_b$ is the mean arity when restricted to elements only from block~$b \in [K]$. Then $1/\mu_b < \alpha_b < 1$ implies $b$-local sparsity for~$b \in [K]$. Furthermore, $E[ v(\bfY_m) ] = \sum_{b=1}^K E[ v^{(b)} (\bfY_m)] \simeq m^{\alpha_\star}$ where~$\alpha_\star = \max_b \alpha_b$. Therefore, $1/\mu < \alpha_\star < 1$ implies global sparsity.
\end{thm}

The proof can be found in Section A.2 of the supplementary materials.  Theorem~\ref{eq:sparsity} establishes that the B-VCM can capture degrees of sparsity.  Note that $\mu_b \leq \mu$ implies a dense network must be $b$-locally dense for all blocks (i.e.~$\alpha_b \leq \alpha_\star \leq 1/\mu \leq 1/\mu_b$).  We next turn to considerations of power-law degree distribution.  %We start with a definition.

\begin{defn}[Global power-law degree distributions]
A sequence $(\bfY_m)_{m \geq 1}$ exhibits \emph{power-law degree distribution} if for some~$\gamma > 1$ the degree distributions $(d(\bfY_m))_{m \geq 1}$ satisfy $d_k (\bfY_m) \sim l(k) k^\gamma$ as $m \to \infty$ for all large $k$ for some slowly varying function $l(x)$; that is, $\lim_{x \to \infty} l(tx) / l(x) = 1$ for all $t > 0$, where $a_m \sim b_m$ indicates that $a_m/b_m \to 1$ as $m \to \infty$.  More precisely, $(\bfY_m)_{m \geq 1}$ has power law degree distribution with index~$\gamma$ if:
\begin{equation}
\label{eq:powerlawdefn}
\lim_{k \to \infty} \lim_{m \to \infty} \frac{d_k (\bfY_m)}{l(k) k^{-\gamma}} = 1.
\end{equation}
We say the sequence has \emph{$b$-local power-law degree distribution} with index~$\gamma_b$ if~\eqref{eq:powerlawdefn} holds with $d_k^{(b)} (\bfY_m)$ replacing the global degree distribution statistic.
\end{defn}

Theorem~\ref{thm:powerlaw} establishes the power-law degree distribution for the block-specific restricted networks built from the B-VCM as well as the global network.
\begin{thm}
\label{thm:powerlaw}
Let~$(\bfY_m)_{m \geq 0}$ obey the B-VCM sequential description in Section~\ref{section:seqdesc}. For each~$m \geq 1$, let $p_m (k) = N_k (\bfY_m)/v(\bfY_m)$ and $p_m^{(b)} (k) = N_k^{(b)} (\bfY_m)/v^{(b)} (\bfY_m)$ for $k \geq 1$ be the empirical global and local degree distributions.  Then, for every~$k \geq 1$, $$
p_m^{(b)} (k) \sim \alpha_b k^{-(\alpha_b + 1)} / \Gamma(1-\alpha_b)
$$
where $\Gamma(t) = \int_0^\infty x^{t-1} e^{-x} dx$ is the gamma function.  That is, $(\bfY_m)_{m \geq 1}$ has $b$-local power law degree distributions with exponents $\gamma_b = 1 + \alpha_b \in (1,2)$ for all $b \in [K]$.  Moreover, $\lim_{k \to \infty} \lim_{m \to \infty} p_m(k) / p_m^{(b_\star)} (k) = 1$ where $b_\star = \arg \max_b \alpha_b$ implying the sequence has global power law degree distribution with exponent $\gamma = 1 + \alpha_\star$. 
\end{thm}

\subsubsection{Consistent recovery of block structure}

In prior sections, the block structure was assumed known and observed as part of the sequential data generating process.  In most real-world settings, the block structure is latent and must be recovered from the observed data. An important theoretical question is whether the estimated block structure is consistent (up to label permutation) with the true block structure.  Theoretical guarantees for DC-SBMs have been well established~\cite{zhao2012consistency, gao2018community, gao2015rate}. 

Here, based on the connection to the DC-SBM as discussed in Remark~\ref{rmk:connectionsbm}, we consider an analogue of a DC-SBM setting~\cite{amini2013pseudo} where the misspecification rate was shown to be bounded as the number of observed vertices in the adjacency graph increases.  In the edge exchangeability framework, however, the asymptotics are with respect to the number of observed interactions leading to different asymptotic behavior.  Specifically, Theorem~\ref{thm:powerlaw} implies that even as the number of interactions grows, a significant fraction of vertices will have low degree.  Therefore, we also bound the misclassification rate as a function of minimal degree. The proof can be found in Section A.1 of the supplementary materials.

\begin{thm}%[Consistency] 
\label{thm:consist}
Let~$(Y_m)_{m \geq 0}$ obey the sequential description in Section~\ref{section:seqdesc} assuming $K=2$, $\alpha_1=\alpha_2$, and $\theta_1=\theta_2$. Let~$B$ denote the true block assignments and~$e_m: [v(\bfY_m)] \mapsto [K]$ denote a current labeling of the non-isolated vertices after observing $m$ interactions. 
Under a symmetry condition provided in Section A.1, an approximate updating rule can be derived such that $\hat B (i) = 1$ if $Deg(i,1) > Deg(i,2)$ and $\hat B(i) = 2$ otherwise, where $Deg(i,b)$ is the number of interactions involving node $i$ and an element from block~$b$ (as determined by the current labeling $e_m$).  Let 
$$
M_{v(\bfY_m)}= \inf_{\rho':[K]\mapsto [K]} \frac{1}{v(\bfY_m)}\sum_{i=1}^{v(\bfY_m)} 1(\hat{B} (i) \neq \rho' B (i))
$$ 
denote the misclassification rate (under potential label switching). As $m\rightarrow\infty$, for any $u \ge\frac{1}{e}$, under certain conditions on the current labeling provided in Section A.1 and the approximate updating rule we have
$$\mathbb{P}[M_{v(\bfY_m)}\ge euP_{out}]\le \exp\left[-ev(\bfy_m)P_{out}u\log u\right]$$
where $e$ is Euler's number, $u > 1/e$ is a constant, and~$P_{out} \to \sum_{d=1}^\infty \alpha B(d, \alpha+1) \exp (- d \mu_{\min}^2 /4)$ as $m \to \infty$ where $\mu_{\min}$ is a constant depending on the current labeling. Specifically, restricting the misclassification rate to nodes of at least degree~$D_m$ denoted~$M_{v(\bfY_m)}^{D_m}$, it is possible to construct a sequence $D_m$, such that $|\{ i \in v(\bfY_m) : \text{Deg}(i) \ge D_m\}|/|v(\bfY_m)|$ is non-empty with probability one and 
$$\lim_{m(Y_m)\rightarrow\infty}\mathbb{P}[ M_{v(\bfY_m)}^{D_m} \ge \epsilon]= 0, \quad \text{for all } \epsilon > 0.$$ 
\end{thm}

\section{Inference}

Theorem~\ref{thm:consist} shows that a simple updating rule can lead to consistent estimation of the latent block structure for high degree nodes. Figure 2 in Section A.5 of the supplementary materials shows empirical misclassification rate decreases rapidly as degree increases in a simulation study.  The theory and empirical evidence thus suggests a potential large misclassification rate for low degree nodes.  This uncertainty motivates the consideration of a Bayesian framework to assess posterior uncertainty.  Specifically, we introduce a Gibbs sampling algorithm for sampling from the posterior distribution of $(\Psi, B)$ given observed interaction-labeled network $\bfY_m$ with latent block structure.  We start with the discussion of the selection of priors, followed by a brief overview of the Gibbs sampling algorithm. 
% We then proceed into the accurate estimate of the power-law parameters and the recoverability of the cluster assignment. At the end of this section, we propose one potential model selection criteria based on the marginal log likelihood. 

\subsection{Selection of Priors}

% The likelihood shown in Eq~\eqref{eq:jointform1} doesn't have closed form solution for the power-law parameters $\{\alpha_b\}$ and $\{\theta_b\}$, the entries in the propensity matrix $\B$, and the latent block assignment $\{\mathbf{C}_m\}$ for each nodes. Therefore, the Bayesian framework was applied to estimate the posterior of these model parameters. 
Here we outline an approach to choosing priors for the B-VCM parameters that admit conjugate Gibbs updates.  First, consider the estimated block assignment $\hat B$. An equivalent mixture representation of the sequential description leads to the natural prior:
$$\hat B(i) \mid \pi \sim \text{Multinomial}(\pi_1,...,\pi_K); \quad \pi:= (\pi_1,...,\pi_K) \sim \text{Dirichlet}(\omega)$$
where $\omega > 0$ is a constant, i.e., $\pi$ follows a symmetric Dirichlet distribution. 
%In terms of the number of cluster $K$, a model selection criteria based on likelihood can be utilized to determine the number of clusters as discussed in Section 4.3.  
Next, considering equation~\eqref{eq:repthme2}, based on a similar argument and the fact that the entries in the same row of propensity matrix $\mathcal{B}$ should sum up to 1, we propose a row-wise Dirichlet prior for the propensity matrix:
$$(\B(b,1)\ldots \B(b,K))\sim \text{Dirichlet}(\zeta),\ b\in[K].$$
This choice of prior does not enforce symmetry, i.e. $\B (b,b')$ may not equal $\B (b',b)$. A discussion of a prior under a symmetry constraint can be found in Section A.9 of the supplementary materials.

%To accommodate the assumption that the within cluster connection is stronger than between cluster assumption, it is reasonable to set a higher propensity for the diagonal elements in the prior than the off-diagonal elements. 

For the block-specific parameters~$\{ \alpha_b, \theta_b \}$, conjugate priors given block assignments~$B$ exists.  Specifically, a conjugate prior for the power-law parameters~$\alpha_b$ is a Beta distribution. For datasets of reasonable size, empirically we have found that the choice of hyper-parameters does not significantly affect the resulting posterior distributions. In the subsequent examples, the size of the datasets was more than sufficient to not be strongly affected by the choice of global priors.  For the power-law parameter, this suggests using~$\alpha_b \sim \text{Beta} (1, 1)$, i.e., the Uniform distribution, for each~$b \in [K]$. For the other parameter~$\theta_b$, we find Gamma(1,1) is an appropriate prior in both the simulations and the real world data after applying our method to the Talklife data. An alternative approach is to fit the Hollywood model~\cite{crane2016edge} to the entire dataset, and get the estimate $\hat{\theta}$. In TalkLife, we found that $\hat{\theta}$ was between 1 and 2. Therefore, $\text{Gamma}(\hat{\theta},1)$ is also a reasonable choice of prior such that the mean is centered at the global fit. %we follow~\cite{dempsey2021hierarchical} and choose a high-variance Gamma distribution.  Specifically, we have found that $\theta_b \sim Gamma(1, 1000)$ is an appropriate diffuse prior that allows for fast mixing.  An alternative approach would be to fit the hollywood model~\cite{crane2016edge} to the entire dataset, and use $\text{Gamma}(\hat \theta/100, 100)$ as a prior for the $\theta_b$ (i.e., a diffuse prior centered at the global fit).  
% In terms of the priors of power-law parameters $\{\alpha_b\}$ and $\{\theta_b\}$, let $\alpha_b \sim P(\mu)$, $\theta_b\sim P(\delta)$, $b\in[K]$. A reasonable choice could be a Beta distribution and a Gamma distribution. 
%\textcolor{red}{Yuhua: how did you choose the hyperparameters for theta?  A diffuse prior?} \textcolor{blue}{Gamma($\hat{\theta}$/10,100) in Talklife data, Gamma(1,1) in simulation data.}

\subsection{Gibbs Sampling Algorithm}
\label{sec:alg}

Here we introduce a Gibbs sampling algorithm for sampling from the posterior distribution of~$(\Psi, B)$ given an observed interaction-labeled network~$\bfY_m$ with latent block structure~$B$.  The algorithm uses auxiliary variable methods~\cite{West1995} to perform conjugate updates for all parameters~$\Psi$.  Pseudo-code is provided in~Algorithm~\ref{alg:cap}.
\begin{algorithm}
\caption{Pseudo-code of Gibbs-sampling based algorithm}\label{alg:cap}
\begin{algorithmic}
\State Specify $\omega$, $\zeta$, $\delta$, $\mu$, $K$
\State Initiate $\B$, $\{\alpha_b\}$, $\{\theta_b\}$, $\{ B(i) \}$ 
\For{\text{iterations}}
    \For{$i \in v(\mathcal{Y}_m)$}
        \State Update $B(i)$ according to Eq (\ref{eq:c})
    \EndFor
    \For{$b\in [K]$}
        \State Update $\alpha_b$ using auxiliary variables according to Eq (\ref{eq:alpha_c})
        \State Update $\theta_b$ using auxiliary variable according to Eq (\ref{eq:theta_c})
    \EndFor
    \For{$b\in [K]$}
        \State Update $\B(b,1)\ldots B(b,K)$ according to Eq (\ref{eq:prop_mat})
    \EndFor
\EndFor
\end{algorithmic}
\end{algorithm}

In Algorithm~\ref{alg:cap}, we first sequentially update block assignments, $\{B (i)\}_{i \in v(\bfY_m)}$. Let $\H_m^{(b)}= v(\bfY_m) \cap B^{-1}(b)$ denote all observed vertices in block~$b \in [K]$.  Then the probability of assigning $i \in v(\bfY_m)$ to block $b \in [K]$ given $\Psi$ and all other block assignments $\{ B(i')\}_{i' \not = i}$ is proportional to:
\begin{equation}
    \label{eq:c}
p_b := \underbrace{\mathscr{B}(\omega+L)}_{(\rom{1})}\times \underbrace{\frac{D_m(i)!}{\prod_b D_m(i,b)!}\prod_{b'=1}^{K}\B(b,b')^{D_m(i,b^\prime)}}_{(\rom{2})}\times \underbrace{P(\{\H_m^{(b)}\}|\alpha_b,\theta_b,\delta,\mu)}_{(\rom{3})}
\end{equation}
where $L=\{L_1,...,L_K\}$ is the counts of the interactions initiated from block $1$ to $K$ respectively, $\mathscr{B}(\cdot)$ is the Beta function, $D_m(i)$ is the degree of node i, and $D_m(s,b)$ is the number of interactions node $i$ has with block~$b$.
Term~($\rom{1}$) is derived by integrating over the latent parameter $\pi$ in the distribution $P(\{B(i)\}|\pi,\omega)$. 
% $$P(\{C^{(s)}_m\}|\omega)=\int_{\pi}P(\{C^{(s)}_m\}|\pi)P(\pi|\omega)d\pi=$$
Term~($\rom{2}$) is derived by considering nodes that have connections with the node $i$ that is being updated in the current iteration.  
% $$P(\B(C^{(s)}_m=b,)|\zeta)\propto \frac{D_m(s)!}{D_m(s,1)!...,D_m(s,K)!}\prod_{b'=1}^{K}\B(b,b')^{D_m(s,b^\prime)}$$
Term~($\rom{3}$) takes the explicit form as shown in Part $\rom{2}$ of Eq~\eqref{eq:jointform1}. 
% Normalizing over all $b \in [K]$, the conditional probability of assigning the node to block $b$ with probability
% \begin{equation}\label{eq:c}
% P( B(i) = b |\alpha_b,\theta_b,\B, \omega, \{ B(i') \}_{i' \not = i} ) = \frac{p_b}{\sum_{k=1}^{K}p_{k}}.
% \end{equation} 

The conditional updates of $\alpha_b$ and $\theta_b$ is shown as follows. Given the block assignment $\{B(i)\}$, the values of $\alpha_b$ and $\theta_b$ can be sampled from the posterior distribution. 
% First, consider the $\theta_b$:
% $$P(\theta_b|\{\mathbf{C}_m\},\delta)\propto\frac{[\theta_b+\alpha_b]^{v(\mathbf{y}_b)-1}_{\alpha_b}}{[\theta_b+1]^{m(\mathbf{y}_b)-1}_{1}}P(\theta_b|\delta)$$
Here, auxiliary variable methods \cite{teh2006bayesian} are used in a similar way as in prior inferential approaches for edge exchangeable model \cite{dempsey2021hierarchical}. 
% Let $\{\alpha_b^{*}\}$ and $\{\theta_b^*\}$ be the values of the power-law parameters from the previous Gibbs updating iteration, and $\{a,b,c,d\}$ be the pre-specified constants. 
Then, from Eq~\eqref{eq:jointform1}, we have
\begin{align}
x_b &\sim Beta \left(\theta_{b}+1,m(\mathbf{Y}_{b})-1 \right), b \in [K] \nonumber \\
y_{i,b}&\sim Bernoulli \left( \frac{\theta_{b}}{\theta_{b}+\alpha_b \cdot i} \right), 
i =1,\ldots,v(\bfY_b)-1, b\in[K] \nonumber \\
z_{s,j,b} &\sim \text{Bernoulli} \left( \frac{j-1}{j-\alpha_b} \right),  
s \in v (\bfY_b), j = 1,\ldots, D_m (s,b)-1 \nonumber \\
\theta_b &\sim \text{Gamma} \left(\sum_{i=1}^{v(\bfY_b)-1}y_{i,b}+a,b-log x_b \right) \label{eq:theta_c} \\
\alpha_b &\sim \text{Beta}\left(b+\sum_{i=1}^{v(\mathbf{Y}_{b})-1}(1-y_{i,b}),d+\sum_{s\in v(\bfY_b)}\sum_{j=1}^{D_m(b,s)-1}(1-z_{s,j,b}) \right) \label{eq:alpha_c}
\end{align}
where~$\{a,b,c,d\}$ are the hyperparameters for the Gamma and Beta priors respectively.
The difference between this algorithm and~\cite{dempsey2021hierarchical} is that the prior work considered hierarchical structure while here we consider latent non-overlapping communities.  
% $$\frac{1}{[\theta_b+1]^{m(\mathbf{y}_{b})-1}}=\frac{1}{\Gamma(m(\mathbf{y}_{b})-1)}\int_{0}^{1}x^{\theta_b}(1-x)^{m(\mathbf{y}_{b})-2}dx$$
% where $x$ is an intermediate variable that can be sampled given $\{\theta_b^{*}\}$, such that $x\sim Beta(\theta_{b^*}+1,m(\mathbf{y}_{b})-1)$. Similarly, 
% $$[\theta_b+\alpha_b]^{v(\mathbf{y}_{b})-1}_{\alpha_b}=\prod_{i=1}^{v(\mathbf{y}_{b})-1}\sum_{y_i\in[0,1]}\theta_b^{y_i}(\alpha_b*i)^{1-y_i}$$
% where $y_i\sim Bernoulli(\frac{\theta_{b}^*}{\theta_b^*+\alpha_b^{*} *i})$. Eventually, the posterior of $\theta_b$ is given by 
% \begin{equation}
% \label{eq:theta_c}
    % \theta_b|C_m\sim \text{Gamma}(\sum_{i=1}^{v(\mathbf{y}_{c})-1}y_i+a,b-log x)
% \end{equation}.
% Similarly, the posterior of $\alpha_b$ is given by \begin{equation}
% \label{eq:alpha_c}
% \alpha_b|C_m\sim \text{Beta}(b+\sum_{i=1}^{v(\mathbf{y}_{b})-1}(1-y_i),d+\sum_{s_c}\sum_{j=1}^{D_m(c,s)-1}(1-z_{s_c,j}))\end{equation} 
% where $y_i\sim \text{Bernoulli}(\frac{\theta_b}{\theta_b+\alpha_b*i})$ and $z_{s_c,j}\sim \text{Bernoulli} (\frac{j-1}{j-\alpha_c^*})$. 
% To summarize, the algorithm first draws the values of intermediate variables $x$, $y$, and $z$ based on the previous round values of $\alpha_b^*$ and $\theta_b^*$. Given the values of these variables, we draw the values of the updated $\alpha_b$ and $\theta_b$ according to Eq~\eqref{eq:alpha_c} and Eq~\eqref{eq:theta_c}.
Last, given the block assignment~$\{ B(i)\}$, the row-wise update of the propensity matrix follows from the conjugate Dirichlet prior:
\begin{equation}
\label{eq:prop_mat}
(\B(b,1)\ldots\B(b,K))\sim \text{Dirichlet}(D_m(b,1)+\zeta_{b,1}\ldots D_m(b,K)+\zeta_{b,K})
\end{equation}
for each~$b \in [K]$. Convergence of Algorithm~\ref{alg:cap} can be checked via traceplots and, in our experiments, occurs within the first hundred or so iterations; see Figure 8 and Figure 9 in the supplementary materials for trace plots from our simulation and case study. 

%Given the cluster allocation $c\in[K]$, we have the row-wise update of the propensity matrix taking the following form: 

%\begin{equation}
%    {\B}(c,)|C_m\sim \text{Dirichlet}(\nu+\vec *) 
%\end{equation}
%where $\vec *$ is the count of interactions between cluster $c$ and cluster $\{1,...,K\}$. The propensity matrix $\B$ can be either symmetric or not, the possible solutions to guarantee the symmetry of the propensity matrix are briefly discussed in the \textcolor{yellow}{Appendix 7.4}.

\begin{rmk}[Scalability]
\label{rmk:scalable}
Algorithm~\ref{alg:cap} may not scale well to large networks.  
% Specifically, there are loops over all nodes and blocks.  
Given $K \ll |v(\bf Y)|$, the critical issue in scaling up our proposal is finding good approximate algorithms for updating block assignments.  Equation~\eqref{eq:c} suggests an approximate sampling algorithm that relies on small, local node neighborhoods.  Moreover, Theorem~\ref{thm:consist} suggests high degree nodes may have low posterior uncertainty.  Both suggest a path forward via a parallel, approximate algorithm that may perform well in practice.  We do not pursue this here, but we consider this important future work.
\end{rmk}

\section{Simulations}
\label{sec:simu}

In this section, we demonstrate via simulation that Algorithm~\ref{alg:cap} performs empirically well in recovering network properties. These include (1) recoverability of the block assignment and (2) accurate estimation of the power-law parameters. At the end of this section, we propose and empirically assess a model selection criteria to identify the number of latent communities, $K$. 

\subsection{Recovering Block Assignment Labels}
\label{sec:recover}

% Theorem \ref{thm:consist} establishes a bound on the mis-specification rate, which decays rapidly to zero when restricting to nodes of increasing minimum degree. To empirically assess this, a generative model as described in Section \ref{section:seqdesc} is used to simulate various networks to assess our ability to recover the correct community labels of each node. We show in simulation that the posterior concentrates on the correct block assignment labels when only a finite number of interactions are observed. 

In this section, we assume the number of blocks is $K=2$. The standardized $L_2$ norm is used as a distance metric between true block assignment and the posterior distribution over block assignments.  
% Recall that for a network consisting of $m$ interactions $\bfY_m$, $v(\bfY_m)$ is the number of nodes in the network. 
In the $i$th iteration of the Gibbs sampler, let the inferred block assignment be $\hat{B}_{m}^{(i)}\in\{0,1\}^{v(\mathcal{Y}_m)}$ and $B_m\in\{0,1\}^{v(\mathcal{Y}_m)}$ be the true block assignment. Define the standardized $L_2$ norm to be:
$$
\mathcal{L}_m = \frac{1}{\sqrt{v(\mathbf{Y}_m)}} \left \|B_m-\frac{1}{n}\sum_{i=1}^{n}{\hat{B}_{m}^{(i)}} \right \|_2
$$
where $n$ is the total number of Gibbs iterations. The norm is standardized by dividing the number of nodes in the network such that the minimum (0) and maximum (1) values represent entirely correct and incorrect node assignments respectively. Note that the block labels are arbitrary; that is, when calculating the $\mathcal{L}_m$, one must be concerned with \emph{label switching} \cite{jasra2005markov}.  Given a complete random guess, the $\mathcal{L}_m$ norm will be approximately 0.5 as $m$ increases. To account for potential label switching, the $\mathcal{L}_m$ norm was calculated for the inferred block assignment $\hat B_m^{(i)}$ as well as it's conjugate $1-\hat B_m^{(i)}$ and the minimum of the two norms was taken.  This leads to a range of 0 to 0.5 for the estimated distance, such that the closer the distance is to 0, the better the recovery of the community structure. 

In our simulations, we consider simulated datasets with different propensity matrices and power-law parameters. We assume the power-law parameters are the same across different blocks, i.e. $\alpha_1=\alpha_2$, and we fix $\theta_1=\theta_2=5$. The number of interactions in each simulated network vary from 1000, 2500, to 10000. The probability of an interaction initiated from either block is the same, i.e. $\pi_1=\pi_2=0.5$. We repeated the simulation in each setting 20 times.  The 
mean values of the $L_2$ norm in different settings are shown in Table~\ref{tab:l2norm_same}. Two conclusions based on the results are: (1) given the same value of the power-law parameter $\alpha$ and the size of the network, the lower the inter-community connection rate, the better the recovery of the block assignments; (2) as the number of observations increase, the block assignments become more accurate in almost all settings. 

\begin{table}[!th]
\centering
%\begin{adjustwidth}{-1.8cm}{}
\begin{tabular}{SSSSS} \toprule
    & {\# Interactions} &{$\B=\{0.1,0.9\}$}  & {$\B=\{0.3,0.7\}$} &{$\B=\{0.5,0.5\}$} \\ \midrule
    {$\alpha=\{0.1,0.1\}$} & 1,000  & {0.043 (0.016)} & {0.183 (0.033)} & {0.458 (0.028)} \\ 
      &2,500 & {0.067 (0.096)} & {0.161 (0.042)} & {0.457 (0.023)} \\
       & 10,000  & {0.053 (0.072)} & {0.226 (0.139)} & {0.454 (0.018)}\\
      \midrule

      {$\alpha=\{0.3,0.3\}$} &{1,000}  & {0.075 (0.013)} & {0.255 (0.067)} & {0.475 (0.020)} \\ 
      &{2,500} & {0.067 (0.008)} & {0.232 (0.058)} & {0.465 (0.037)}\\
       & {10,000}  & {0.058 (0.007)} & {0.232 (0.068)} & {0.462 (0.024)} \\
      \midrule

    {$\alpha=\{0.5,0.5\}$} & {1,000} & {0.104 (0.013)}   & {0.303 (0.028)} &{0.466 (0.035)} \\ 
     &{2,500} & {0.101 (0.008)} & {0.298 (0.044)} & {0.476 (0.019)} \\
     & {10,000} & {0.089 (0.006)}  & {0.295 (0.047)} & {0.474 (0.025)} \\
    \midrule

    {$\alpha=\{0.7,0.7\}$} &{1,000}  & {0.184 (0.041)} & {0.382 (0.039)} & {0.474 (0.030)} \\ 
      &{2,500} & {0.156 (0.012)} & {0.378 (0.022)} & {0.490 (0.009)}\\
       & {10,000}  & {0.143 (0.008)} & {0.348 (0.014)} & {0.480 (0.023)} \\
      \midrule

    {$\alpha=\{0.9,0.9\}$} &{1,000}   & {0.388 (0.068)}  & {0.482 (0.023)} & {0.484 (0.032)} \\ 
    &{2,500}  & {0.335 (0.034)} & {0.465 (0.023)} & {0.488 (0.012)} \\
     & {10,000} & {0.279 (0.028)}  & {0.454 (0.017)} & {0.489 (0.011)} \\ 
     \bottomrule
\end{tabular}
%\end{adjustwidth}
 \caption{Standardized L2 norm of the inferred block assignments and the underlying truth in different settings. For each set with the same values of the connectivity propensity $\B$, the power-law parameters $\{\alpha_1,\alpha_2\}$, and the number of interactions ($\{1,000, 2,500, 10,000\}$), we repeated the simulation 20 times. The mean values (SD) over 20 simulations are shown here.} 
 \label{tab:l2norm_same}
\end{table}

\begin{figure}[!th]
\centering
\begin{subfigure}{0.3\textwidth}
  \centering
  \includegraphics[width=.9\linewidth]{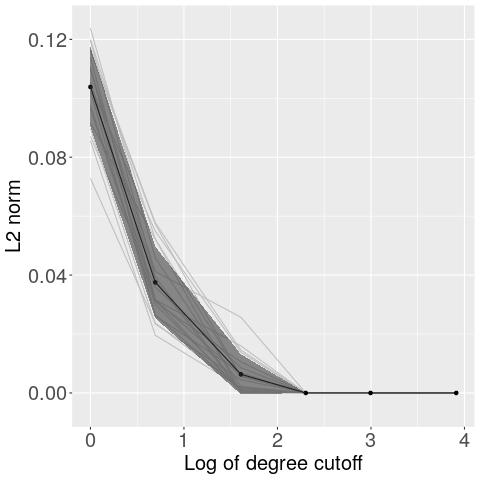}
  \caption{}
  \label{fig:sub.5.2.1.1}
\end{subfigure}%
\begin{subfigure}{0.3\textwidth}
  \centering
  \includegraphics[width=.9\linewidth]{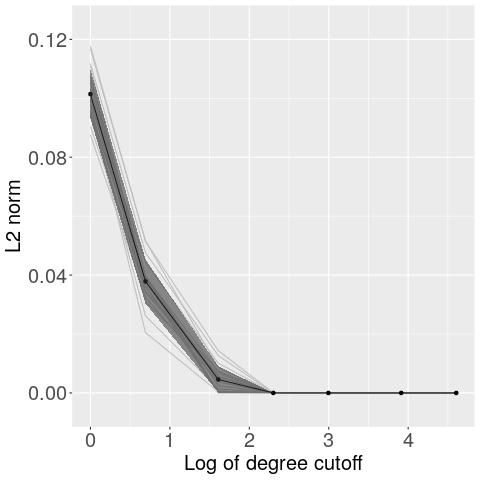}
  \caption{}
  \label{fig:sub.5.2.1.2}
\end{subfigure}
\begin{subfigure}{0.3\textwidth}
  \centering
  \includegraphics[width=.9\linewidth]{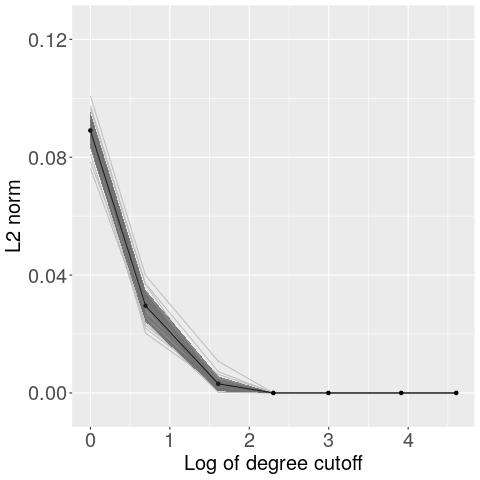}
  \caption{}
  \label{fig:sub.5.2.1.3}
\end{subfigure}
\caption{The $L_2$ norm as a function of the degree cutoff, in the setting where $\alpha_1=\alpha_2=0.5$, $\B(1,1)=\B(2,2)=0.9$, with (a) 1,000, (b) 2,500, and (c) 10,000 interactions presented in the network. The solid dark line is the average $L_2$ norm over 20 replications. The $L_2$ norm is calculated based on the nodes whose degree is greater or equal to the degree cutoff as indicated by the x-axis. The grey shadow indicates the SD.}
\label{fig:spg_high_deg}
\end{figure}

To empirically demonstrate the conclusion from Theorem \ref{thm:consist} on high-degree nodes, we calculated the $L_2$ norm as a function of the degree cutoff, as shown in Figure~\ref{fig:spg_high_deg}. The visualization is based on the setting where $\alpha_1=\alpha_2=0.5$, and the within block interaction propensity $\B(1,1)=\B(2,2)=0.9$. And the $L_2$ norm is calculated with nodes whose degree is greater or equal to the value indicated by the x-axis. Regardless of the size of the observed network, the higher the degree, the smaller the $L_2$ norm, which indicates a lower misspecification rate. Meanwhile, as the number of interactions increases, the $L_2$ norm decreases to 0 given the same degree cutoff. See Section A.5 of the supplementary materials for similar plots for the misspecification rate as a function of the degree cutoff, which yields similar conclusions. 
% Further, we compare the results in the other settings with a certain degree cutoff value that is proportional to $m(\mathbf{y}_m)^{\alpha}$. The results are shown in Table~\ref{tab:l2norm_same_deg} in the Appendix.

Note that the $L_2$ norm is not applicable when there are multiple blocks ($K>2$). To overcome this potential problem, we also use the Cross Entropy Loss as a criteria to evaluate the accuracy of the block assignment (See Table 3 in Section A.8 of the supplementary materials), which gives similar conclusions. Results from simulation settings where $\alpha_1\neq\alpha_2$ are presented in Table 4 in Section A.7.

\subsubsection{Estimates of power-law parameters}
\label{sec:pl_est}
Another important question is whether the posterior distributions adequately concentrate around the true block-specific parameters $\{ \alpha_b \}_{b \in [K]}$ which controls the within-block sparsity and power-law structure. Following the same generative steps as described in Section \ref{section:seqdesc}, we simulate various networks to assess our ability to accurately estimate these parameters.

In the generative model, we set the number of blocks $K$ equal to 2 and choose $\{\alpha_1,\alpha_2\}$ to be from the set $\{\{0.1,0.9\}$, $\{0.2,0.8\}$, $\{0.3,0.7\}$, $\{0.4,0.6\}\}$ and set $\theta_1=\theta_2=5$. %The  within-group interaction propensity is set to 0.9 (i.e., $a: = \B(1,1)=\B(2,2)=0.9$), and between-group is set to 0.1 (i.e., $b := \B(1,2)=\B(2,1)=0.1$). 
Blocks are assumed equally likely to initiate an interaction (i.e., $\{\pi_1,\pi_2\}=\{0.5,0.5\}$). Networks of size 1,000, 10,000 and 100,000 were simulated, repeated 20 times. Given the priors specified in the previous section, the means of the posterior of the power-law parameters and the entries in propensity matrix are shown in Table~\ref{tab:powerlaw}.  

First, for fixed within-block propensities and sample size, estimation accuracy is higher for larger power-law parameter values.
For example, when $m=1000$, $\{ a, b \} = \{0.9, 0.1\}$ and $\{\alpha_1,\alpha_2\}=\{0.3,0.7\}$, the power-law estimate for $\alpha_2 = 0.7$ was more accurate than the estimate for $\alpha_1 = 0.3$ (highlighted in red).  This statement empirically holds even when latent communities are known. Second, power-law estimation accuracy degrades as the inter-community connection rate increases. For example, in the 1000 interaction setting where $\{\alpha_1,\alpha_2\}=\{0.2,0.8\}$, the power-law estimate decreases in accuracy as the likelihood of inter-community connections goes up from $0.1$ to $0.5$ (highlighted in yellow). 
Third, estimation accuracy increases as the number of interactions increase. When $\alpha_1=0.1$, and $\{a,b\}=\{0.9,0.1\}$, for example, the power-law estimate becomes more accurate as the sample size grows from $N=1,000$ to $100,000$ (highlighted in blue). In Section A.10 of the supplementary materials, simulation results are presented for varying values of $\{\theta_1,\theta_2\}$.

\begin{table}[!th]
\begin{adjustwidth}{-2.2cm}{}
\vspace*{0cm}
\begin{tabular}{SSSSSS} \toprule
    {\textbf{1,000 interactions}} 
    &{Parameters}&{$\{\alpha_c\}=\{0.1,0.9\}$}  & {$\{\alpha_c\}=\{0.2,0.8\}$} &{$\{\alpha_c\}=\{0.3,0.7\}$} & {$\{\alpha_c\}=\{0.4,0.6\}$} \\ \midrule
    {$\{a,b\}=\{0.9,0.1\}$} &{$\alpha_1$}  & {\textcolor{blue}{\textbf{0.301 (0.182)}}}& {\textbf{\textcolor{yellow}{0.296 (0.116)}}} & {\textbf{\textcolor{red}{0.374 (0.08)}}}& {\textbf{0.449 (0.083)}}  \\ 
      &{$\tilde{\alpha}_1$ }  & {0.200 (0.100)} & {0.274 (0.084)} & {0.354 (0.068)}& {0.460 (0.086)}  \\ 
      &{$\alpha_2$}& {\textbf{0.886 (0.079)}} & {\textbf{0.804 (0.022)}}& {\textbf{\textcolor{red}{0.712 (0.037)}}} & {\textbf{0.612 (0.056)}} \\
       &{$\tilde{\alpha}_2$ }  & {0.904 (0.015)} & {0.805 (0.021)} & {0.714 (0.037)}& {0.604 (0.057)} \\
       & {Diagonal} & {0.907 (0.020)}& {0.905 (0.019)} & {0.903 (0.019)} & {0.898 (0.018)} \\
      \midrule
    {$\{a,b\}=\{0.7,0.3\}$} &{$\alpha_1$}  & {\textbf{0.335 (0.196)}}& {\textbf{\textcolor{yellow}{0.335 (0.139)}}}& {\textbf{0.412 (0.143)}}& {\textbf{0.501 (0.099)}} \\ 
      &{$\tilde{\alpha}_1$ }  & {0.226 (0.094)}& {0.256 (0.095)} & {0.384 (0.113)} &  {0.470 (0.100)} \\ 
      &{$\alpha_2$}& {\textbf{0.886 (0.069)}} & {\textbf{0.801 (0.029)}} & {\textbf{0.704 (0.061)}} & {\textbf{0.604 (0.064)}} \\
       &{$\tilde{\alpha}_2$ }  & {0.901 (0.015)}& {0.806 (0.025)} & {0.712 (0.061)} & {0.617 (0.058)} \\
       & {Diagonal} & {0.709 (0.031)} & {0.710 (0.028)} & {0.708 (0.030)} & {0.708 (0.034)} \\
      \midrule
    {$\{a,b\}=\{0.5,0.5\}$} &{$\alpha_1$}  & {\textbf{0.627 (0.162)}} & {\textbf{\textcolor{yellow}{0.624 (0.185)}}} & {\textbf{0.613 (0.115)}} & {\textbf{0.523 (0.124)}} \\ 
      &{$\tilde{\alpha}_1$ }  & {0.212 (0.096)}& {0.290 (0.079)} & {0.368 (0.078)}& {0.479 (0.105)} \\ 
      &{$\alpha_2$}& {\textbf{0.800 (0.267)}} & {\textbf{0.726 (0.125)}} & {\textbf{0.604 (0.112)}} & {\textbf{0.555 (0.106)}} \\
       &{$\tilde{\alpha}_2$ }  & {0.898 (0.017)}& {0.804 (0.025)} & {0.710 (0.030)} & {0.601 (0.064)}  \\
       & {Diagonal} & {0.521 (0.042)} & {0.532 (0.051)} & {0.540 (0.053)} & {0.526 (0.050)} \\
     \bottomrule
    {\textbf{10,000 interactions}} 
    &{Parameters}&{$\{\alpha_c\}=\{0.1,0.9\}$}  & {$\{\alpha_c\}=\{0.2,0.8\}$} &{$\{\alpha_c\}=\{0.3,0.7\}$} & {$\{\alpha_c\}=\{0.4,0.6\}$} \\ \midrule
    {$\{a,b\}=\{0.9,0.1\}$} &{$\alpha_1$}  & {\textcolor{blue}{\textbf{0.201 (0.089)}}} & {\textbf{0.273 (0.083)}}& {\textbf{0.336 (0.051)}}& {\textbf{0.426 (0.042)}} \\ 
      &{$\tilde{\alpha}_1$ }  & {0.175 (0.063)}& {0.244 (0.063)} & {0.336 (0.043)}& {0.420 (0.039)} \\ 
      &{$\alpha_2$}& {\textbf{0.899 (0.006)}} & {\textbf{0.799 (0.011)}}& {\textbf{0.700 (0.017)}} & {\textbf{0.605 (0.020)}} \\
       &{$\tilde{\alpha}_2$ }  & {0.900 (0.006)}& {0.800 (0.011)} & {0.701 (0.017)} & {0.607 (0.019)} \\
       & {Diagonal} & {0.900 (0.006)}& {0.900 (0.006)} & {0.901 (0.006)} & {0.900 (0.006)} \\

     \bottomrule
    {\textbf{100,000 interactions}} 
    &{Parameters}&{$\{\alpha_c\}=\{0.1,0.9\}$}  & {$\{\alpha_c\}=\{0.2,0.8\}$} &{$\{\alpha_c\}=\{0.3,0.7\}$} & {$\{\alpha_c\}=\{0.4,0.6\}$} \\ \midrule
    {$\{a,b\}=\{0.9,0.1\}$} &{$\alpha_1$}  & {\textcolor{blue}{\textbf{0.157 (0.063)}}}& {\textbf{0.242 (0.054)}}& {\textbf{0.322 (0.034)}}& {\textbf{0.413 (0.025)}} \\ 
      &{$\tilde{\alpha}_1$ }  & {0.136 (0.045)} & {0.220 (0.040)} & {0.319 (0.027)} & {0.412 (0.022)}  \\ 
      &{$\alpha_2$}& {\textbf{0.900 (0.002)}}& {\textbf{0.799 (0.004)}} & {\textbf{0.700 (0.006)}}& {\textbf{0.604 (0.012)}} \\
       &{$\tilde{\alpha}_2$ }  & {0.900 (0.002)} & {0.799 (0.004)} & {0.700 (0.006)} & {0.605 (0.011)}  \\
       & {Diagonal} & {0.900 (0.002)} & {0.900 (0.002)} & {0.900 (0.002)} & {0.900 (0.002)} \\

     \bottomrule
\end{tabular}
\end{adjustwidth}
\caption{Posterior means (SD) of power-law parameters $\{\alpha_c\}$ and within/between block propensity $\{a,b\}$ in different settings. To assess uncertainty due to latent communities, we also infer the posteriors of power-law parameters given true block assignments, denoted $\tilde{\alpha}_c$, $c\in[K]$. Three main conclusions are highlighted in red, yellow, and blue correspondingly.} \label{tab:powerlaw}
\end{table}

\subsubsection{The selection of K}
\label{sec:slct_K}
So far we have assumed the number of communities $(K)$ known. Here we address the selection of K. Prior solutions include but are not limited to likelihood ratio test (LRT) methods (\cite{vuong1989likelihood}, \cite{wang2017likelihood}), cross validation (CV) methods (\cite{grun2011topicmodels},\cite{grimmer2010bayesian}), and Bayesian selection criteria methods (\cite{yan2016bayesian},\cite{teh2007collapsed}). Though applicable in many cases, these methods do not fit well into our current setting. LRT methods require a good distributional approximation, which is hard to find in our setting. CV methods lead to additional sampling issues and lack interpretability; and Bayesian selection criteria can be computationally expensive.

Here, we consider a model selection criteria based on the marginal posterior maximization that is similar to techniques used in the topic modeling~\cite{taddy2012estimation}. The number of blocks is determined by choosing $K$ that maximizes the marginal likelihood.
% $P(\bfY_m|\{\alpha_b, \theta_b\}, \{ B(i) \} \in [K]^{(v(\bfY_m)})$.  
We simulate 10,000 interactions and set the number of blocks K to be 3, 5, and 10 correspondingly. The true values of $\alpha_b$ are randomly chosen from $\text{Uniform}(0.4,0.8)$, and $\{\theta_b\}$ are all set to be 5. The intra-block connectivity is set to be $a=0.9$, while the inter-block connectivity is set to be $b=\frac{0.1}{K-1}$. We calculate the marginal log likelihood for different K and repeat the procedure 20 times. 

\begin{figure}
%\centering
\begin{subfigure}{0.3\textwidth}
  \centering
  \includegraphics[width=1\linewidth]{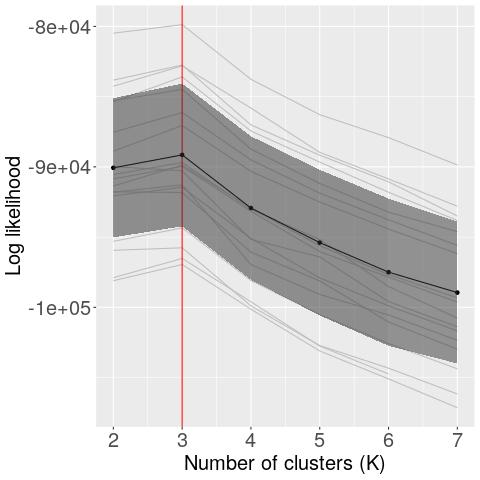}
      \caption{}
%  \caption{B=0.8,random selected alpha, varying K}
  \label{fig:sub1}
\end{subfigure}
\begin{subfigure}{0.3\textwidth}
  \centering
  \includegraphics[width=1\linewidth]{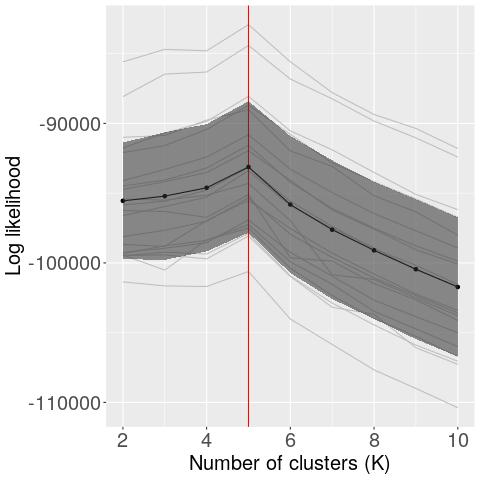}
      \caption{}
%  \caption{B=0.8,random selected alpha, varying K}
  \label{fig:sub2}
\end{subfigure}
\begin{subfigure}{0.3\textwidth}
  \centering
  \includegraphics[width=1\linewidth]{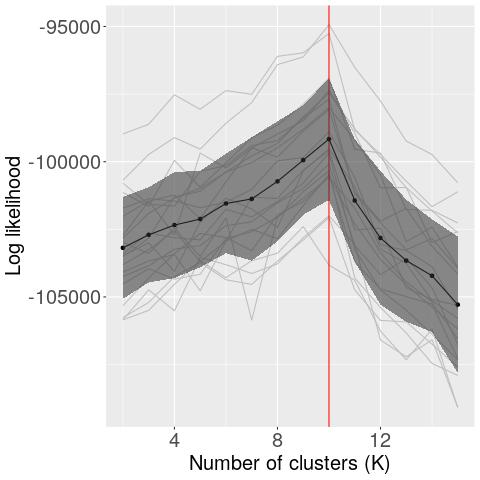}
%  \caption{B=0.8,random selected alpha, varying K}
    \caption{}
  \label{fig:sub3}
\end{subfigure}\\
\caption{The trace plot of the log likelihood when the underlying true number of blocks is (a) K=3, (b) K=5, and (c) K=10. The grey lines represent results from 20 simulated datasets. The black line is their average. The red vertical line indicates the underlying true number of blocks in the generative model.}
\label{fig:spa}
\end{figure}

Figure~\ref{fig:spa} presents traceplots of the marginal log-likelihood as a function of~$K$ for each simulated dataset. In all settings, the true~$K$ gives the largest marginal likelihood, which serves as empirical evidence in support of our proposed method.  Note that the likelihood is also affected by the underlying power-law properties of the network and the strength of the inter/intra connectivity ($a$ and $b$) of different blocks as shown in Figure 3 in the supplementary materials. 
% We consider a formal study of the model selection criteria in the block edge exchangeable setting important future work.
%In general, the plateau of likelihood is observed before hitting the truth. As the K increases, the likelihood drops immediately after passing the truth. Based on what is observed in the simulations, we suggest a careful checking of the marginal likelihood. And a second criteria is to choose the K right before the likelihood begin to drop. 

\section{Case Study}
\label{sec:Talklife}

\subsection{Data overview}

In this section, we apply the B-VCM to the TalkLife dataset which consists of millions of user posts and comments. 
%Recall that TalkLife is a large-scale peer support network for mental health. Individuals on the network receive support for a variety of mental health issues including anxiety, depression, eating disorders, and self-harm. Users can post to the platform and respond to other users' posts. The social network has collected millions of interactions among users to date. To better understand how users organize on their platform, TalkLife is interested in identification of any underlying communities and each community's network properties.
Here we consider all non-deleted posts on TalkLife during the year 2019.  %Since users can delete their own posts and comments, we limit our analysis to those posts that were not later deleted by the user. 
%that is a total of 1,508,895 posts with an average of 2.82 comments per post, leading to 4,258,792 post-comment pairs. 
TalkLife deploys a series of classification algorithms to determine whether a post can be flagged as including language related to one of several mental health topics. For example, a post could be flagged as \emph{Anxiety Panic Fear Suspected} if the corresponding classifier was triggered by the text of the post. Posts that are not tagged by any one of the 33 classifiers are removed, leading to a final dataset consisting 1,481,296 posts with an average of 2.83 comments, summing up to 4,194,609 post-comment pairs. Figures 10 and 11 in the supplementary materials highlight overlap among classifiers, which shows that most posts are only flagged by a single classifier and very few users have posts that are flagged by multiple  classifiers.

\begin{figure}
\centering
\begin{subfigure}{0.5\textwidth}
  \centering
  \includegraphics[width=.8\linewidth]{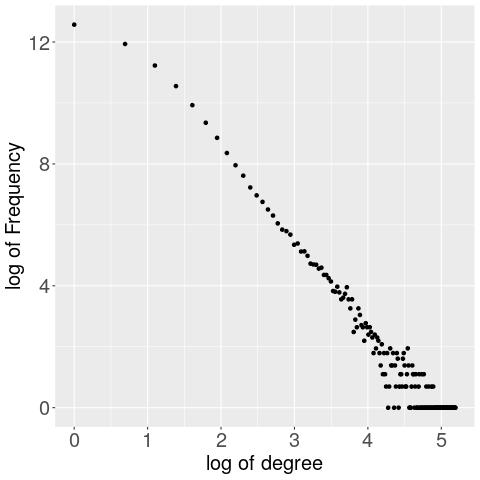}
  \caption{}
  \label{fig:sub1rep}
\end{subfigure}%
\begin{subfigure}{0.5\textwidth}
  \centering
  \includegraphics[width=.8\linewidth]{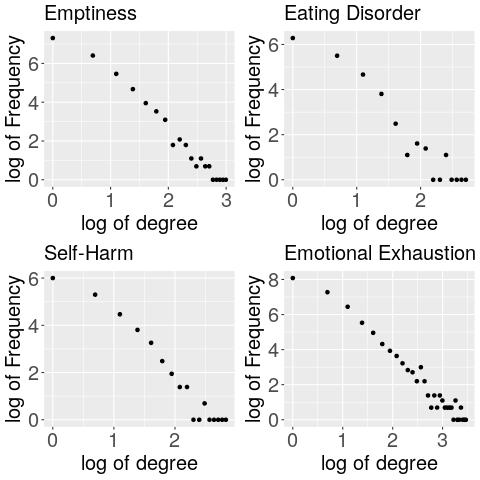}
  \caption{}
  \label{fig:sub2rep}
\end{subfigure}
\caption{(a) Overview of the degree distribution of 2019 senders; (b) Degree distribution of the 2019 senders in sub-networks.}
\label{fig:test}
\end{figure}

Figure~\ref{fig:test} shows the global degree distribution for posts, and the distribution for subsets of posts flagged by 4 out of 33 different classifiers. 
% including Emptiness Suspected, Nausea With Eating Disorder Suspected, SelfHarm Remission Or Relapse Suspected, and Emotional Exhaustion Suspected. 
The power-law degree distribution is apparent in the overall network of 2019, as well as the classifier-specific sub-networks. 
%The empirical phenomenon exactly fits the power-law assumption of the proposed model. We thus applied our model to the data in the following sections.

\subsection{Inferring Block-labels}

%\textcolor{red}{The logic of this section:}

%\noindent \textcolor{red}{
%0) Introduction and data processing;\\
%1) Show the algorithm is learning the real communities; likelihood %and the communities visualization\\
%2) Show for the varying choice of K, the within community connection is greater than the between community;\\
%3) Compare the method with the Spectral clustering/ ML tag, make sure the communities detected by the two methods make sense;\\
%4) Calculate the Hellinger distance for our model and the other methods; show the communities detected by both methods are the same?\\
%5) The power-law estimate?}

In this section, we aim to identify the community structure of TalkLife's users. While the sequential description in Section~\ref{section:seqdesc} assumed a single commentator, TalkLife posts consists of multiple commentators per post.  To address this, we naturally extend the B-VCM model by generating additional commentators according to the exact same procedure.  Using~\eqref{eq:repthme2}, this is equivalent to a conditional independence assumption given asymptotic propensities:
\begin{equation}
    \label{eq:talklife}
P( E =  (s, \{ r_1, \ldots, r_k \} ) \mid \pi_c, f, \B) = 
\pi_b \times f_{s}^{(b)} \times \prod_{l=1}^k \mathcal{B}(b,b_l^\prime) \times  \times f_{r_l}^{(b^\prime_l)}.
\end{equation}

Based on preliminary data analysis, the machine learning generated tags partition users into non-overlapping sub-communities.  Therefore, we subset the data based on these machine learning generated tags and run the Gibbs-sampling algorithm to learn the latent community structure within each sub-community. We apply the maximal marginal likelihood method in Section~\ref{sec:slct_K} to infer the block assignments of each user over a range of the number of blocks,~$K$.

\begin{figure}
\centering
\begin{subfigure}{0.3\textwidth}
  \centering
  \includegraphics[width=.9\linewidth]{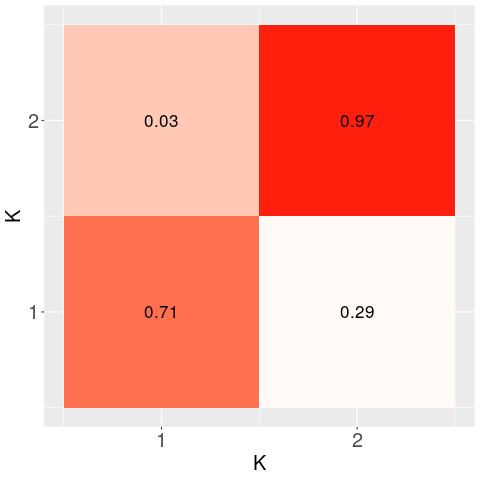}
  \caption{}
  \label{fig:sub.5.2.2.1}
\end{subfigure}%
\begin{subfigure}{0.3\textwidth}
  \centering
  \includegraphics[width=.9\linewidth]{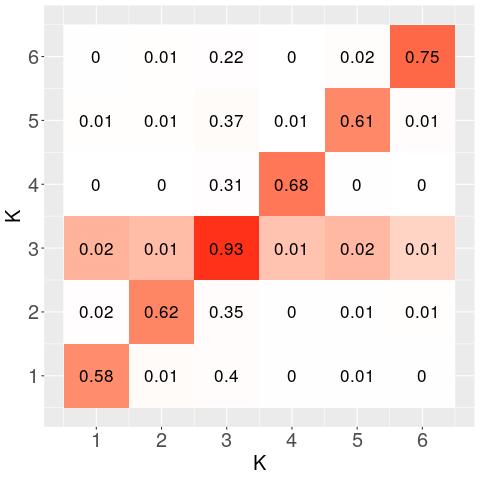}
  \caption{}
  \label{fig:sub.5.2.2.2}
\end{subfigure}
\begin{subfigure}{0.3\textwidth}
  \centering
  \includegraphics[width=.9\linewidth]{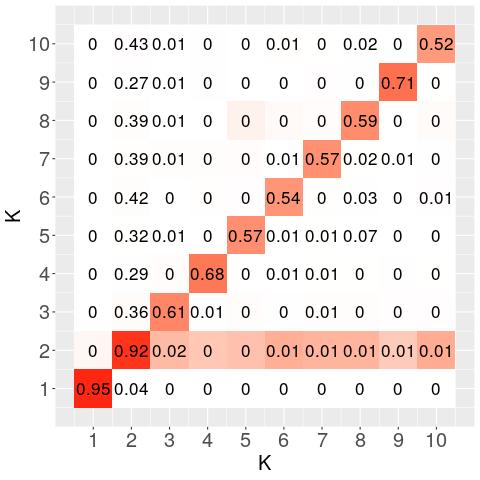}
  \caption{}
  \label{fig:sub.5.2.2.3}
\end{subfigure}

\begin{subfigure}{0.3\textwidth}
  \centering
  \includegraphics[width=.9\linewidth]{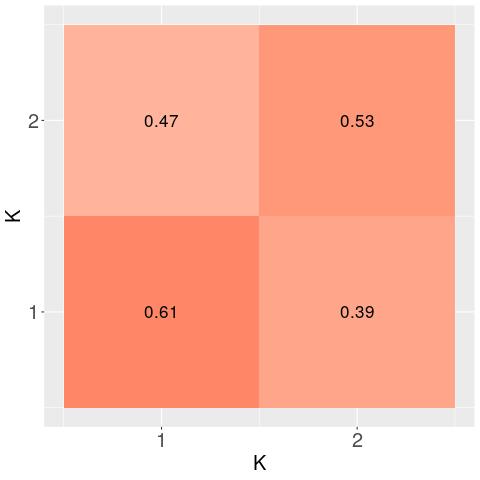}
  \caption{}
  \label{fig:sub.5.3.1.1}
\end{subfigure}%
\begin{subfigure}{0.3\textwidth}
  \centering
  \includegraphics[width=.9\linewidth]{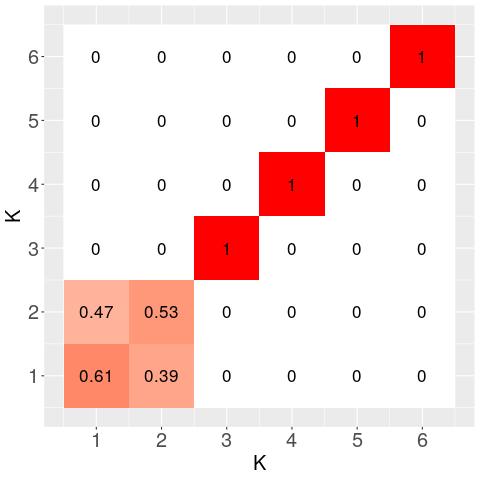}
  \caption{}
  \label{fig:sub.5.3.1.2}
\end{subfigure}
\begin{subfigure}{0.3\textwidth}
  \centering
  \includegraphics[width=.9\linewidth]{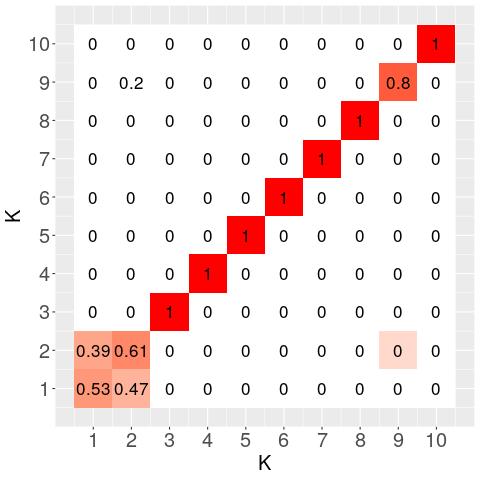}
  \caption{}
  \label{fig:sub.5.3.1.3}
\end{subfigure}

\begin{subfigure}{0.3\textwidth}
  \centering
  \includegraphics[width=.9\linewidth]{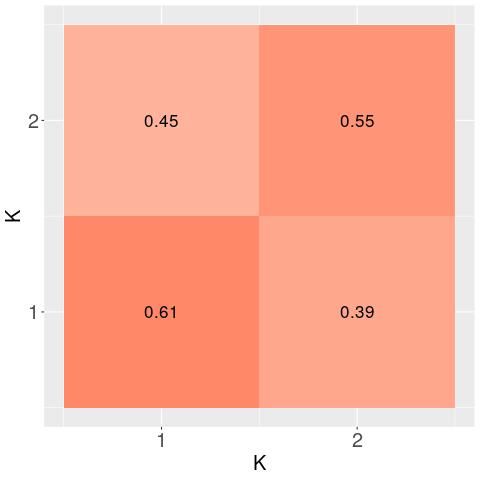}
  \caption{}
  \label{fig:sub.5.3.2.1}
\end{subfigure}%
\begin{subfigure}{0.3\textwidth}
  \centering
  \includegraphics[width=.9\linewidth]{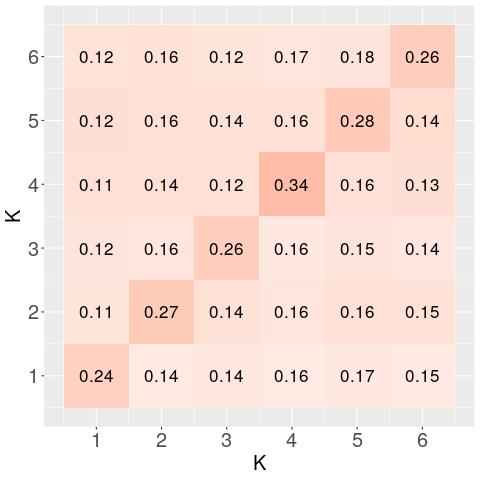}
  \caption{}
  \label{fig:sub.5.3.2.2}
\end{subfigure}
\begin{subfigure}{0.3\textwidth}
  \centering
  \includegraphics[width=.9\linewidth]{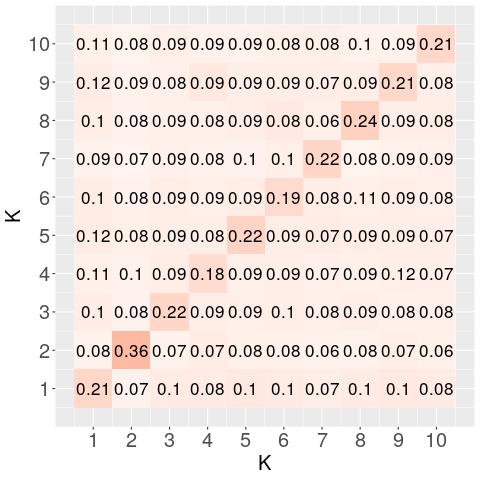}
  \caption{}
  \label{fig:sub.5.3.2.3}
\end{subfigure}
\caption{The inter/intra connectivity of the communities detected by B-VCM ((a)(b)(c)), the spectral clustering ((d)(e)(f)), and the degree corrected stochastic block model ((g)(h)(i)), ranging from 0 to 1, indicates the proportion of the interactions that initiated from one block to the other. The number within each cell is the proportion of the interactions initiated from one cluster (y-axis) to another (x-axis), normalized by each row.}
\label{fig:5.3.1}
\end{figure}

We next investigate the within and the between block connectivity of the inferred communities. Let the block assignment of each node be a binary variable indicating the maximal hits of the blocks in the Gibbs iterations. Using Alcohol and Substance Abuse network as an example, the connectivity of the detected communities when K is set to be 2, 6, and 10 is shown in Figure~\ref{fig:sub.5.2.2.1},~\ref{fig:sub.5.2.2.2},and~\ref{fig:sub.5.2.2.3}. It is consistently observed that within-community connectivity is greater than the between-community connectivity.  Note that B-VCM infers one larger community across different K values. To check if this is the same community across K values, we compared the overlapping of unique nodes when K=2, 6, and 10. The proportion is defined as $\frac{\text{Number of nodes present in both groups}}{\text{Number of nodes present in either of the groups}}$.
The proportion of the overlapping nodes are 0.58, 0.58, and 0.98 when comparing K=2 to K=6; K=2 to K=10; and K=6 to K=10 correspondingly. This provides empirical evidence that the algorithm is finding sub-communities from this larger block as $K$ increases.
% many nodes  the variation of the underlying community structure when setting K to different values.

\subsection{Comparison with Other Methods}
\label{sec:casestudy}
Here we compare our method with the spectral clustering method \cite{van2019scalable, ng2002spectral} and DC-SBM \cite{karrer2011stochastic, funke2019stochastic}. To apply spectral clustering, the interaction data is converted into a binary graph where an edge exists if there was an interaction between the two individuals. The connectivity of the communities detected by the spectral clustering method in Alcohol and Substance Abuse network is shown in Figure~\ref{fig:sub.5.3.1.1},~\ref{fig:sub.5.3.1.2}, and~\ref{fig:sub.5.3.1.3}. The spectral clustering method consistently learns two large blocks. The proportion of the overlapping nodes in the largest community detected by spectral clustering are 0.99, 0.99, and 0.99 when comparing K=2 to K=6; K=2 to K=10; and K=6 to K=10 respectively. This indicates little variation in the inferred community structure across values of $K$. We compare the overlapping of nodes clustered in the largest block by B-VCM and the largest two blocks by spectral clustering. The overlapping proportion of the nodes are 0.62, 0.91, 0.90 when K=2, 6, and 10 respectively.
%The overlapping proportion of the nodes are 0.39, 0.47, 0.47 when K=2, 6, and 10 correspondingly. 

We then fit the degree corrected stochastic block model (DC-SBM) to the same binary graph .
% where the weight between user $i$ and $j$ is the number of interactions involve both of users.
% , we construct the input network that counts each poster-commentator pair as one edge, and the corresponding poster and commentator as two nodes. 
The connectivity of the communities detected by the DC-SBM in Alcohol and Substance Abuse network is shown in Figure~\ref{fig:sub.5.3.2.1},~\ref{fig:sub.5.3.2.2}, and~\ref{fig:sub.5.3.2.3} in the supplementary materials. While the sizes of the DC-SBM inferred communities are much more balanced, the within- compared to between-community connectivity is much weaker. 
% Since B-VCM tends to cluster the users into a large community, 
Users that were part of the largest B-VCM inferred block are distributed across different DC-SBM inferred blocks, indicating strong differences in inferred community structure. 
% Note that DC-SBM is designed for dense networks by assumption. 
To assess sensitivity to the projection onto a binary graph based on existence of a single interaction, the DC-SBM was fit to a projected network where an edge exists if the number of interactions between the two nodes is greater than or equal to two. This projection had 13\% of the edges of the original binary graph.  The results are shown in Section A.14 of the supplementary materials.
%\textcolor{red}{The average proportion of these users in different groups identified by DC-SBM, for example, is 0.62 (0.0008), 0.91 (0.0084), 0.90 (0.011) for K=2, 6, and 10.} 
% This indicates that the DC-SBM gives very different community structures from B-VCM.

%Also note there is little variations in the nodes clustered to the largest two groups by spectral clustering over different K values. This makes the communities detected by the spectral clustering method less informative than our method.
%Notice that the spectral clustering method gives very different community structures as compared to our method in the sense that it tends to cluster nodes into one or two large chunks. This makes the communities detected by the spectral clustering method less informative than our method.

%\begin{figure}
%\centering
%\begin{subfigure}{0.3\textwidth}
%  \centering
%  \includegraphics[width=.9\linewidth]{comm_spec_2.jpeg}
%  \caption{}
%  \label{fig:sub.5.3.1.1}
%\end{subfigure}%
%\begin{subfigure}{0.3\textwidth}
%  \centering
%  \includegraphics[width=.9\linewidth]{comm_spec_6.jpeg}
%  \caption{}
%  \label{fig:sub.5.3.1.2}
%\end{subfigure}
%\begin{subfigure}{0.3\textwidth}
%  \centering
%  \includegraphics[width=.9\linewidth]{comm_spec_10.jpeg}
%  \caption{}
%  \label{fig:sub.5.3.1.3}
%\end{subfigure}
%\caption{The inter/intra connectivity of the communities detected by spectral clustering, ranging from 0 to 1, indicating the proportion of the nodes that have within/between cluster connectivity. The number within each cell is the number of nodes.}
%\label{fig:5.3.1}
%\end{figure}

To further assess the quality of the inferred latent community structure, we investigate the inferred block consistency over time. We split the data into two halves based on whether the post occurred before June 30th, and infer the latent community structure within each time window using all three approaches. The Hellinger distance is used to quantify the difference between the block assignments, which is defined as:
$$HL=\frac{1}{v(\bfY)}\sum_{i=1}^{v(\bfy)}\frac{1}{\sqrt{2}}\sqrt{\sum_{k=1}^{K}(\sqrt{p_{ki}}-\sqrt{q_{ki}})^2}$$
where $K$ is the number of presumed blocks, and $v(\bfY)$ is the number of nodes; $p$ and $q$ correspond to the probability vector of a node belonging to each of the $K$ blocks. To overcome the label switching problem, 
% we observe that when the largest one or two communities are correctly matched, the variation in Hellinger distance is small regardless of the matching status of the rest clusters. \textcolor{red}{In particular, 
% we randomly sample 20 possible label combinations and take the minimal Hellinger distance. The step is repeated 20 times, and the mean value is set to be the final value.
we perform a greedy matching of blocks by size, starting with the largest communities, followed by the second largest, and so on so forth. 
% The conclusions are similar.

For illustrative purposes we use Alcohol and Substance Abuse network and Behavioral Symptoms network as two examples. The results are shown in Figure~\ref{fig:5.3.2.new} (See Section A.13 of the supplementary materials for results in other example networks). For the two subnetworks, the marginal likelihood is maximized when $K=2$ and $4$ respectively. Due to uncertainty in K, we present a range of values. The Hellinger Distance of inferred blocks assignments using our proposed method is smallest at these of optimal $K$ valued based on marginal likelihood.
% which demonstrates the usefulness of our proposed strategy in a real data setting. 
Moreover, when comparing to inferred block assignments using spectral clustering and DC-SBM, the Hellinger Distance is smaller using our proposed method for all choices of $K$, which indicates the communities detected by our method are more consistent over time than the communities detected by the other two methods. 
% That is, users tend to have more connections with people from the same inferred communities than people outside the inferred communities over time. 

% of our model is consistent with the marginal likelihoods. 
% It serves to prove the efficacy of the heuristic strategy mentioned in Section \ref{sec:slct_K} in real data. 

\begin{figure}
\centering
\begin{subfigure}{0.6\textwidth}
  \centering
  \includegraphics[width=.9\linewidth]{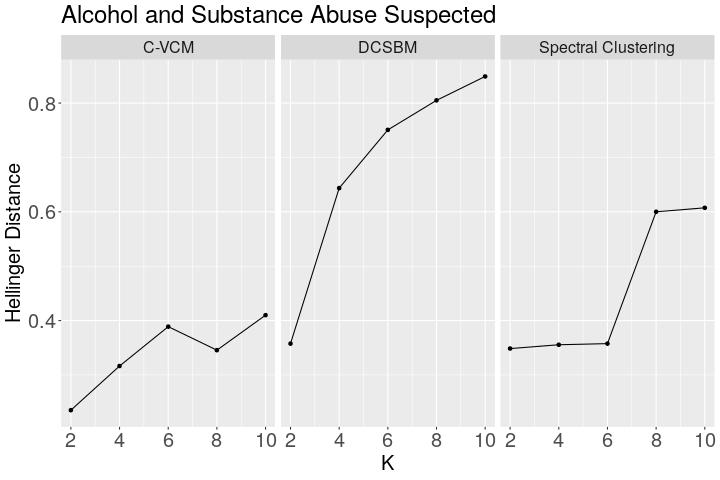}
  \caption{}
  \label{fig:sub.HD_58}
\end{subfigure}%
\begin{subfigure}{0.3\textwidth}
  \centering
  \includegraphics[height=160pt,width=.9\linewidth]{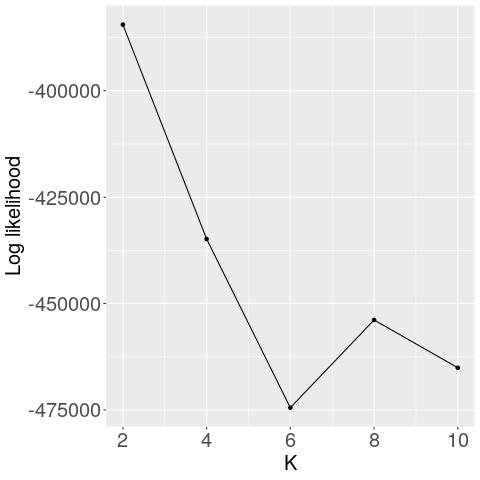}
  \caption{}
  \label{fig:sub.llk_58}
\end{subfigure}

\begin{subfigure}{0.6\textwidth}
  \centering
  \includegraphics[width=.9\linewidth]{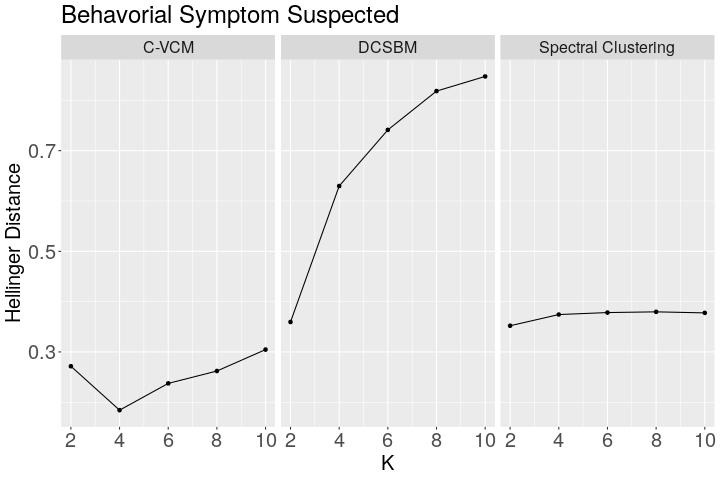}
  \caption{}
  \label{fig:sub.HD_60}
\end{subfigure}%
\begin{subfigure}{0.3\textwidth}
  \centering
  \includegraphics[height=160pt,width=.9\linewidth]{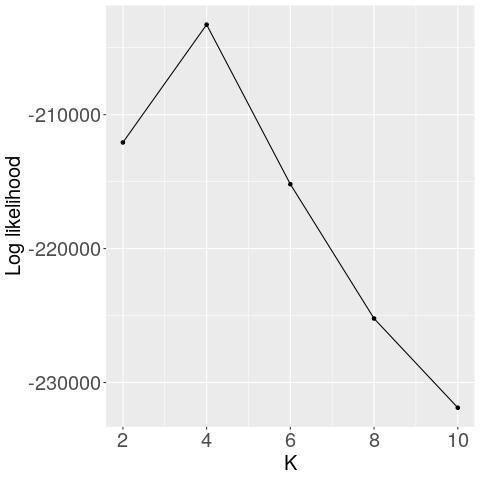}
  \caption{}
  \label{fig:sub.llk_60}
\end{subfigure}

\caption{The Hellinger distances of the block assignment between the first half and the second half of the 2019 data in (a) Alcohol and Substance Abuse suspected network and (c) Behavorial Symptoms Suspected network. The marginal likelihood over different K values in (b) Alcohol and Substance Abuse suspected network and (d) Behavorial Symptoms Suspected network.}
\label{fig:5.3.2.new}
\end{figure}

%\begin{figure}[!Htp]
%    \centering
%    \includegraphics[width=14cm]{5.3.2.PNG}
%    \caption{The Hellinger distances of the cluster assignment between the first half and the second half of the 2019 data in (a) Alcohol and Substance Abuse suspected network and (c) Behavorial Symptoms Suspected network. The marginal likelihood over different K values in (b) Alcohol and Substance Abuse suspected network and (d) Behavorial Symptoms Suspected network.}
%    \label{fig:5.3.2.new}
%\end{figure}

\section{Discussion}
\label{section:discussion}
%1. Conclusion 

In this paper, we have proposed a new class of exchangeable interaction models that allow for node-level community structure. The framework models networks arising from the interaction process perspective and allows for networks that exhibit power-law degree distributions and network sparsity.  We provide theoretical guarantees that the network sparsity can be retained in both the block-specific sub-networks and the entire network. We also prove that the misspecification rate of the inferred block assignments can be bounded, with the bound decreasing rapidly for high degree nodes. We implement the proposed B-VCM using a Gibbs sampling approach and demonstrate the efficacy of our algorithm through simulation. We end by applying our method to the TalkLife data to identify online user communities.  We compare our results with inferred communities based on spectral clustering and degree-corrected stochastic block models, demonstrating more interpretable clustering results and consistency of the clusters over time. 

% The contribution of our work is to account for the community structure when modeling the interaction process. Using TalkLife data as an example, our method can identify the underlying users' groups so that the platform will know to whom they can boost the visibility of the posts from certain users. Besides, our method provides the estimates of the power-law parameters for each of the group. The revelation of the network properties can help the platform make proper decisions on what kind of interventions to make. For example, a smaller power-law parameter indicates a looser connection between users. The platform will need extra efforts to ensure certain users within the community getting enough attention and social supports. Though we demonstrate our method using the TalkLife data, it can be well applied to any sparse interaction data sets \cite{sapiezynski2019interaction}.    

%2. Drawback of this paper
% Note that the consistent conclusion only apply to high-degree nodes. We acknowledge it's the generic problem of our model, but it is generically hard to solve given that low degree nodes tend to have little information from the network structure.

%3. Future plan, hierarchical extension, and what?
There are numerous directions for future work. First, our proposed framework considered node-level community detection. Interaction processes may also exhibit clustering of different types of interactions (e.g., latent interaction topics). Second, given the size of modern interaction data, as discussed in Remark~\ref{rmk:scalable} there is a need to consider scalable implementations of the proposed algorithm or alternative algorithms that can handle hundreds of millions to billions of interactions~\cite{chiquet2019variational}. Third, our model assumes conditional independence of interactions given block assignments and node-level propensities. Recent work has considered community detection with dependent connectivity \cite{yuan2018community}. A natural question how to incorporate such dependency into the interaction exchangeable framework.

\printbibliography

\end{document}

\maketitle

\appendix

\section{Appendix}

\subsection{Proof of Theorem 2.12--Two cluster scenario}
\label{app:thmconsist}
% copied version on 12/5 below just in case.
Given an initial labeling $e: v(\bfY_m) \to [K]$ of all observed nodes in the network~$\bfY_m$, our goal is to show 
% $\sup_{e}M_{v(\mathcal{Y}_m)}=\frac{1}{v(\mathcal{Y}_m)}\sum_{i=1}^{v(\mathcal{Y}_m)}1(\hat{C}_i\neq C_i)$ 
the misclassification rate (accounting for label switching) is bounded after a single iteration of an updating algorithm that approximately maximizes the likelihood given initial labeling.  Specifically, we will show the misclassification rate decays rapidly and is therefore small for high-degree nodes. 

Conditional on the asymptotic frequency of initiating an interaction and the propensity matrix,~$\pi$ and $\B$ respectively, the B-VCM likelihood~$P(\bfY_m=\bfy_m|\{\alpha_b, \theta_b\}, \pi, \B, \{ B(i) \}_{i \in v(\bfY_m)})$ is given by:
\begin{equation}
\prod_{b=1}^{K}\pi_b^{L_b}\times 
\frac{[\alpha_{b}+\theta_{b}]_{\alpha_b}^{v(\mathbf{y}_{b})-1}}{[\theta_{b}+1]_1^{m(\mathbf{y}_b)-1}}\prod_{j\in \H_{m}^{(b)}}[1-\alpha_{b}]_1^{D_m(j,b)-1}
\times\prod_{b'=1}^{K} \B(b,b')^{D_m(b,b')} 
\label{eq:bvcm}
\end{equation}
where~$L_b$ is the number of interactions initiated by cluster~$b \in [K]$, $v(\bfy_b)$ is the number of non-isolated vertices from cluster $b$, $m(\bfy_b) = \sum_{j \in v(\bfy_b)} D_m (j,b)$ is the number of times (with multiplicity) that the cluster~$b$ nodes are observed, $D_m (j,b)$ is the degree of node~$j$ in cluster~$b$, and $D_m (b,b')$ is the number of interactions between cluster~$b$ and~$b'$. 
%For simplicity of the notation, denote $(*)_b= \frac{[\alpha_{b}+\theta_{b}]_{\alpha_b}^{v(\mathbf{y}_{b})-1}}{[\theta_{b}+1]_1^{m(\mathbf{y}_b)-1}}\prod_{j\in \H_{m}^{(b)}}[1-\alpha_{b}]_1^{D_m(j,b)-1}$. 

\begin{assm}
\label{assm:thme2}
\normalfont
In the proof below, we consider the setting where $K=2$, $a=\B(1,1)=\B(2,2)$, $b=\B(1,2)=\B(2,1)$, $\alpha_1=\alpha_2$, $\theta_1=\theta_2$, $\pi_1=\pi_2$, and $a > b$.   We call this a \emph{balanced setting} since the likelihood of initiating an interaction and power-law degree distributions are equal across clusters.
\end{assm}

\paragraph{Directed network setting}
We start by considering a \emph{directed} version of the B-VCM in which out- and in-degree labels are distinct.  
% Suppose an arbitrary labeling $e_{in}: v_{in} (\bfY_m) \to [K]$ of all observed receiver nodes, i.e., $i \in v_{in} (\bfY_m)$ if there exists $E_j \in E_{[m]}$ such that $i$ the receiver of the interaction, $E_j = (\bullet, i)$.  
Let $Deg_{out}(i)$ be the out-degree of node $i$, and $Deg_{out}(i,1)$ and $Deg_{out}(i,2)$ be the degree of node $i$ being connected to nodes in cluster 1 and 2 correspondingly according to labeling~$e$. Then, based on a directed version of~\eqref{eq:bvcm}, the log likelihood of assigning the node $i$ to cluster 1 is:
\begin{align*}
l_{i,1} = &(Deg_{out}(i)+L_{1,-i})\log \pi_1+\log [\B(1,1)^{D_m(1,1)}\B(1,2)^{D_m(1,2)}] \\
&+(L_{2,-i})\log \pi_2+\log [\B(2,1)^{D_m(2,1)}\B(2,2)^{D_m(2,2)}] \\
&+\log\left[ \frac{[\alpha_{1}+\theta_{1}]_{\alpha_1}^{v_{-i}(\mathbf{y}_{1})}}{[\theta_{1}+1]_1^{m_{-i}(\mathbf{y}_1)+Deg_{out}(i)-1}}\prod_{j\in \H_{m}^{(1)},j\neq i}[1-\alpha_{1}]_1^{D_m(j,1)-1}[1-\alpha_1]^{Deg_{out}(i)-1}_{1}\right] \\
&+\log \left[ \frac{[\alpha_{2}+\theta_{2}]_{\alpha_2}^{v_{-i}(\mathbf{y}_{2})-1}}{[\theta_{2}+1]_1^{m_{-i}(\mathbf{y}_2)-1}}\prod_{j\in \H_{m}^{(2)},j\neq i}[1-\alpha_{1}]_1^{D_m(j,2)-1}\right]
\end{align*}
%$$=(Deg_{out}(i)+L_{1,-i})\log \pi_1+L_{2,-i}\log\pi_2+\log (*)_1 (*)_2+[Deg_{out}(i,1)+D_{m,-i}(1,1)]\log\B(1,1)$$
%$$+[Deg_{out}(i,2)+D_{m,-i}(1,2)]\log\B(1,2)+D_{m,-i}(2,1)\log\B(2,1)+D_{m,-i}(2,2)\log\B(2,2)$$
where the subscript~$-i$ denotes the statistic with node $i$ removed. That is, ~$L_{b,-i}$ is the number of interactions initiated by a node in cluster~$b=1,2$ according to the labeling~$e$, excluding node~$i$;  $v_{-i}(\mathbf{y}_b)$ is the number of nodes in block $b$ excluding node $i$; and $m_{-i}(\mathbf{y}_b)$ is the number of interactions in block $b$ excluding node $i$. Similarly, the log likelihood of assigning node~$i$ to cluster $2$ is;
\begin{align*}
l_{i,2}= &L_{1,-i}\log \pi_1+\log [\B(1,1)^{D_m(1,1)}\B(1,2)^{D_m(1,2)}] \\
&+[Deg_{out}(i)+(L_{2,-i})]\log \pi_2+\log [\B(2,1)^{D_m(2,1)}\B(2,2)^{D_m(2,2)}] \\
&+\log\left[ \frac{[\alpha_{1}+\theta_{1}]_{\alpha_1}^{v_{-i}(\mathbf{y}_{1})-1}}{[\theta_{1}+1]_1^{m_{-i}(\mathbf{y}_1)-1}}\prod_{j\in \H_{m}^{(1)},j\neq i}[1-\alpha_{1}]_1^{D_m(j,1)-1}\right] \\
&+\log\left[ \frac{[\alpha_{2}+\theta_{2}]_{\alpha_2}^{v_{-i}(\mathbf{y}_{2})}}{[\theta_{2}+1]_1^{m_{-i}(\mathbf{y}_2)+Deg_{out}(i)-1}}\prod_{j\in \H_{m}^{(2)},j\neq i}[1-\alpha_{1}]_1^{D_m(j,2)-1}[1-\alpha_2]^{Deg_{out}(i)-1}_{1}\right].
\end{align*}
%$$=L_{1,-i}\log\pi_1+(Deg_{out}(i)+L_{2,-i})\log \pi_2+\log (*)_1 (*)_2+[Deg_{out}(i,1)+D_{m,-i}(2,1)]\log\B(2,1)$$
%$$+[Deg_{out}(i,2)+D_{m,-i}(2,2)]\log\B(2,2)+D_{m,-i}(1,1)\log\B(1,1)+D_{m,-i}(1,2)\log\B(1,2)$$
% where $L_{b,-i}$ is the out-degree initiated from block $b\in[1,2]$;  
% Let $a=\B(1,1)=\B(2,2)$, and $b=\B(1,2)=\B(2,1)$. Assume $a>b$, the power-la parameters $\alpha_1=\alpha_2$, $\theta_1=\theta_2$, and $\pi_1=\pi_2$, such that the difference is:
Under Assumption~\ref{assm:thme2}, the difference in the log likelihoods simplifies:
\begin{equation}
\label{eq:llikdiff}
\begin{multlined}
    l_{i,1}-l_{i,2}=(\log a-\log b)(Deg_{out}(i,1)-Deg_{out}(i,2))
    \\
    +\log \frac{\alpha_1+\theta_1+(v_{-i}(y_1)-1)\alpha_1}{\alpha_2+\theta_2+(v_{-i}(y_2)-1)\alpha_2}-\sum_{j=1}^{Deg_{out}(i)}\log \frac{\theta_1+1+m_{-i}(y_1)+j-2}{\theta_2+1+m_{-i}(y_2)+j-2}
\end{multlined}
\end{equation}

Under Assumption~\ref{assm:thme2}, $\theta_1 = \theta_2$ and $\alpha_1 = \alpha_2$ implies $v_{-i}(\bfy_1)\simeq v_{-i}(\bfy_2)$.  Moreover, the balanced design, i.e.,~$b:= \B(1,2) = \B(2,1)$, $a = \B(1,1) = \B(2,2)$ and $\pi_1 = \pi_2$, implies $m_{-i}(\bfy_1)\simeq m_{-i}(\bfy_2)$. 
% Let $M_{v(\mathbf{y}_b)}(e_{out})=\frac{1}{v(\mathbf{y}_b)}\sum_{i\in\mathbf{v(y_b)}}1(\hat{B} (i)\neq  B(i))$ be the misclassification rate of nodes  that belong to block $b$ under the ground truth. Given the balanced design in terms of the number of nodes in each block, the overall mismatch is given by
% $$M_{v(\mathbf{y}_m)}(e_{out})=\frac{1}{2}[M_{v(\mathbf{y}_1)}(e_{out})+M_{v(\mathbf{y}_2)}(e_{out})]$$
% We first consider the nodes that belongs to block 1 under the ground truth.
% and the corresponding $M_{v(y_1)}(e_{out})$. 
Based on the Eq.~\ref{eq:llikdiff} and the above discussion, a natural approximate updating rule is $\hat{B}(i)=1$ if $Deg_{out}(i,1)>Deg_{out}(i,2)$ for each node $i$ and $\hat{B}(i) = 2$ otherwise. Next, let $\xi_j(e_{out})$ for~$j \in v(\bfy)$ such that $\xi_{j}(e_{out})=-1$ if $e_{out,j}=1$ and $\xi_{j}(e_{out})=1$ if $e_{out,j}=2$.  Then define $\epsilon_{out,i}:=\sum_j D_m(i,j) \xi_{j}(e_{out})$, where $D_m(i,j)$ is the number of directed interactions $(i,j)$ in $\bfy$, i.e., initiated by node $i$ who interacts with node $j$. 
% \textcolor{blue}{Since only the out degree is considered for now, $\epsilon_{out,i}|Deg(i)\perp Deg(j)$ for any $j$ that connects to $i$.} ## WD: I wouldn't put this here since it's not needed yet.
Then the updating rule leads to correct specification of the node (i.e., the assigned label~$\B(1)$ equals the true label~$\B(1) = 1$) if $\epsilon_{out,i}<0$.  Thus bounding the probability of misclassifying node~$i$ is equivalent to bounding~$P(\epsilon_{out, i} > 0)$.

To bound this probability, we rely on the representation theorem applied to the B-VCM which guarantees conditional independence of the interactions given the node propensities.  Let~$W^{(b)}_i$ denote the stick-breaking propensities for observed vertices~$i \in v(\bfy_b)$.  Then, given the observed set of vertices, we can construct the propensity of observing a specific node $i$ in block $b$ as $f_i^{(b)} := W_i^{(b)} / \sum_{i' \in v(\bfy_b)} W_{i'}^{(b)}$. 
% By design $\sum_{i\in v(y_b)} f_i^{(b)}=1$, and $\mathbb{E}(\frac{Deg(i)}{\sum_{i}Deg(i)})=f_i^{(b)}$. 
Let $\{J_1\}$ be the set of nodes that match to the truth under $e$, and $\{J_2\}$ be the set of nodes that do not match to the truth under $e$. Given the out degree of each node $i$,~$Deg_{out}(i)$ and the propensities $\{f_i^{(b)}\}$, 
\begin{align*}
\mathbb{E}(\epsilon_{out,i})&=-Deg_{out}(i)\left(a\sum_{j\in \{J1\}}f_{j}^{(1)}+b\sum_{j\in \{J2\}}f_{j}^{(2)}-b\sum_{j\in \{J1\}}f_{j}^{(2)}-a\sum_{j\in \{J2\}}f_{j}^{(1)}\right) \\
&=-Deg_{out}(i)\left[a\left(\sum_{j\in \{J1\}}f_{j}^{(1)}-\sum_{j\in \{J2\}}f_{j}^{(1)}\right)-b\left(\sum_{j\in \{J1\}}f_{j}^{(2)}-\sum_{j\in \{J2\}}f_{j}^{(2)}\right)\right].
\end{align*}
% \textcolor{red}{WD: is this correct?  Probability $a$ of connecting to a true cluster 1 and $b$ connecting to true cluster 2.  So wouldn't we get $a ( 2 \sum_{j \in {J1}} f_j^{(1)} - 1) - b ( 2 \sum_{j \in {J1}} f_j^{(2)} -1)$. This seems right cause if you assume these are both $\gamma$ then you get $(2 \gamma - 1) (a-b)$. I changed below but let me know if I missed something.} \textcolor{blue}{Yes, I'm wrong, and yours is right.}

Let~$\gamma_b = \sum_{j \in {J1}} f_j^{(b)} \in [0,1]$ denote the weighted fraction of nodes that are correctly specified and note that $\sum_{j \in {J1}} f_j^{(b)} + \sum_{j \in {J2}} f_j^{(b)} = 1$.  Let $\mu_{1,2} := a(2 \gamma_1 - 1) - b(2 \gamma_2 - 1)$ then
% Let $(**)=a(\sum_{j\in \{J1\}}f_{j}^{(1)}-\sum_{j\in \{J2\}}f_{j}^{(2)})-b(\sum_{j\in \{J1\}}f_{j}^{(2)}-\sum_{j\in \{J2\}}f_{j}^{(1)})$, such that we have
$$Var(\epsilon_{out,i})=4Deg_{out}(i)[\mu_{1,2}(1-\mu_{1,2})]\le Deg_{out}(i)$$
Bernstein's inequality states that for a sequence of \emph{independent} random variables $(X_1,...,X_n)$ with means $(\mu_1,...,\mu_n)$, such that $|X_i|\le M$, for $t \geq 0$ we have:
$$
P\left(\sum_{i=1}^{n}(X_i-\mu_i)\ge t \right)\le \exp\left(-\frac{t^2}{2(\sum_{i=1}^{n}Var(X_i)+tM^{\prime}/3)}\right)
$$
where $M^{\prime}=M+\max_i|\mu_i|$. Recall $\epsilon_{out,i}=\sum_j D_m(i,j)\xi_{j}(e_{out})$ which consists of independent random variables given the propensities~$\{ f_j^{(b)} \}$ with mean~$\mu$ and variance bounded by~$1$. Then, by Bernstein's inequality, we have
\begin{equation}
\label{eq:bernstein}
P\left(\epsilon_{out,i}>\mathbb{E}(\epsilon_{out,i})+t\right)\le\exp \left(-\frac{t^2}{2(Deg_{out}(i)+2t/3)}\right)
\end{equation}
for any $t \geq 0$. 

\paragraph{Construction of a necessary assumption on $\gamma_b$.}
To set the RHS of the LHS inequality in~\eqref{eq:bernstein} equal to $0$ requires $t=-\mathbb{E}(\epsilon_{out,i})=Deg_{out}(i) \mu_{1,2}$ which requires~$\mu_{1,2} > 0$ since $t \geq 0$.  When considering node~$i$ in cluster~$2$, this implies a positivity condition~$\mu_{2,1} > 0$.  Combining  Assumption~\ref{assm:thme2} that $a > b$ and $a + b = 1$ with the above positivity conditions, we arrive at the following set of conditions on~$\gamma_b$:
\begin{assm}[Positivity]
\label{assm:mupos}
Assume (a) $\gamma_b \in (1/2,1]$ for $b=1,2$, i.e., the weighted fraction of nodes that are correctly specified is greater than $1/2$; and (b) that~$\{ \gamma_1, \gamma_2\}$ satisfy
\begin{equation}
\mu_{1,2} \wedge \mu_{2,1} =: \mu_{\min} = 2 \cdot a \left( 2 \frac{\gamma_{\min} + \gamma_{\max}}{2} - 1 \right) - \left( 2 \gamma_{\max} - 1 \right) > 0
\label{eq:balance}    
\end{equation}
where~$\gamma_{\min} = \gamma_1 \wedge \gamma_2$ and $\gamma_{\max} = \gamma_1 \vee \gamma_2$.
\end{assm}
Assumption~\ref{assm:mupos}(a) can be guaranteed by label switching, i.e., if $e$ has $\gamma_b < 1/2$ then switching all labels in block $b$ will yield an $e'$ with $\gamma_b' := 1-\gamma_b > 1/2$. 
Assumption~\ref{assm:mupos}(b) guarantees positivity of both~$\mu_{1,2}$ and~$\mu_{2,1}$ which is necessary for the labelling~$e$ to guarantee a bound via Bernstein's inequality.  In~\cite{amini2013pseudo}, the proof required a \emph{balance condition}, i.e., the fraction of nodes correctly labelled by~$e$ is constant across the two clusters).  The equivalent balance condition in our setting would be~$\gamma_1 = \gamma_2$, i.e., the weighted fraction of nodes correctly labelled by~$e$ is constant across the two clusters. This balance condition immediately implies~\eqref{eq:balance} but is a much stronger assumption.  Here we rely on the weaker assumption of approximate balance which allows these fractions to be unequal but the level of imbalance depends on the connectivity parameter~$a$.  

\begin{lemma}
\label{lemma:epsbound}
For each~$i \in v(\bfy)$, given $Deg_{out}(i)$ and propensities~$\{ f_i^{(b)} \}$, then under Assumption~\ref{assm:thme2} and~\ref{assm:mupos} we have
$$P(\epsilon_{out,i}>0)\le\exp\left(-\frac{Deg_{out}(i) \mu_{\min}^2}{4}\right)$$
\end{lemma}
% Lemma~\ref{lemma:epsbound} similarly holds for~$i \in v(\bfy_2)$. 

\paragraph{Bounding misclassification rate.}
% \textcolor{red}{WD: Why bound just the blocks separately?  The two are independent so you can target across all, no?} \textcolor{blue}{I think we need to bound the two blocks separately because when a node belongs to block $1$ under the true labeling, the mis-specification happens when $Deg(i,1)<Deg(i,2)$, which is equivalent to $\epsilon_i=Deg(i,2)-Deg(i,1)>0$, as shown above $\mathbb{E}(\epsilon_i)=a(2\gamma_1-1)-b(2\gamma_2-1)$. When a node belongs to block 2 under the truth, the mis-specification happens when $Deg(i,1)>Deg(i,2)$, which is equivalent to $\epsilon_i=Deg(i,1)-Deg(i,2)>0$. In this case, $\mathbb{E}(\epsilon_i)=a(2\gamma_2-1)-b(2\gamma_1-1)$, which is slightly different from what we have before. I do think the overall conclusion should be the same, it's just the notation might be a little confusing if we merge the two conditions together.}
For a labeling~$e: v(\bfy) \to [K]$, the misclassification rate is
$$
M_{v(\bfy)}(e) = \inf_{\rho': [K] \to [K]} \frac{1}{v(\bfy)} \sum_{i=1}^{v(\bfy)} 1[ \hat B (i) = \rho' B(i) ]
$$
where~$\hat B(i) = e(i)$.  Let $N(\xi(e))=\sum_{i=1}^{v(\bfy)}1(\epsilon_{out,i}\ge 0)$. 
% $$
% N_1(\xi(e_{out}))=\sum_{i=1}^{v(\mathcal{Y}_1)}1(\epsilon_{out,i}\ge 0)
% $$
% Our goal is to show through maximizing the likelihood, the mis-specification rate $M_{v(y_1)}(e_{out})$ will be bounded, especially for high-degree nodes. To achieve this goal, we need to connect $N_1(\xi(e_{out}))$ to $M_{v(y_1)}(e_{out})$. Let $\gamma_{e(out)}$ be the proportion of nodes that matches labels in community 1 given $e_{out}$, such that:
% $$\sum_{j=1}^{v(\mathcal{Y}_1)}1(\xi_{j}(e_{out})=1)=\gamma_{e(out)} v(\mathcal{Y}_1)$$
% There exists a permutation of $\{\xi_j(e_{out})\}$ in which the above equation can be satisfied. 
Then the misspecification rate is bounded by :
\begin{equation}
    \label{eq2}
    M_{v(\bfy)}(e)\le \frac{N(\xi(e))}{v(\bfy)}
\end{equation}
%\begin{equation}
%    \label{eq2}
%    \sup_{e} M_{v(\bfy)}\le\sup_{\xi} \frac{N(\xi(e))}{v(\bfy)}
%\end{equation}
%where~$\sup_e$ is a supremum over all labels~$e$ that fix~$\gamma_1, \gamma_2$. 
% \textcolor{red}{WD: Can you define what you mean by the sup of e and $\xi$?  The $e$ is a fixed labeling, no?  And the other is fixed given $\xi$, no?}  \textcolor{blue}{The previous proof didn't require the exact labeling of all nodes under labeling $e$, but only the proportion of the nodes that are being correctly specified, i.e. $\gamma_1$ and $\gamma_2$. This gives a total number of $n\choose \gamma n$ of possible combinations of node labeling. We thus defined the mis-specification rate to be upper-bounded by $\sup_{e}M_{v(y_1)}$, which stands for the largest possible mis-specification rate among all the combinations.}
The inequality is due to treating the ambiguous case $\epsilon_{out,i}=0$ as an error. Note that we prove the case for a specific labeling $e$ rather than an arbitrary labeling $e$ that satisfying Assumption~\ref{assm:mupos}. It is sufficient to show the RHS of inequality~\eqref{eq2} is bounded. 
% We use the following steps to bound the RHS. 
% Note that in all the proof shown below, we condition on the observed networks, such that the number of observed nodes in the network is not random.
Lemma~\ref{lemma:approxp1} shows that the RHS converges almost surely to 

\begin{lemma}
\label{lemma:approxp1}
Given the out degree sequence~$\{ Deg_{out}(i) \}_{i \in v(\bfy)}$ and propensities~$\{ f_j^{(b)} \}$ generated via the B-VCM, then
$$P_{out}:=\frac{1}{v(\bfy)}\sum_{i=1}^{v(\bfy)}P(\epsilon_{out,i}>0) \overset{a.s.}{\to}\sum_{d=1}^\infty \alpha B(d,\alpha+1)\exp(-d \mu_{\min}^2 /4)$$
%$$P(\epsilon_i>0)\le\exp(-\frac{Deg(i)(**)^2}{4})$$
\end{lemma}

\begin{proof}
Let $P_{out}=\frac{1}{v(\mathbf{y}_1)}\sum_{i=1}^{v(\mathbf{y}_1)}P(\epsilon_{out,i}>0)$. The RHS can be rewritten as $\sum_{d=1}^\infty p_d \exp(-\frac{d(\mu_{\min})^2}{4})$, where $d$ is the degree, and $p_d$ is the fraction of the nodes with degree $d$.
By Theorem 2.11,
% \textcolor{red}{WD: cite our local and global powerlaw thms}, 
$p_d \overset{a.s.}{\to} \alpha B(d,\alpha+1)$, i.e., the degree distribution converges almost surely to the Yule-Simon Distribution.  The conclusion follows immediately.
% , which can be approximated by Yule's law. This gives:
% $$P_{out}^1\simeq\sum_{d}\alpha B(d,\alpha+1)\exp(-d(**)/4)$$
\end{proof}

\paragraph{Bounding the RHS of~\eqref{eq2}}. 
We next show $\frac{1}{v(\bfy)}N(\xi(e))$ is bounded. The proof relies on the following Lemma 5 from~\cite{amini2013pseudo} which we write here for convenience: 
\begin{lemma} \label{Lemma 2} For independent Bernoulli R.V. $X_i$, $i\in [n]$ and any $u>\frac{1}{e}$,
$$
P\left(\overline{X}\ge eu \frac{1}{n}\sum_{i=1}^{n}\mathbb{E}(X_i)\right)\le \exp \left(-e\left(\sum_{i=1}^{n}\mathbb{E}(X_i) \right)u\log u\right) 
$$ 
\end{lemma}

%\begin{lemma} \label{Lemma 6} Assume $\gamma\in (0,1)$, and that $n$ and $\gamma n$ are positive integers, then
%$${n\choose\gamma n} \le\exp \{n(h(\gamma))+\kappa_{\gamma}(2n)\}$$
%where $\kappa_{\gamma}(n)=\frac{1}{n}[\log \frac{n}{4\pi\gamma(1-\gamma)}+\frac{1}{3n}]$, and $h(\gamma)$ is the binary entropy function.
%\end{lemma}

% \textcolor{red}{I think you need to have the lemmas and explain that they require independent bernoulli r.v.s.  Right now it's hard to understand why we go through the whole out degree argument.}
Note that $1(\epsilon_{out,i}\ge 0)$ is a Bernoulli random variable. Given $\{f_i^{(b)}\}$, $1(\epsilon_{out,i}>0)$ are independent random variables for $i\in [v(\bfy)]$.  Then by Lemma~\ref{Lemma 2}:
$$P\left[\frac{1}{v(\bfy)}N(\xi(e))\ge euP_{out}\right]\le \exp \left(-ev(\bfy)P_{out} u\log u\right)$$
which guarantees that as $m \to \infty$ the misclassification rate can be bounded by $u \cdot e \cdot P_{out}$.  By Lemma~\ref{lemma:approxp1}, $P_{out}$ converges to a constant.  We next restrict to higher-degree nodes which allows us to have the misclassification rate go to zero as $m \to \infty$. 

\begin{lemma}
\label{Lemma deg}
Let $v_D(\bfy)$ denote the set of nodes with degree greater than $D$. Then define 
$$P_D=\frac{1}{v_D(y)}\sum_{i=1}^{v_D(\bfy)}P(\epsilon_{out,i}>0) \to C_{\alpha}^{-1} \sum_{d>D}\alpha B(d,\alpha+1)\exp(-d\mu_{min}^2)/4)$$ 
where $C_{\alpha} = \sum_{d>D}\alpha B(d,\alpha+1) \leq 1$. 
Then one can construct a sequence $D_m$, such that as $m(\bfy)\rightarrow \infty$, $P_{D_m}\rightarrow 0$, and $v_{D_m}(\bfy)P_{D_m} \rightarrow \infty$.  Any such sequence guarantees
$$\lim_{m(y)\rightarrow\infty}P\left( \frac{1}{v_{D_m} (\bfy)}N_{D_m}(\xi(e_{out}))> 0\right)=0.$$
\end{lemma}

\begin{figure}[!th]
\centering
\begin{subfigure}{ 0.5\textwidth}
  \centering
  \includegraphics[width=1\linewidth]{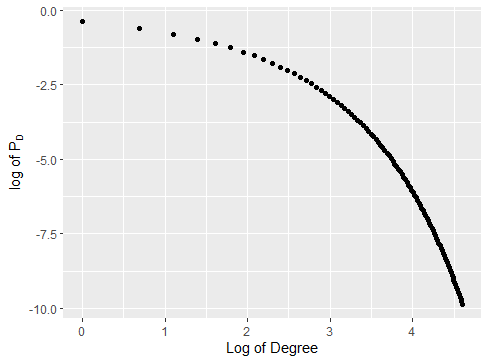}
  \caption{}
  \label{fig:sub1}
\end{subfigure}%
\begin{subfigure}{ 0.5\textwidth}
  \centering
  \includegraphics[width=1\linewidth]{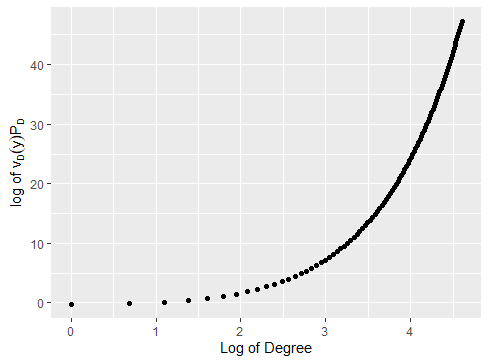}
  \caption{}
  \label{fig:sub2}
\end{subfigure}
\caption{(a) $P_{D_m}$ as a function of the degree cutoff; (b) $v_{D_m}(\bfy)P_{D_m}$ as a function of the degree cutoff. $P_{D_m}$ is approximated by $C_{\alpha}^{-1} \sum_{d>D_m}\alpha B(d,\alpha+1)\exp(-d\mu_{min}^2/4)$, and $v_{D_m}(\bfy)$ is approximated by $v(\bfy) \cdot D_m B(D_m,\alpha+1)$ by applying Yule's law. The $v(\bfy)$ is approximated by $m(\bfy)^{\alpha}$.  Specifically, $D_m=\log (m(\bfy))$, which leads to $v_{D_m}(\bfy)\simeq \exp(D_m)^{\alpha}D_m B(D_m,\alpha+1)$.}
\label{fig:th1}
\end{figure}

Figure~\ref{fig:th1} plots the $P_{D_m}$ and $v_D(\bfy)P_{D_m}$ as a function of the degree cutoff when $D=\log(m(\bfy))$.

%One of the example is letting $D=m(\bfy)^{\alpha}$, such that as $m(\bfy)\rightarrow\infty$, $D\rightarrow\infty$, and $P_D\rightarrow 0$. On the other hand, $v(\bfy)P_D=m(y_1)*\frac{\alpha B(D,\alpha+1)}{D B(D,\alpha+1)}=\frac{\alpha m(y_1)}{m(y_1)^{\alpha}}$, which goes to infinity as $m(y_1)\rightarrow\infty$.

%\textcolor{red}{Again, I'd just do it all at once.  Not sure why we need to split.  Specifically the issue is that $1/v(\bfy) \not = 1/v(\bfy_b)$. So the proofs you show below use this it seems.  Now I think they're fine cause $1/v(\bfy) \leq 1/v(\bfy_b)$ almost surely so it should work.  But I don't see why it's necessary. }
\paragraph{From Directed to the Undirected Setting.}
By symmetry we have the same bound for the mis-specification rate of the nodes when considering the in-degree of the nodes.  The reason we needed to prove these bounds separately is that the concentration inequalities can only be applied to \emph{independent} random variables.  When considering the total degree of each node, the same interaction impacts out- and in-degree of the two nodes involved in the interaction.  Therefore, the independence assumption is violated.  To overcome this, we show how to use these separate in- and out- misspecification bounds to bound the misspecification in the undirected case. 
% That is, if we only use out- or in-degree information, then we can bound the misclassification of the node labels.  

Here, we consider the degree of the complete network. We define $\epsilon_i$ in a similar way as we did in the directed case. The misclassification of the node can happen only if the node is misclassified either in out-degree or in-degree. In other words, $1(\epsilon_i>0)\le 1(\epsilon_{out,i}>0)+1(\epsilon_{in,i}>0)$.

Note that in the presence of self interaction loop, we can rewrite $\epsilon_i=2*\epsilon_{self,i}+\epsilon_{out,-i}+\epsilon_{in,-i}$. A node is mis-classified if it is mis-labelled in either $e_{out}$ or $e_{in}$, or it is mis-classified given the in degree or out degree. In other words, $1(\epsilon_i>0)\le 1(\epsilon_{self,i}>0)+1(\epsilon_{out,-i}>0)+1(\epsilon_{in,-i}>0)$. This is bounded by $1(\epsilon_{out,i}>0)+1(\epsilon_{in,i}>0)$.

With this in mind, we can take the average sum of all the nodes in the network, which gives:
$$\frac{1}{v(y)}N(\xi(e))\le \frac{1}{v(y)}N(\xi(e_{out}))+\frac{1}{v(y)}N(\xi(e_{in}))$$
%$$=\frac{1}{2}[\frac{1}{v(\mathbf{y}_1)}N_1(\xi(e_{out}))+\frac{1}{v(\mathbf{y}_2)}N_2(\xi(e_{out}))]+\frac{1}{2}[\frac{1}{v(\mathbf{y}_1)}N_1(\xi(e_{in}))+\frac{1}{v(\mathbf{y}_2)}N_2(\xi(e_{in}))]$$

which leads to:

$$P\left(\frac{1}{v(y)}N(\xi(e)\right)\ge euP_{out}+euP_{in}$$
$$\le \exp(-ev(\bfy)P_{out}u\log u)+\exp(-ev(\bfy)P_{in}u\log u)$$

%$$P(\sup_{\xi}\frac{1}{v(y_m)}N(\xi(e))\ge e[u_{1,m}P_{out}^1]+u_{2,m}P_{out}^2]+u_{1,m}P_{in}^1]+u_{2,m}P_{in}^2])$$
%$$\le \text{exp}(v(\mathcal{Y}_1)[2h(\gamma_{e(out)})+2\kappa_{\gamma_{e(out)}}(2v(\mathcal{Y}_1))]-ev(\mathcal{Y}_1)P_{out}^1\mu_{1,m}\log \mu_{1,m})$$
%$$+\text{exp}(v(\mathcal{Y}_2)[2h(\gamma_{e(out)})+2\kappa_{\gamma_{e(out)}}(2v(\mathcal{Y}_2))]-ev(\mathcal{Y}_2)P_{out}^2\mu_{2,m}\log \mu_{2,m})$$
%$$+\text{exp}(v(\mathcal{Y}_1)[2h(\gamma_{e(in)})+2\kappa_{\gamma_{e(in)}}(2v(\mathcal{Y}_1))]-ev(\mathcal{Y}_1)P_{in}^1\mu_{1,m}\log \mu_{1,m})$$
%$$+\text{exp}(v(\mathcal{Y}_2)[2h(\gamma_{e(in)})+2\kappa_{\gamma_{e(in)}}(2v(\mathcal{Y}_2))]-ev(\mathcal{Y}_2)P_{in}^1\mu_{2,m}\log \mu_{2,m})$$

Follow a similar arguments as shown before, we can get to the conclusion that for high degree nodes in the network considering both in and out degree:
$$\lim_{m(\bfy)\rightarrow\infty}P\left(\frac{1}{v_{D_m} (y_m)}N_{D_m} (\xi(e)) > 0\right)=0$$

\subsection{Proof of Theorem 2.9}
\label{sec:pf2}
Let $\bfY_m$ be an interaction network consisting of $m$ interactions generated by the sequential description from Section 2.4.  Let $v(\bfY_m)$ be the number of vertices in the network. Within each community, denote $v^{(b)}(\bfY_m)$, $b\in[K]$ as the number of nodes in $b$th sub-network. Recall the block-specific parameters $0<\alpha_b<1$, and $\theta_b>-\alpha_b$. Since each block-specific sub-network is generated following an equivalent sequential description as the Hollywood model, the first half of Theorem 2.9 follows by Theorem 4.3 in~\cite{crane2016edge}. Specifically, by Theorem 3.8 in \cite{pitman2002combinatorial} we have
%$m(Y_b)^{-\alpha_b}v(Y_b)\rightarrow S_{\alpha_b}$ a.s., where $S_{\alpha_b}$ is a strictly positive and finite random variable. Therefore, $v(Y_b)\sim m(Y_b)S_{\alpha_b}$, such that 
$$\mathbb{E}(v^{(b)}(\bfY_m))\sim\frac{\Gamma(\theta_b+1)}{\alpha_b\Gamma(\theta_b+\alpha_b)}(\mu_b m)^{\alpha_b},\ n\rightarrow\infty$$
for $b\in[K]$, where $a_n \sim b_n$ for two sequences $a_n$ and $b_n$ implies $a_n/b_n \to 1$ as $n \to \infty$, and $\mu_b$ is the average arity of interactions when restricting to nodes from block~$b$.

% Suppose for the block structured network, the within block propensity and the between block propensity are the same across different blocks, that is $\B(i,j)=b$ for $i,j\in[K]$ and $i\neq j$, and $\B(i,i)=a$ for $i\in[K]$. For an arbitrary interaction $E_i$, $i\in[m(Y_m)]$, given $\{a,b,\pi_b\}$, the probability of selecting the initiating node from a certain block $b\in[K]$ is given by:
% $$\mathbb{P}(\mathbf{C}_i^{(s)}=b)=\pi_b$$
% The probability of selecting the receiving node from a certain block $b\in [K]$ is given by:
% $$P(\mathbf{C}_i^{(r)}=b)=\sum_{b'}\pi_{b'}\B (b',b)$$
We can express the arity~$\mu_b$ in terms of the within- and between-block propensities~$\{ \B(b,b')\}_{b,b' \in [K]}$ and probabilities of initiating an interaction~$\{ \pi_b\}_{b \in [K]}$ so the above statement can be re-expressed as
$$\mathbb{E}(v^{(b)}(\bfY_m)) \sim \frac{\Gamma(\theta_b+1)}{\alpha_b\Gamma(\theta_b+\alpha_b)} \left[ m \times \left(\pi_b+\sum_{k=1}^\infty \nu_k \sum_{b'}\pi_{b'} k \B(b',b) \right) \right]^{\alpha_b}$$
where~$\nu_k$ is the distribution over the number of commentators (e.g, $\nu_1 = 1$ is the simple case of a single commentator in every interaction).  By linearity of expectations we then have:
$$\mathbb{E}(v(\bfY_m))=\sum_{b}\mathbb{E}(v^{(b)}(\bfY_m))\sim \sum_b\frac{\Gamma(\theta_b+1)}{\alpha_b\Gamma(\theta_b+\alpha_b)} \left[ m \times \left(\pi_b+ (\mu-1) \sum_{b'}\pi_{b'}\B(b',b) \right)\right]^{\alpha_b}$$
where~$\mu = 1 + \sum_{k\ge 1} k \nu_k$ is the average global arity.

To establish global sparsity of $(\bfY_m)_{m \geq 1}$, consider $(m^{-1} v(\bfY_m)^{m_\bullet (\bfY_m)})$ where $m_\bullet (\bfY_m)$ is the average total degree of $\bfY_m$.  By the strong law of large numbers, $m_{\bullet} (\bfY_m) \to \mu$ a.s. as $m \to \infty$ where $\mu = 1 + \sum_{k\ge 1} \nu_k k$ is the average arity of an interaction. Let $\alpha_{\star}=\max_b \alpha_b$, and $b_\star$ indicate the corresponding block. 
By the above discussion,~$v(\bfY_m) \sim (\mu m)^{\alpha_\star} S$ a.s. as $m \to \infty$ for a strictly positive and finite random variable~$S$.  It follows that $m^{-1} v(\bfY_m)^{m_\bullet(\bfY_m)} \sim m^{-1} (\mu n)^{\mu \alpha_\star} S$ a.s. as $m \to \infty$ which goes to infinity as long as $\mu \alpha_\star > 1$.  Thus, $(\bfY_m)$ is sparse with probability 1 provided $1/\mu < \alpha_\star < 1$.

\subsection{Proof of Theorem 2.11}
\label{sec:powerlawproof}

Let $\bfY_m$ be an interaction network consisting of $m$ interactions generated by the sequential description from Section 2.4.  Let $v(\bfY_m)$ be the number of vertices in the network. Within each community, denote $v^{(b)}(\bfY_m)$, $b\in[K]$ as the number of nodes in $b$th sub-network. Recall the block-specific parameters $0<\alpha_b<1$, and $\theta_b>-\alpha_b$. Since each block-specific sub-network is generated following an equivalent sequential description as the Hollywood model, the first half of Theorem 2.11 follows by Theorem 4.2 in~\cite{crane2016edge}. Specifically, by Theorem 3.11 in \cite{pitman2002combinatorial} we have
$N_d^{(b)} (\bfY_m) / v^{(b)}(\bfY_m) \to \alpha_b B(d, \alpha_b +1)$ a.s. for every $d \geq 1$ as $m \to \infty$. 

To prove global power-law, we note that $N_d (\bfY_m) = \sum_{b=1}^K N_d^{(b)} (\bfY_m)$, then $N_d (\bfY_m)/v(Y_m) = \sum_b N_d^{(b)} (\bfY_m)/ v^{(b)} (\bfY_m) \cdot v^{(b)}(\bfY_m)/v (\bfY_m) \to \alpha_{b_\star} B(d, \alpha_{b_\star} +1)$ since $v^{(b)}(\bfY_m) / v(\bfY_m) \to 1[\alpha_b = \alpha_{b_\star}]$ a.s. as $m \to \infty$.

\subsection{Proof of Theorem 2.5}
\label{sec:repthmproof}

We prove here a general characterization of edge exchangeable networks with node-level community structure with undirected binary edges.  Let $\fin_2 (\P)$ denote the set of all multisets of size~$2$.  We assume~$\P = \Nat$, i.e., the population is countably infinite.  The directed and multiple arity edge cases follows a similar argument with some additional notation.

Given a binary interaction-labelled network~$\bfY$ with $e(\bfY) = n$, let~$S: [n] \to \fin_2 (\mathbb{N})$ denote a \emph{selection function for}~$\bfY$ if $S$ is an interaction process whose induced interaction-labeled network agrees with $\bfY$.  Let~$S_B [n] \to \fin_2 ([K])$ denote the \emph{block selection function for}~$\bfY$ which induces the block network from~$\bfY$.  For simplicity, we say selection function to refer to the pair $\bfS := (S, S_B)$ since the two jointly define an interaction-labelled network with block structure.  Two selection functions~$\bfS, \bfS'$ are equivalent, $\bfS \equiv \bfS'$, if they correspond to the same interaction-labelled network with the same block structure. To every interaction-labelled network with block structure, we associate a canonical selection function defined by labeling the vertices \emph{and the blocks} in order of appearance.
% Canonical. Add Figure to show difference from the E2 paper (need colored nodes).

The $\fin_2 ([K] \times \Nat)$ simplex consists of all $(f_{(b,i), (b', j)})$ such that for all $j,i \geq -1$ $f_{(b,-1), (b', i)} = 0$ for all $b,b' \in [K]$ and $i \not = 0$, and $\sum_{b,b' \in [K], i,j \geq -1} f_{(b,i), (b', j)} = 1$.  For any $f$ in the simplex and $b \in [K]$ define
\begin{align*}
    f_{\bullet}^{(b)} &= \sum_{i,j \geq -1, b' \in [K]} f_{(b,i), (b', j)} \\
    f_{\bullet}^{(b,i)} &= \sum_{j \geq -1, b' \in [K]} f_{(b,i), (b', j)}.
\end{align*}
Every~$f = (f_{(b,i), (b', j)} )_{i,j \geq -1, b,b' \in [K]}$ in the joint $\fin_2 ([K] \times \Nat)$ simplex determines a probability distribution on interaction-labelled networks with block structure, denoted~$\epsilon_f$, as follows.  Let $E_1,E_2,\ldots$ be random iid random pairs $\{ (b,i), (b',j)\}$ with
\begin{equation}
\label{eq:repthm_maineq}
P( E_i = \{ (b,i), (b',j)\} | f) = f_{\{ (b,i), (b',j)\}}, \quad
j, i \geq -1 \quad \text{ and } \quad b,b' \in [K]
\end{equation}
Given the sequence $E_1,E_2,\ldots$ we define the selection function $S: \mathbb{N} \to \fin_2 ([K] \times \mathbb{Z})$ as follows. Let $m_{0,b} = 0$ for each $b \in [K]$.  For $n \geq 1$, suppose $m_{n-1, b} = z_b \leq 0$ for each $b \in [K]$.  If $E_n$ contains no 0s, then $S(n) = E_n$ and $m_{n,b} = m_{n-1,b}$.  If $E_n = \{ (b,0), (b', j) \}$ for some $j \geq 1$ then $S(n) = \{ (b, z_b-1), (b', j) \}$ and update $m_{n,b} = z_b -1$.  If $E_n = \{ (b,0), (b', 0) \}$ then $S(n) = \{ (b, z_b-1), (b', z_{b'} - 1) \}$ and update $m_{n,b} = z_b -1$ and $m_{n,b'} = z_{b'} - 1$.  If $E_n = \{ (b,0), (b, 0)\}$ then $S(n) = \{ (b, z_b-1), (b, z_{b} - 1) \}$ and $m_{n,b} = z_b - 1$. If $E_n = \{ (b,0), (b, -1)\}$ then $S(n) = \{ (b, z_b-1), (b, z_{b} - 2) \}$ and $m_{n,b} = z_b - 2$. These events are `blips' that involve vertices in different blocks that appear once and never again. Define $\bfY_B$ to be the interaction-labelled network with block structure~$B$ induced by $S$.  
\begin{prop}
The block-labelled network~$\bfY_B$ corresponding to $E_1,E_2,\ldots$ iid from~\eqref{eq:repthm_maineq} is block interaction exchangeable for all $f$ in the $\fin_2 ([K] \times [N])$-simplex.
\end{prop}

For identifiability, we define the \emph{rank ordering of $f$} by $f^{\downarrow} = \left( f^\downarrow_{\{ (b,i), (b',j)\}} \right)$ which is obtained by first reordering blocks~$1,\ldots,K$ so that $f_{\bullet}^{(b)} \geq f_{\bullet}^{(b')}$ for all $b'>b$.  Then we reorder elements within each block $1,2,\ldots$ so that $f_{\bullet}^{(b,i)} \geq f_{\bullet}^{(b,i+1)}$ for $i \geq 1$.  Ties can be handled in a similar fashion as in~\cite{crane2016edge}. Write~$\mathcal{F}^{\downarrow}$ to denote the space of rank reordered elements of the $\fin_2 ([K] \times \Nat)$-simplex.

As vertex labels other than $-1$ and $0$ are inconsequential, $\epsilon_f$ and $\epsilon_{f'}$ determine the same distribution for any $f,f'$ for which $f^{\downarrow} = f^{' \downarrow}$.  For any edge-labeled network $\bfY$, then $|\bfY|^\downarrow \in \mathcal{F}^\downarrow$ denote its signature, if it exists, as follows. Let~$S_{\bfY}$ be the canonical selection function for~$\bfY$.  For every $\{ (b,i), (b', j)\}$ $j \geq i \geq 1$ define
\begin{align*}
    f_{\{ (b,i), (b', j)\}} (\bfY) &= \lim_{n \to \infty} n^{-1} \sum_{k=1}^n 1\left[ S_{\bfY} (k)  = \{ (b,i), (b', j)\} \right] \\
    f^{(b,b')}_\bullet (\bfY) &= \lim_{n \to \infty} n^{-1} \sum_{k=1}^n 1\left[ (b,b') \in S_{\bfY} (k) \right] \\
    f^{(b,(b',j))}_\bullet (\bfY) &= \lim_{n \to \infty} n^{-1} \sum_{k=1}^n 1\left[ (b,(b',j)) \in S_{\bfY} (k) \right] \\
    f_{\bullet}^{(i,j)} (\bfY) &= \lim_{n \to \infty} \frac{\sum_{k=1}^n 1\left[ S_{\bfY} (k)  = \{ (b,i), (b', j)\} \right]}{\sum_{k=1}^n 1\left[ (b,b') \in S_{\bfY} (k) \right]}.
\end{align*}
if the limits exist, where $f_{\bullet}^{(i,j)} (\bfY)$ implicitly depends on $(b,b')$, i.e., it is the asymptotic fraction of interactions between $b$ and $b'$ that include $i$ and $j$. We also define
\begin{align*}
    f_{\{ (b,0), (b', j)\}} (\bfY) &= f_\bullet^{(b,(b',j))} (\bfY)  - \sum_{i=1}^\infty f_{\{ (b,i), (b', j)\} } \\
    f_{\{ (b,0), (b', 0)\}} (\bfY) &= f_\bullet^{(b,b')} (\bfY) -  \sum_{i,j=1}^\infty f_{\{ (b,i), (b', j)\}} (\bfY) \\
    f_{\{ (b,0), (b,0)\}} (\bfY) &= \lim_{n \to \infty} n^{-1} \sum_{k=1}^n \left( \sum_{l\geq 1} 1\left[ \{ (b,l), (b,l) \} = S_{\bfY} (k) \right] \right) -  \sum_{i=1}^\infty f_{\{ (b,i), (b, i)\}} (\bfY) \\
    f_{\{ (b,-1), (b, 0)\}} (\bfY) &= \lim_{n \to \infty} n^{-1} \sum_{k=1}^n \left( \sum_{l,r\geq 1; l \not = r} 1\left[ \{ (b,l), (b,r) \} = S_{\bfY} (k) \right] \right) -  \sum_{j>i\geq 1}^\infty f_{\{ (b,i), (b, j)\}} (\bfY).
\end{align*}
We can similarly define~$f^{(0,j)}_{\bullet}(\bfY)$ and other blip terms that are specific to pairs~$(b,b')$ as above.

\begin{thm}
\label{thm:bliprepthm}
Let~$\bfY$ be an interaction exchangeable network with block structure.  Then there exists a unique probability measure on $\phi = (\phi_K, \{ \phi_{\bar b} \})$ on $\mathcal{F}^\downarrow$ such that $\bfY \sim \epsilon_\phi$, where
\begin{equation}
    \label{eq:realrep}
\epsilon_\phi = \int_{\mathcal{F}^\downarrow} \epsilon_f(\cdot) \phi (df)
\end{equation}
That is, every interaction exchangeable network with block structure~$B$ can be generated by first sampling $f_K \sim \phi_K$ then sampling $\{ f_{\bar b}\} \sim \phi_{\bar b}$ and then, given $f := (f_k, \{ f_{\bar b}\} )$, generating the interaction process according to~\eqref{eq:repthm_maineq}.
\end{thm}

Theorem 2.5 follows as a corollary to Theorem~\ref{thm:bliprepthm} by ruling out blips.

\begin{proof}[Proof of Theorem~\ref{thm:bliprepthm}]

Equip the space of interaction labelled networks with block structure with the product-discrete topology induced by the metric
$$
d(\bfY, \bfY') = (1 + \sup \{ n \in \Nat: \bfY_n = \bfY'_n\} )^{-1}
$$
with $1/\infty = 0$, and~$\mathcal{F}^\downarrow$ with the topology induced by
$$
d_{\mathcal{F}^\downarrow} (f,f') =  \sum_{b,b'} \sum_{j \geq i \geq -1} | f_{\{ (b,i), (b', j)\}} - f_{\{ (b,i), (b', j)\}}'|
$$
We work with the Borel $\sigma$-fields induced by these topologies.

Let $\bfY$ be an interaction exchangeable random network with block structure~$B$.  Let~$S_{\bfY}$ be the canonical selection function, and $(\xi_1,\ldots, \xi_K)$ and $\xi^{(b)}_{1},\xi^{(b)}_{2},\ldots$ for each $b \in [K]$ be K iid sequences of $\text{Uniform}[0,1]$ random variables, which are independent of $\bfY$.  Given $\bfY$ and $\xi:= \left( \{\xi_k\}_{k=1}^K, \{ \xi^{(b)}_i \}_{i \geq 1, b\in [K]} \right)$, we define  $Z: \Nat \to \fin_2 ([0,1] \times [0,1])$ by $Z(n) = \{ (\xi_b, \xi_i^{(b)}), (\xi_{b'}, \xi_j^{(b')}) \}$ on the event $S_{\bfY} (n) = \{ (b,i), (b', j)\}$ for $n \geq 1$.

By independence of $\bfY$ and $\xi$ and interaction exchangeability of $\bfY$, $(Z(n))_{n \geq 1}$ is an exchangeable sequence taking values in the Polish space $\fin_2 ([0,1] \times [0,1])$.  By de Finetti's theorem, there exists a unique measure $\mu$ on the space of probability measures on $fin_2 ([0,1] \times [0,1])$ such that $Z =_D Z^\star = (Z^\star (n))_{n \geq 1}$ with 
$$
pr (Z^\star \in \dot) = \int m^{\infty} (\cdot) \mu(dm),
$$
where $m^\infty$ denotes the infinite produce measure of $m$. Specifically there exists a random measure $\nu$ on $\fin_2 ([0,1] \times [0,1])$ such that $P(Z \in \dot | \nu) = \nu^\infty$ almost surely.  Given~$\nu$, we define
\begin{align*}
    f_{\{(b,i), (b',j)\}} &= \nu (\{(\xi_b,\xi^{(b)}_i), (\xi_{b'},\xi^{(b')}_j)\}), \quad i,j \geq 1 \\
    f^{(b,b')}_{\bullet} &= \nu (\{(x_1,y_1), (x_2,y_2)\} \in \fin_2 ([0,1] \times [0,1]) : (\xi_b, \xi_{b'}) = \{ x_1, x_2\}), \\
    f^{\{ b,(b',j)\}}_{\bullet} &= \nu (\{(x_1,y_1), (x_2,y_2)\} \in \fin_2 ([0,1] \times [0,1]) : (\xi_b, \xi_{b'}) = \{ x_1, x_2\}, \xi^{(b')}_j \in (y_1, y_2)), \\
    f^{(i,j)}_{\bullet} &= f_{\{(b,i), (b',j)\}}/f^{(b,b')}_{\bullet},
\end{align*}
and the blip-related terms:
\begin{align*}
    f_{\{ (b, 0), (b', j)\}} &= f^{\{ b,(b',j)\}}_{\bullet} - \sum_{i=1}^\infty f_{\{(b,i), (b,j)\}} \\
    f_{\{ (b, 0), (b', 0)\}} &= f^{(b,b')}_{\bullet} - \sum_{i,j=1}^\infty f_{\{(b,i), (b',j)\}} \\
    f_{\{ (b, 0), (b, 0)\}} &= \nu \bigg(\{(x_1,y_1), (x_2,y_2)\} \in \fin_2 ([0,1] \times [0,1]): \\
    &(\xi_b, \xi_{b}) = \{ x_1, x_2\}, \{ u,u\} \in \{ y_1, y_2\} \bigg) - 
    \sum_{i=1}^\infty f_{\{(b,i), (b,i)\}} \\
    f_{\{ (b, 0), (b, -1)\}} &= \nu \bigg(\{(x_1,y_1), (x_2,y_2)\} \in \fin_2 ([0,1] \times [0,1]): \\
    &(\xi_b, \xi_{b}) = \{ x_1, x_2\}, \{ u,v\} \in \{ y_1, y_2\}, u \not = v \bigg) - 
    \sum_{j > i \ge 1}^\infty f_{\{(b,i), (b,j)\}}.
\end{align*}
We can then define blip-related terms,e.g.,~$f^{(0,j)} = f_{\{ (b, 0), (b', j)\}}/ f_\bullet^{(b,b')}$, in a similar fashion.

By construction~$\left(f_{\{(b,i), (b',j)\}}\right)$ is in the $\fin_2([K] \times \Nat)$-simplex and therefore,~$f^{\downarrow} \in \mathcal{F}^{\downarrow}$. Given $\nu$, let $(Z', S')$ be an iid copy of $(Z, S_{\bfY})$ and let $\bfY'$ be the interaction-labelled network induced by $S'$.  We next show that $\pr (\bfY' \in \cdot | \nu) = \epsilon_{f^\downarrow}$ for $f^\downarrow$ defined above from $\nu$.

Define~$A_{b,b'} := \{ (i,j) \in \Nat^2 : f_\bullet^{(i,j)} > 0 \}$ and $\xi_{A_{b,b'}} := \{ (\xi^b_i, \xi^{b'}_j) : (i,j) \in A_{b,b'} \}$.  Then it follows that
$$
\pr ( Z'(1) \cap \xi_{A_{b,b'}} = \emptyset | \nu, (b,b') \in Z'(1)) = f_{\{ (b, 0), (b', 0)\}} + 1[b = b'] f_{\{ (b, 0), (b, -1)\}}.
$$
and
$$
\pr ( Z'(1) \cap \xi_{A_{b,b'}} = \{ i \} | \nu, (b,b') \in Z'(1) ) = f_{\{ (b, i), (b', 0)\}}
$$
By exchangeability, $(i,j) \not \in A_{b,b'}$ implies $\{ (\xi_b, \xi^{(b)}_i), (\xi_{b'}, \xi^{(b')}_j) \}$ appears at most once in $Z$ with probability $1$.  Moreover, 
\begin{align*}
\pr ( Z'(1) \cap \{ (b, \xi_{A_{b,b}}) \} &= \emptyset \text{ and } Z'_B(1) = \{(b,u), (b,u)\} \text{ for some } u \in [0,1]  | \nu ) = f_{\{(b,0), (b,0)\}}, \\
\pr ( Z'(1) \cap \{ (b, \xi_{A_{b,b}}) \} &= \emptyset \text{ and } Z'_B(1) = \{(b,u), (b,v)\} \text{ for some } u\not = v \in [0,1]  | \nu ) = f_{\{(b,0), (b,-1)\}}, \\
\end{align*}
Now define $X' : \Nat \to \fin_2 ([K] \times \Nat \cup \{-1,0\} )$ and the random selection function~$S_{X'}$ as follows.  Let $m^{(b)}_0 = 0$ for all $b \in [K]$. For $n \geq 1$, suppose $m^{(b)}_{n-1} = z_b \leq 0$.  If $Z'(n) \cup \xi_{A_b} = \{ \xi_i \}$ and $Z'(n) \cup \xi_{A_{b'}} = \{ \xi_j \}$ for some $i,j \in \Nat$ and $b,b' \in [K]$ then put $X'(n) = S_{X'} (n) = \{ (b,i), (b', j)\}$. If $(b,b') \in Z'(n)$ with $b \not = b' \in [K]$ and $Z'(n) \cup \xi_{A_b} = \emptyset$ but $Z'(n) \cup \xi_{A_{b'}} = \{ \xi_j \}$ then put $X'(n) = S_{X'} (n) = \{ (b,z_b-1), (b', j)\}$ and set $m^{(b)}_n = z_b - 1$. If $(b,b) \in Z'(n)$ and $Z'(n) \cup \xi_{A_b} = \emptyset$ and $Z'(n) = \{ (b,u), (b,u)\}$ for some $u \in [0,1]$ then put $X'(n) = S_{X'} (n) = \{ (b,z_b-1), (b, z_b-1)\}$ and set $m^{(b)}_n = z_b - 1$. If $(b,b) \in Z'(n)$ and $Z'(n) \cup \xi_{A_b} = \emptyset$ and $Z'(n) = \{ (b,u), (b,v)\}$ for some $u \not = v \in [0,1]$ then put $X'(n) = S_{X'} (n) = \{ (b,z_b-1), (b, z_b-2)\}$ and set $m^{(b)}_n = z_b - 2$. By construction, $S_{X'} \equiv S'$ and, given $f$, $X'$ is conditionally i.i.d from~\eqref{eq:repthm_maineq}.  The integral representation in~\eqref{eq:realrep} follows by de Finetti's theorem, which completes the proof.
\end{proof}

\subsection{Misspecification Rate as a function of degree cutoff}
\label{sec:mis_rate}

\begin{figure}
\centering
\begin{subfigure}{0.3\textwidth}
  \centering
  \includegraphics[width=.9\linewidth]{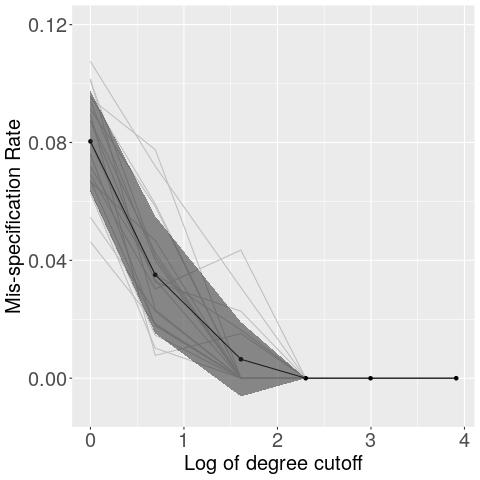}
  \caption{}
  \label{fig:sub.1}
\end{subfigure}
\begin{subfigure}{0.3\textwidth}
  \centering
  \includegraphics[width=.9\linewidth]{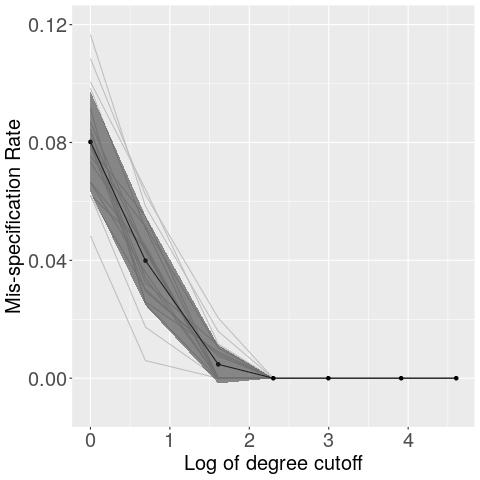}
  \caption{}
  \label{fig:sub.2}
\end{subfigure}
\begin{subfigure}{0.3\textwidth}
  \centering
  \includegraphics[width=.9\linewidth]{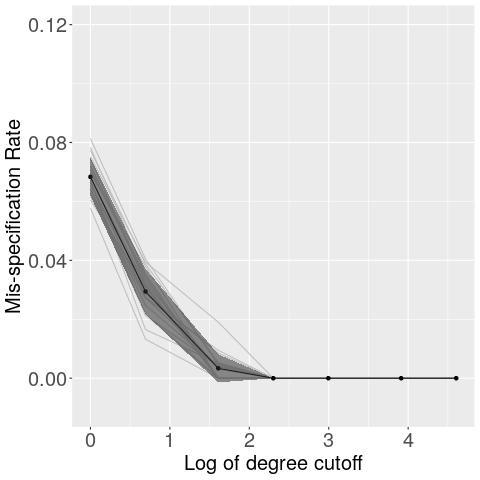}
  \caption{}
  \label{fig:sub.3}
\end{subfigure}
\caption{The mis-specification rate as a function of the degree cutoff, in the setting where $\alpha_1=\alpha_2=0.5$, $\B(1,1)=\B(2,2)=0.9$, with (a) 1,000, (b) 2,500, and (c) 10,000 interactions presented in the network. The solid dark line is the average $L_2$ norm over 20 repeats. The $L_2$ norm is calculated based on a binary classifier for each node.}
\label{fig:spg_high_deg_05}
\end{figure}

Section 4.0.1 showed the $L_2$ norm as a function of the degree cutoff where the inferred block assignments are calculated based on the average block assignments over the Gibbs samplers. Here, we consider another criteria to assess the inferred block assignments that is based on the mis-specification rate defined as $\inf_{\rho: [K] \to [K]} \frac{1}{v(\bfy_m)}\sum_{i=1}^{v(\bfy_m)}1(\B(i)\neq \rho \hat{\B}(i))$. A binary classifier $\hat{\B}(i)$ for each node is determined by whether the node has more than 0.5 of chance being assigned to a certain block in all valid Gibbs iterations. The results are shown in Figure~\ref{fig:spg_high_deg_05}. The conclusions are similar to the conclusions in Section 4.0.1, where the misspecification rate decreases to 0 as increase of the degree cutoff.

\subsection{L2 norm for high-degree nodes}

To demonstrate the consistency conclusion on high-degree nodes, we calculate the $L_2$ norm for nodes with degree greater than certain threshold. We select the cutoff value at $0.1*m(\mathbf{y}_m)^{\alpha}$. The multiplication of constant $0.1$ is to guarantee the existence of nodes with degree greater than the cutoff value, especially when $\alpha$ is large. The results are shown in Table~\ref{tab:l2norm_same_deg}. A significant decrease in the $L_2$ norms are observed. For example, in the setting where $\alpha_1=\alpha_2=0.7$, $\B(1,1)=\B(2,2)=0.9$, $m(\mathbf{y}_m)=10,000$, the $L_2$ norm decreases from 0.143 (0.008) to 0 (0). Note that when the power-law parameter is small, i.e. $\alpha_1=\alpha_2=0.1$, the degree cutoff threshold is smaller than 1. That is, all the nodes are included when calculating the $L_2$ norm. Therefore, the corresponding results don't change from Table 1 in main context.

\begin{table}
\centering
%\begin{adjustwidth}{-1.8cm}{}
\begin{tabular}{SSSSS} \toprule
    & {\# Interactions} &{$\B=\{0.1,0.9\}$}  & {$\B=\{0.3,0.7\}$} &{$\B=\{0.5,0.5\}$} \\ \midrule
    {$\alpha=\{0.1,0.1\}$} & 1,000  & {0.043 (0.016)} & {0.183 (0.033)} & {0.458 (0.028)} \\ 
      &2,500 & {0.067 (0.096)} & {0.161 (0.042)} & {0.457 (0.023)} \\
       & 10,000  & {0.053 (0.072)} & {0.23 (0.14)} & {0.454 (0.018)}\\
      \midrule

      {$\alpha=\{0.3,0.3\}$} &{1,000}  & {0.075 (0.013)} & {0.255 (0.067)} & {0.475 (0.020)} \\ 
      &{2,500} & {0.024 (0.009)} & {0.162 (0.072)} & {0.461 (0.035)}\\
       & {10,000}  & {0.019 (0.006)} & {0.166 (0.090)} & {0.459 (0.026)} \\
      \midrule

    {$\alpha=\{0.5,0.5\}$} & {1,000} & {0.010 (0.006)}   & {0.131 (0.028)} &{0.454 (0.038)} \\ 
     &{2,500} & {0.002 (0.003)} & {0.087 (0.081)} & {0.463 (0.027)} \\
     & {10,000} & {3.76e-6 (6.59e-6)}  & {0.055 (0.084)} & {0.460 (0.017)} \\
    \midrule

    {$\alpha=\{0.7,0.7\}$} &{1,000}  & {0.022 (0.092)} & {0.091 (0.099)} & {0.457 (0.031)} \\ 
      &{2,500} & {0 (0)} & {0.042 (0.064)} & {0.455 (0.026)}\\
       & {10,000}  & {0 (0)} & {3.44e-6 (1.14e-5)} & {0.433 (0.027)} \\
      \midrule

    {$\alpha=\{0.9,0.9\}$} &{1,000}   & {0.074 (0.155)}  & {0.314 (0.157)} & {0.360 (0.128)} \\ 
    &{2,500}  & {0.007 (0.030)} & {0.099 (0.153)} & {0.308 (0.143)} \\
     & {10,000} & {0 (0)}  & {0.056 (0.101)} & {0.171 (0.131)} \\ 
     \bottomrule
\end{tabular}
%\end{adjustwidth}
 \caption{Standardized average L2 norm (SD) of the inferred block assignments and the underlying truth for high-degree nodes with the degree cutoff set to be proportional to $m(\mathbf{y}_m)$.} 
 \label{tab:l2norm_same_deg}
\end{table}

\subsection{Recovering Block structure with different power-law parameter values}
\label{sec:label_diff_alpha}

In Section 4, we show the results from the setting where the power-law parameters $\alpha_1=\alpha_2$. Table 1 shows the $L_2$ norm calculated from two blocks of different power-law parameter values. The generative model is the same as we used in the Section 4.0.2. 

From the results shown in Table 1, we get the following conclusions. First, for fixed $\{\alpha_1,\alpha_2\}$, the higher the within block connection probability, the better the recovery of the block assignments. 
Second, as the number of observations increases, accuracy of the inferred block assignments increases across almost all settings.   
Third, for fixed $\B$ as the difference in power-law parameters decreases between the two blocks, the block assignment accuracy decreases. 

\begin{table}
\centering
%\begin{adjustwidth}{-1.8cm}{}
\begin{tabular}{SSSSSS} \toprule
    & {Interactions} &{$\alpha=\{0.1,0.9\}$}  & {$\alpha=\{0.2,0.8\}$} &{$\alpha=\{0.3,0.7\}$}  & {$\alpha=\{0.4,0.6\}$}\\ \midrule
    {$\B=\{0.1,0.9\}$} & 1,000  & {0.085 (0.049)} & {0.111 (0.070)} & {0.151 (0.077)} & {0.172 (0.079)}\\ 
      &10,000 & {0.043 (0.024)} & {0.047 (0.038)} & {0.081 (0.064)} &{0.151 (0.079)}\\
       & 100,000  & {0.014 (0.011)} & {0.030 (0.026)} & {0.084 (0.056)}&{0.109 (0.068)}\\
      \midrule

      {$\B=\{0.2,0.8\}$} &{1,000}  & {0.134 (0.031)} & {0.161 (0.063)} & {0.200 (0.086)} & {0.270 (0.085)}\\ 
      &{10,000} & {0.061 (0.009)} & {0.111 (0.017)} & {0.202 (0.049)}& {0.304 (0.040)}\\
       & {100,000}  & {0.035 (0.029)} & {0.061 (0.007)} & {0.144 (0.012)} &{0.259 (0.020)}\\
      \midrule

    {$\B=\{0.3,0.7\}$} & {1,000} & {0.132 (0.054)}   & {0.153 (0.071)} &{0.250 (0.115)} & {0.304 (0.076)}\\ 
     &{10,000} & {0.102 (0.110)} & {0.092 (0.059)} & {0.159 (0.097)} & {0.271 (0.106)}\\
     & {100,000} & {0.041 (0.076)}  & {0.075 (0.053)} & {0.093 (0.054)} & {0.276 (0.096)}\\
    \midrule

    {$\B=\{0.4,0.6\}$} &{1,000}  & {0.256 (0.094)} & {0.255 (0.105)} & {0.345 (0.089)} & {0.446 (0.054)}\\ 
      &{10,000} & {0.187 (0.091)} & {0.239 (0.122)} & {0.275 (0.084)}& {0.419 (0.052)}\\
       & {100,000}  & {0.242 (0.140)} & {0.134 (0.084)} & {0.198 (0.077)} & {0.380 (0.073)}\\
      \midrule

    {$\B=\{0.5,0.5\}$} &{1,000}   & {0.369 (0.093)}  & {0.390 (0.078)} & {0.452 (0.042)} & {0.461 (0.042)}\\ 
    &{10,000}  & {0.241 (0.092)} & {0.261 (0.088)} & {0.335 (0.091)} &{0.451 (0.030)}\\
     & {100,000} & {0.345 (0.067)}  & {0.247 (0.100)} & {0.269 (0.112)} &{0.391 (0.074)}\\ 
     \bottomrule
\end{tabular}
%\end{adjustwidth}
 \caption{Standardized $L_2$ norm of the inferred block assignments and the underlying truth in different settings. For each set with the same values of the connectivity propensity $\{a,b\}$, the power-law parameters $\{\alpha_1,\alpha_2\}$, and the number of interactions ($\{1,000, 10,000, 100,000\}$), we repeated the simulation 20 times. The mean values (SD) over 20 simulations are shown here. We do observe some instability when the propensity of connection is close to 0.5.} 
 \label{tab:l2norm}
\end{table}

\subsection{Cross Entropy Loss}
\label{sec:6.3}
In Section 4.0.1, the $L_2$ norm was used as one metric to check if inferred block assignment agrees with the true block assignments (up to label switching). Here, we show another metric that can be applied to $K>2$ scenarios. The \emph{cross entropy loss} is defined for a specific block $b\in[K]$ as: $$\mathscr{L}_b=\sum_{j\in X_b}-p(\B(j)=b)\log q(\hat{\B}(j)=b)$$
where the $p(\cdot)$ is the probability of the true cluster assignment, and $q(\cdot)$ is the posterior mean of the inferred cluster assignment. In the two clusters scenario, for example, a node $s$ has block labeling $\B(s)=1$ in the generative model, while in the inferred block assignment, the mean probability node $s$ being assigned to block $1$ is 0.6, the cross entropy loss for node $s$ of cluster 1 is $\mathscr{L}_s=-1\times\log(0.6)$. In the best case where all the nodes are assigned correctly to the true block, the entropy loss is 0; in the case where the block assignments are completely wrong, the entropy loss is infinity; in the case of random guess, the expected entropy loss of a specific node is $-\log(0.5)$. Further define the overall cross entropy loss $\mathscr{L}$ and the average per node entropy loss $\mathscr{L}_s$ as:
$$\mathscr{L}=\sum_{b=1}^{K}\mathscr{L}_b;\mathscr{L}_s = \frac{1}{v(\mathbf{y}_m)} \mathscr{L}$$

The results of the per node entropy loss in the two block scenario is shown in Table~\ref{tab:CEL_same} and Table~\ref{CEL}. We labelled the block by selecting the smallest value of the entropy loss over all possible combinations. The results are consistent with those of the $L_2$ norm. Conclusions based on Table~\ref{tab:CEL_same} and Table~\ref{CEL} are similar to the conclusions based on $L_2$ norms.

%\textcolor{blue}{We compared our model with the DC-SBM/SBM/Spectral Clustering in the simulation sets?}

\begin{table}
\centering
%\begin{adjustwidth}{-1.8cm}{}
\begin{tabular}{SSSSS} \toprule
    & {\# Interactions} &{$\B=\{0.1,0.9\}$}  & {$\B=\{0.3,0.7\}$} &{$\B=\{0.5,0.5\}$} \\ \midrule
    {$\alpha=\{0.1,0.1\}$} & 1,000  & {0.097 (0.056)} & {0.336 (0.075)} & {0.686 (0.044)} \\ 
      &2,500 & {0.143 (0.123)} & {0.298 (0.059)} & {0.665 (0.030)} \\
       & 10,000  & {0.112 (0.121)} & {0.357 (0.165)} & {0.659 (0.028)}\\
      \midrule

      {$\alpha=\{0.3,0.3\}$} &{1,000}  & {0.161 (0.045)} & {0.460 (0.114)} & {0.696 (0.041)} \\ 
      &{2,500} & {0.143 (0.025)} & {0.412 (0.071)} & {0.465 (0.051)}\\
       & {10,000}  & {0.130 (0.027)} & {0.488 (0.265)} & {0.696 (0.040)} \\
      \midrule

    {$\alpha=\{0.5,0.5\}$} & {1,000} & {0.222 (0.034)}   & {0.521 (0.028)} &{0.701 (0.033)} \\ 
     &{2,500} & {0.223 (0.036)} & {0.498 (0.060)} & {0.703 (0.033)} \\
     & {10,000} & {0.198 (0.016)}  & {0.501 (0.081)} & {0.689 (0.026)} \\
    \midrule

    {$\alpha=\{0.7,0.7\}$} &{1,000}  & {0.360 (0.068)} & {0.597 (0.034)} & {0.708 (0.038)} \\ 
      &{2,500} & {0.318 (0.028)} & {0.593 (0.017)} & {0.694 (0.013)}\\
       & {10,000}  & {0.296 (0.017)} & {0.561 (0.015)} & {0.692 (0.026)} \\
      \midrule

    {$\alpha=\{0.9,0.9\}$} &{1,000}   & {0.606 (0.074)}  & {0.691 (0.016)} & {0.711 (0.064)} \\ 
    &{2,500}  & {0.550 (0.030)} & {0.679 (0.032)} & {0.696 (0.008)} \\
     & {10,000} & {0.487 (0.032)}  & {0.663 (0.017)} & {0.702 (0.013)} \\ 
     \bottomrule
\end{tabular}
%\end{adjustwidth}
 \caption{Cross Entropy Loss the inferred block assignments when comparing to the underlying truth in different settings. For each set with the same values of the connectivity propensity $\B$, the power-law parameters $\{\alpha_1,\alpha_2\}$, and the number of interactions ($\{1,000, 2,500, 10,000\}$), we repeated the simulation 20 times. The mean values (SD) over 20 simulations are shown here.} 
 \label{tab:CEL_same}
\end{table}

\begin{table}
%\caption{Cross Entropy Loss over different settings} 
\begin{adjustwidth}{-1.5cm}{}
\begin{tabular}{SSSSSS} 
    \toprule
    &{Interactions}&{$\{\alpha_b\}=\{0.1,0.9\}$}  & {$\{\alpha_b\}=\{0.2,0.8\}$} &{$\{\alpha_b\}=\{0.3,0.7\}$}  & {$\{\alpha_b\}=\{0.4,0.6\}$}\\ \midrule
    {$\{a,b\}=\{0.1,0.9\}$} &{1,000}  & {0.056 (0.014)} & {0.095 (0.028)} & {0.152 (0.028)} & {0.212 (0.040)}\\ 
      &{10,000} & {0.014 (0.004)} & {0.039 (0.013)} & {0.094 (0.012)} &{0.170 (0.032)}\\
       & {100,000}  & {0.0027 (0.0006)} & {0.013 (0.004)} & {0.054 (0.012)}&{0.134 (0.018)}\\
      \midrule

      {$\{a,b\}=\{0.2,0.8\}$} &{1,000}  & {0.084 (0.020)} & {0.151 (0.032)} & {0.239 (0.065)} & {0.343 (0.050)}\\ 
      &{10,000} & {0.019 (0.005)} & {0.055 (0.015)} & {0.157 (0.084)}&{0.298 (0.074)}\\
       & {100,000}  & {0.013 (0.033)} & {0.018 (0.004)} & {0.097 (0.040)}& {0.223 (0.029)}\\
      \midrule

    {$\{a,b\}=\{0.3,0.7\}$} & {1,000} & {0.102 (0.152)}   & {0.182 (0.050)} &{0.346 (0.100)} & {0.445 (0.061)}\\ 
     &{10,000} & {0.052 (0.118)} & {0.078 (0.020)} & {0.210 (0.089)} & {0.411 (0.123)}\\
     & {100,000} & {0.039 (0.086)}  & {0.048 (0.060)} & {0.099 (0.016)} & {0.342 (0.124)}\\
    \midrule

    {$\{a,b\}=\{0.4,0.6\}$} &{1,000}  & {0.161 (0.082)} & {0.345 (0.117)} & {0.489 (0.097)} & {0.625 (0.057)}\\ 
      &{10,000} & {0.120 (0.129)} & {0.279 (0.186)} & {0.311 (0.120)} &{0.546 (0.098)}\\
       & {100,000}  & {0.291 (0.202)} & {0.120 (0.125)} & {0.216 (0.118)}& {0.508 (0.103)}\\
      \midrule

    {$\{a,b\}=\{0.5,0.5\}$} &{1,000}   & {0.311 (0.202)}  & {0.517 (0.110)} & {0.621 (0.052)} & {0.658 (0.067)}\\ 
    &{10,000}  & {0.254 (0.158)} & {0.291 (0.140)} & {0.434 (0.115)} &{0.711 (0.136)}\\
     & {100,000} & {0.422 (0.107)} & {0.283 (0.146)} & {0.332 (0.140)} &{0.604 (0.049)}\\ 
     \bottomrule
\end{tabular}
 \end{adjustwidth}
 \caption{Cross Entropy Loss averaged by number of nodes in different settings. For each set with the same values of the connectivity propensity $\{a,b\}$, the power-law parameters $\{\alpha_1,\alpha_2\}$, and the number of interactions ($\{1,000, 10,000, 100,000\}$), we repeated the simulation 20 times. The mean values (SD) over 20 simulations are shown here.} 
  \label{CEL}
\end{table}

\subsection{The symmetry of the propensity matrix}
\label{sec:6.4}
In the main text, we discuss the setting where the propensity matrix is not symmetric, i.e., the interaction network is directed. A natural question is ``What if the interaction is not directed?'' One solution is presented here. 

Let $k\in[K]$ be the current dimension that's being updated. When $k=1$, we follow the same step as described in Section 4.2:
$${\B}(1,)|C_m\sim Dirichlet(\nu+\vec *) $$
where the Dirichlet distribution is characterized by $K$ parameters, $\vec *$ stands for the count of the interactions between block 1 and the other blocks. After the first row of the propensity matrix being sampled, we let $\B(,1)=\B(1,)$. For an arbitrary $k\ge 2$, we follow the sequential updates as described below:
$${\B}(k,k:K)^*|\{\B(1,),...,\B(k-1,)\},C_m\sim Dirichlet(\nu_{k:K}+\vec *_{k:K}) $$
$$\B(k,k:K)=\B(k,k:K)^**(1-\sum_{i=1}^{k-1}\B(k,i))$$
where $\nu_{k:K}$ stands for the last $K-k$ elements of the original vector $\nu$. The notation is the same for $\vec *_{k:K}$ and $\vec *$. For the last elements in row, we have the following update:
$$\B(K,K)=1-\sum_{i=1}^{K-1}\B(K,1:K-1)$$

We compare the model parameter estimates with and without the assumption that B is symmetric in the simulation data. We assume $\pi_k$s are all the same across different blocks; the inter block connectivity $b=0.05$, while the intra block connectivity $a=1-b*(K-1)$. The K is set to be 3 and 5. The number of interactions are 1,500 and 2,500. The power-law parameter are set to be $\{\alpha_b\}=\{0.2,0.5,0.8\}$ and $\{\alpha_b\}=\{0.1,0.3,0.5,0.7,0.9\}$ correspondingly. We repeated 20 times for each setting. The mean value of the estimates for the diagonal elements of the propensity matrix for $K=3$ an $K=5$ are $0.902\ (0.015)$ and $0.773\ (0.066)$ correspondingly as compared to $0.899\ (0.02)$ and $0.770\ (0.069)$ with symmetry assumption and without symmetry assumption. The difference is almost negligible assuming $\B$ is symmetric as compared to when $\B$ is not.

\subsection{Simulation results for varying \texorpdfstring{$\theta$}{Lg}s}
\label{sec:6.5}

In Section 4.0.2, we have shown the posteriors given $\theta_1=\theta_2=5$. Though both $\{\alpha_b\}$ and $\{\theta_b\}$ are the power-law distribution parameters, the values of $\{\alpha_b\}$ have more contribution to the observed power-law properties. For the completeness, we show here the parameter estimates of the model parameters with varying values of $\theta$s. to be more specific, we have $\theta$s to be chosen from the following values: $\{\{10,10\},\{1,1\},\{10,1\},\{1,10\}\}$. In all these settings, the conclusions based on the means of the posteriors are the similar to those when $\theta=\{5,5\}$, with minor variations in the specific values.

\begin{table}
\begin{adjustwidth}{-2cm}{}
\vspace*{0cm}
\begin{tabular}{SSSSSS} \toprule
    {\textbf{1,000 interactions}} 
    & {Parameters} & {$\{\alpha_b\}=\{ 0.1, 0.9 \}$}  & {$\{\alpha_b\}=\{0.2,0.8\}$} &{$\{\alpha_b\}=\{0.3,0.7\}$} & {$\{\alpha_b\}=\{0.4,0.6\}$} \\ \midrule
    {$\{a,b\}=\{0.9,0.1\}$} &{$\alpha_1$}  & {0.378 (0.110)}& {0.360 (0.106)} & {0.451 (0.113)}& {0.495 (0.077)}  \\ 
      %&{$\tilde{\alpha}_1$ }  & {} & {} & {}& {}  \\ 
      &{$\alpha_2$}& {0.908 (0.015)} & {0.809 (0.023)}& {0.694 (0.069)} & {0.611 (0.057)} \\
       %&{$\tilde{\alpha}_2$ }  & {} & {} & {}& {} \\
       & {Diagonal} & {0.910 (0.021)}& {0.910 (0.020)} & {0.906 (0.019)} & {0.902 (0.019)} \\
      \midrule
    {$\{a,b\}=\{0.7,0.3\}$} &{$\alpha_1$}  & {0.409 (0.206)}& {0.511 (0.167)}& {0.479 (0.102)}& {0.520 (0.096)} \\ 
      %&{$\tilde{\alpha}_1$ }  & {}& {} & {} &  {} \\ 
      &{$\alpha_2$}& {0.863 (0.124)} & {0.748 (0.112)} & {0.705 (0.051)} & {0.593 (0.066)} \\
       %&{$\tilde{\alpha}_2$ }  & {}& {} & {} & {} \\
       & {Diagonal} & {0.716 (0.030)} & {0.728 (0.031)} & {0.714 (0.031)} & {0.705 (0.032)} \\
      \midrule
    {$\{a,b\}=\{0.5,0.5\}$} &{$\alpha_1$}  & {0.736 (0.214)} & {0.677 (0.128)} & {0.619 (0.091)} & {0.567 (0.082)} \\ 
      %&{$\tilde{\alpha}_1$ }  & {}& {} & {}& {} \\ 
      &{$\alpha_2$}& {0.757 (0.172)} & {0.718 (0.115)} & {0.629 (0.084)} & {0.573 (0.081)} \\
       %&{$\tilde{\alpha}_2$ }  & {}& {} & {} & {}  \\
       & {Diagonal} & {0.540 (0.052)} & {0.541 (0.057)} & {0.554 (0.061)} & {0.535 (0.057)} \\
     \bottomrule
    {\textbf{10,000 interactions}} 
    &{Parameters}&{$\{\alpha_b\}=\{0.1,0.9\}$}  & {$\{\alpha_b\}=\{0.2,0.8\}$} &{$\{\alpha_b\}=\{0.3,0.7\}$} & {$\{\alpha_b\}=\{0.4,0.6\}$} \\ \midrule
    {$\{a,b\}=\{0.9,0.1\}$} &{$\alpha_1$}  & {0.225 (0.072)} & {0.289 (0.062)}& {0.364 (0.04)}& {0.443 (0.038)} \\ 
      %&{$\tilde{\alpha}_1$ }  & {}& {} & {}& {} \\ 
      &{$\alpha_2$}& {0.901 (0.006)} & {0.803 (0.009)}& {0.707 (0.013)} & {0.614 (0.02)} \\
       %&{$\tilde{\alpha}_2$ }  & {}& {} & {} & {} \\
       & {Diagonal} & {0.901 (0.006)}& {0.901 (0.006)} & {0.902 (0.005)} & {0.901 (0.006)} \\
      \midrule
    {$\{a,b\}=\{0.7,0.3\}$} &{$\alpha_1$}  & {0.258 (0.092)} & {0.408 (0.161)}& {0.411 (0.11)}& {0.448 (0.056)} \\ 
      %&{$\tilde{\alpha}_1$ }  & {} & {} & {}& {}  \\ 
      &{$\alpha_2$}& {0.902 (0.005)} & {0.785 (0.066)}& {0.695 (0.054)}& {0.603 (0.032)}\\
       %&{$\tilde{\alpha}_2$ }  & {} & {} & {}& {}  \\
       & {Diagonal} & {0.704 (0.009)}& {0.674 (0.068)} & {0.691 (0.044)} & {0.694 (0.035)} \\
      \midrule
    {$\{a,b\}=\{0.5,0.5\}$} &{$\alpha_1$}  & {0.865 (0.163)} & {0.724 (0.162)}& {0.626 (0.095)}& {0.549 (0.059)} \\ 
      %&{$\tilde{\alpha}_1$ }  & {} & {} & {}& {}  \\ 
      &{$\alpha_2$}& {0.703 (0.127)}& {0.657 (0.126)}& {0.637 (0.081)}& {0.555 (0.051)}\\
       %&{$\tilde{\alpha}_2$ }  & {}& {} & {} & {} \\
       & {Diagonal} & {0.51 (0.015)} & {0.51 (0.015)} & {0.513 (0.017)} & {0.51 (0.017)} \\
     \bottomrule
\end{tabular}
\end{adjustwidth}
\caption{Posterior means (SD) of power-law parameters $\{\alpha_b\}$ and within/between cluster propensity $\{a,b\}$ in different settings, with $\{\theta_1,\theta_2\}=\{10,10\}$} 
\end{table}

\begin{table}
\begin{adjustwidth}{-2cm}{}
\vspace*{0cm}
\begin{tabular}{SSSSSS} \toprule
    {\textbf{1,000 interactions}} 
    &{Parameters}&{$\{\alpha_b\}=\{0.1,0.9\}$}  & {$\{\alpha_b\}=\{0.2,0.8\}$} &{$\{\alpha_b\}=\{0.3,0.7\}$} & {$\{\alpha_b\}=\{0.4,0.6\}$} \\ \midrule
    {$\{a,b\}=\{0.9,0.1\}$} &{$\alpha_1$}  & {0.218 (0.157)}& {0.224 (0.140)} & {0.304 (0.141)}& {0.354 (0.142)}  \\ 
      %&{$\tilde{\alpha}_1$ }  & {} & {} & {}& {}  \\ 
      &{$\alpha_2$}& {0.898 (0.015)} & {0.796 (0.031)}& {0.681 (0.052)} & {0.585 (0.064)} \\
       %&{$\tilde{\alpha}_2$ }  & {} & {} & {}& {} \\
       & {Diagonal} & {0.908 (0.016)}& {0.885 (0.014)} & {0.892 (0.057)} & {0.897 (0.019)} \\
      \midrule
    {$\{a,b\}=\{0.7,0.3\}$} &{$\alpha_1$}  & {0.367 (0.281)}& {0.276 (0.221)}& {0.374 (0.203)}& {0.341 (0.157)} \\ 
      %&{$\tilde{\alpha}_1$ }  & {}& {} & {} &  {} \\ 
      &{$\alpha_2$}& {0.870 (0.100)} & {0.782 (0.100)} & {0.689 (0.074)} & {0.580 (0.079)} \\
       %&{$\tilde{\alpha}_2$ }  & {}& {} & {} & {} \\
       & {Diagonal} & {0.667 (0.078)} & {0.691 (0.06)} & {0.678 (0.063)} & {0.694 (0.046)} \\
      \midrule
    {$\{a,b\}=\{0.5,0.5\}$} &{$\alpha_1$}  & {0.386 (0.268)} & {0.442 (0.266)} & {0.414 (0.237)} & {0.348 (0.186)} \\ 
      %&{$\tilde{\alpha}_1$ }  & {}& {} & {}& {} \\ 
      &{$\alpha_2$}& {0.900 (0.050)} & {0.753 (0.153)} & {0.623 (0.128)} & {0.526 (0.132)} \\
      % &{$\tilde{\alpha}_2$ }  & {}& {} & {} & {}  \\
       & {Diagonal} & {0.519 (0.033)} & {0.521 (0.037)} & {0.519 (0.037)} & {0.517 (0.033)} \\
     \bottomrule
    {\textbf{10,000 interactions}} 
    &{Parameters}&{$\{\alpha_b\}=\{0.1,0.9\}$}  & {$\{\alpha_b\}=\{0.2,0.8\}$} &{$\{\alpha_b\}=\{0.3,0.7\}$} & {$\{\alpha_b\}=\{0.4,0.6\}$} \\ \midrule
    {$\{a,b\}=\{0.9,0.1\}$} &{$\alpha_1$}  & {0.157 (0.111)} & {0.218 (0.121)}& {0.274 (0.101)}& {0.395 (0.075)} \\ 
      %&{$\tilde{\alpha}_1$ }  & {}& {} & {}& {} \\ 
      &{$\alpha_2$}& {0.900 (0.006)} & {0.803 (0.012)}& {0.691 (0.025)} & {0.591 (0.032)} \\
       %&{$\tilde{\alpha}_2$ }  & {}& {} & {} & {} \\
       & {Diagonal} & {0.900 (0.006)}& {0.899 (0.005)} & {0.900 (0.005)} & {0.900 (0.006)} \\
      \midrule
    {$\{a,b\}=\{0.7,0.3\}$} &{$\alpha_1$}  & {0.378 (0.351)} & {0.216 (0.120)}& {0.268 (0.132)}& {0.354 (0.117)} \\ 
      %&{$\tilde{\alpha}_1$ }  & {} & {} & {}& {}  \\ 
      &{$\alpha_2$}& {0.873 (0.067)} & {0.802 (0.012)}& {0.687 (0.031)}& {0.586 (0.050)}\\
       %&{$\tilde{\alpha}_2$ }  & {} & {} & {}& {}  \\
       & {Diagonal} & {0.653 (0.078)}& {0.693 (0.038)} & {0.660 (0.078)} & {0.680 (0.052)} \\
      \midrule
    {$\{a,b\}=\{0.5,0.5\}$} &{$\alpha_1$}  & {0.732 (0.262)} & {0.597 (0.259)}& {0.521 (0.199)}& {0.351 (0.191)} \\ 
      %&{$\tilde{\alpha}_1$ }  & {} & {} & {}& {}  \\ 
      &{$\alpha_2$}& {0.853 (0.121)}& {0.721 (0.138)}& {0.672 (0.081)}& {0.567 (0.073)}\\
       %s&{$\tilde{\alpha}_2$ }  & {}& {} & {} & {} \\
       & {Diagonal} & {0.506 (0.011)} & {0.507 (0.012)} & {0.505 (0.011)} & {0.505 (0.010)} \\
     \bottomrule

\end{tabular}
\end{adjustwidth}
\caption{Posterior means (SD) of power-law parameters $\{\alpha_b\}$ and within/between cluster propensity $\{a,b\}$ in different settings, with $\{\theta_1,\theta_2\}=\{1,1\}$ } 
\end{table}

\begin{table} []
\begin{adjustwidth}{-2cm}{}
\vspace*{0cm}
\begin{tabular}{SSSSSS} \toprule
    {\textbf{1,000 interactions}} 
    &{Parameters}&{$\{\alpha_b\}=\{0.1,0.9\}$}  & {$\{\alpha_b\}=\{0.2,0.8\}$} &{$\{\alpha_b\}=\{0.3,0.7\}$} & {$\{\alpha_b\}=\{0.4,0.6\}$} \\ \midrule
    {$\{a,b\}=\{0.9,0.1\}$} &{$\alpha_1$}  & {0.212 (0.151)}& {0.297 (0.196)} & {0.354 (0.181)}& {0.384 (0.14)}  \\ 
      %&{$\tilde{\alpha}_1$ }  & {} & {} & {}& {}  \\ 
      &{$\alpha_2$}& {0.903 (0.016)} & {0.808 (0.038)}& {0.718 (0.054)} & {0.626 (0.049)} \\
       %&{$\tilde{\alpha}_2$ }  & {} & {} & {}& {} \\
       & {Diagonal} & {0.900 (0.019)}& {0.887 (0.079)} & {0.885 (0.08)} & {0.899 (0.018)} \\
      \midrule
    {$\{a,b\}=\{0.7,0.3\}$} &{$\alpha_1$}  & {0.381 (0.312)}& {0.367 (0.232)}& {0.382 (0.201)}& {0.358 (0.152)} \\ 
      %&{$\tilde{\alpha}_1$ }  & {}& {} & {} &  {} \\ 
      &{$\alpha_2$}& {0.895 (0.054)} & {0.781 (0.119)} & {0.704 (0.081)} & {0.633 (0.048)} \\
      % &{$\tilde{\alpha}_2$ }  & {}& {} & {} & {} \\
       & {Diagonal} & {0.668 (0.079)} & {0.672 (0.074)} & {0.679 (0.065)} & {0.689 (0.05)} \\
      \midrule
    {$\{a,b\}=\{0.5,0.5\}$} &{$\alpha_1$}  & {0.548 (0.308)} & {0.502 (0.25)} & {0.568 (0.202)} & {0.5 (0.169)} \\ 
      %&{$\tilde{\alpha}_1$ }  & {}& {} & {}& {} \\ 
      &{$\alpha_2$}& {0.871 (0.122)} & {0.806 (0.076)} & {0.671 (0.119)} & {0.589 (0.101)} \\
      % &{$\tilde{\alpha}_2$ }  & {}& {} & {} & {}  \\
       & {Diagonal} & {0.513 (0.034)} & {0.526 (0.04)} & {0.518 (0.039)} & {0.525 (0.039)} \\
     \bottomrule
    {\textbf{10,000 interactions}} 
    &{Parameters}&{$\{\alpha_b\}=\{0.1,0.9\}$}  & {$\{\alpha_b\}=\{0.2,0.8\}$} &{$\{\alpha_b\}=\{0.3,0.7\}$} & {$\{\alpha_b\}=\{0.4,0.6\}$} \\ \midrule
    {$\{a,b\}=\{0.9,0.1\}$} &{$\alpha_1$}  & {0.143 (0.101)}& {0.186 (0.104)} & {0.31 (0.103)}& {0.396 (0.075)}  \\ 
      %&{$\tilde{\alpha}_1$ }  & {} & {} & {}& {}  \\ 
      &{$\alpha_2$}& {0.901 (0.006)} & {0.803 (0.011)}& {0.709 (0.013)} & {0.61 (0.018)} \\
      % &{$\tilde{\alpha}_2$ }  & {} & {} & {}& {} \\
       & {Diagonal} & {0.899 (0.006)}& {0.900 (0.006)} & {0.900 (0.006)} & {0.899 (0.007)} \\
      \midrule
    {$\{a,b\}=\{0.7,0.3\}$} &{$\alpha_1$}  & {0.354 (0.319)}& {0.391 (0.262)}& {0.433 (0.195)}& {0.42 (0.113)} \\ 
      %&{$\tilde{\alpha}_1$ }  & {}& {} & {} &  {} \\ 
      &{$\alpha_2$}& {0.900 (0.028)} & {0.794 (0.041)} & {0.694 (0.047)} & {0.609 (0.024)} \\
       %&{$\tilde{\alpha}_2$ }  & {}& {} & {} & {} \\
       & {Diagonal} & {0.656 (0.081)} & {0.649 (0.081)} & {0.647 (0.075)} & {0.666 (0.069)} \\
      \midrule
    {$\{a,b\}=\{0.5,0.5\}$} &{$\alpha_1$}  & {0.751 (0.271)} & {0.629 (0.229)} & {0.596 (0.164)} & {0.531 (0.122)} \\ 
      %&{$\tilde{\alpha}_1$ }  & {}& {} & {}& {} \\ 
      &{$\alpha_2$}& {0.844 (0.109)} & {0.787 (0.06)} & {0.683 (0.08)} & {0.59 (0.063)} \\
      % &{$\tilde{\alpha}_2$ }  & {}& {} & {} & {}  \\
       & {Diagonal} & {0.506 (0.011)} & {0.508 (0.012)} & {0.506 (0.012)} & {0.508 (0.013)} \\
     \bottomrule
    
\end{tabular}
\end{adjustwidth}
\caption{Posterior means (SD) of power-law parameters $\{\alpha_b\}$ and within/between cluster propensity $\{a,b\}$ in different settings, with $\{\theta_1,\theta_2\}=\{1,10\}$} 
\end{table}

\begin{table}
\begin{adjustwidth}{-2cm}{}
\vspace*{0cm}

\begin{tabular}{SSSSSS} \toprule
    {\textbf{1,000 interactions}} 
    & {Parameters} &{$\{\alpha_b\}=\{0.1,0.9\}$}  & {$\{\alpha_b\}=\{0.2,0.8\}$} &{$\{\alpha_b\}=\{0.3,0.7\}$} & {$\{\alpha_b\}=\{0.4,0.6\}$} \\ \midrule
    {$\{a,b\}=\{0.9,0.1\}$} &{$\alpha_1$}  & {0.323 (0.099)}& {0.368 (0.091)} & {0.402 (0.064)}& {0.485 (0.078)}  \\ 
      %&{$\tilde{\alpha}_1$ }  & {} & {} & {}& {}  \\ 
      &{$\alpha_2$}& {0.893 (0.028)} & {0.789 (0.036)}& {0.691 (0.052)} & {0.582 (0.094)} \\
       %&{$\tilde{\alpha}_2$ }  & {} & {} & {}& {} \\
       & {Diagonal} & {0.914 (0.02)}& {0.904 (0.02)} & {0.906 (0.019)} & {0.896 (0.022)} \\
      \midrule
    {$\{a,b\}=\{0.7,0.3\}$} &{$\alpha_1$}  & {0.369 (0.129)}& {0.459 (0.154)}& {0.46 (0.097)}& {0.477 (0.083)} \\ 
      %&{$\tilde{\alpha}_1$ }  & {}& {} & {} &  {} \\ 
      &{$\alpha_2$}& {0.896 (0.024)} & {0.75 (0.099)} & {0.636 (0.101)} & {0.553 (0.097)} \\
      % &{$\tilde{\alpha}_2$ }  & {}& {} & {} & {} \\
       & {Diagonal} & {0.702 (0.054)} & {0.714 (0.03)} & {0.711 (0.03)} & {0.701 (0.043)} \\
      \midrule
    {$\{a,b\}=\{0.5,0.5\}$} &{$\alpha_1$}  & {0.49 (0.222)} & {0.556 (0.19)} & {0.508 (0.157)} & {0.466 (0.142)} \\ 
      %&{$\tilde{\alpha}_1$ }  & {}& {} & {}& {} \\ 
      &{$\alpha_2$}& {0.859 (0.099)} & {0.665 (0.145)} & {0.576 (0.095)} & {0.546 (0.084)} \\
      % &{$\tilde{\alpha}_2$ }  & {}& {} & {} & {}  \\
       & {Diagonal} & {0.528 (0.044)} & {0.524 (0.048)} & {0.53 (0.047)} & {0.52 (0.045)} \\
     \bottomrule
    {\textbf{10,000 interactions}} 
    &{Parameters}&{$\{\alpha_b\}=\{0.1,0.9\}$}  & {$\{\alpha_b\}=\{0.2,0.8\}$} &{$\{\alpha_b\}=\{0.3,0.7\}$} & {$\{\alpha_b\}=\{0.4,0.6\}$} \\ \midrule
    {$\{a,b\}=\{0.9,0.1\}$} &{$\alpha_1$}  & {0.203 (0.067)} & {0.297 (0.056)}& {0.345 (0.042)}& {0.44 (0.03)} \\ 
      %&{$\tilde{\alpha}_1$ }  & {}& {} & {}& {} \\ 
      &{$\alpha_2$}& {0.900 (0.006)} & {0.798 (0.012)}& {0.697 (0.022)} & {0.586 (0.034)} \\
      % &{$\tilde{\alpha}_2$ }  & {}& {} & {} & {} \\
       & {Diagonal} & {0.902 (0.006)}& {0.900 (0.007)} & {0.900 (0.007)} & {0.878 (0.085)} \\
      \midrule
    {$\{a,b\}=\{0.7,0.3\}$} &{$\alpha_1$}  & {0.286 (0.177)} & {0.302 (0.082)}& {0.343 (0.093)}& {0.442 (0.064)} \\ 
      %&{$\tilde{\alpha}_1$ }  & {} & {} & {}& {}  \\ 
      &{$\alpha_2$}& {0.882 (0.082)} & {0.795 (0.018)}& {0.682 (0.041)}& {0.563 (0.056)}\\
      % &{$\tilde{\alpha}_2$ }  & {} & {} & {}& {}  \\
       & {Diagonal} & {0.692 (0.044)}& {0.686 (0.051)} & {0.668 (0.065)} & {0.639 (0.087)} \\
      \midrule
    {$\{a,b\}=\{0.5,0.5\}$} &{$\alpha_1$}  & {0.72 (0.265)} & {0.403 (0.181)}& {0.507 (0.156)}& {0.466 (0.094)} \\ 
      %&{$\tilde{\alpha}_1$ }  & {} & {} & {}& {}  \\ 
      &{$\alpha_2$}& {0.726 (0.172)}& {0.776 (0.059)}& {0.603 (0.092)}& {0.531 (0.054)}\\
      % &{$\tilde{\alpha}_2$ }  & {}& {} & {} & {} \\
       & {Diagonal} & {0.507 (0.012)} & {0.508 (0.012)} & {0.507 (0.013)} & {0.509 (0.013)} \\
     \bottomrule

\end{tabular}
\end{adjustwidth}
\caption{Posterior means (SD) of power-law parameters $\{\alpha_b\}$ and within/between cluster propensity $\{a,b\}$ in different settings, with $\{\theta_1,\theta_2\}=\{10,1\}$} 
\end{table}

\subsection{Selection of K in settings with varying parameters}
\label{sec:6.6}
\begin{figure}
\centering
\begin{subfigure}{1\textwidth}
  \centering
  \includegraphics[width=.8\linewidth,height=5cm]{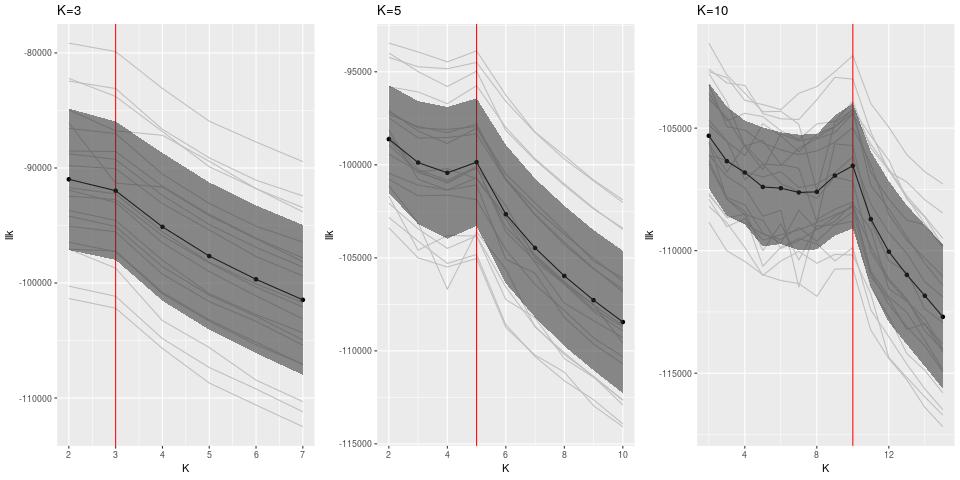}
  \caption{}
  \label{fig:sub.7.6.1}
\end{subfigure}\\
\begin{subfigure}{1\textwidth}
  \centering
  \includegraphics[width=.8\linewidth,height=5cm]{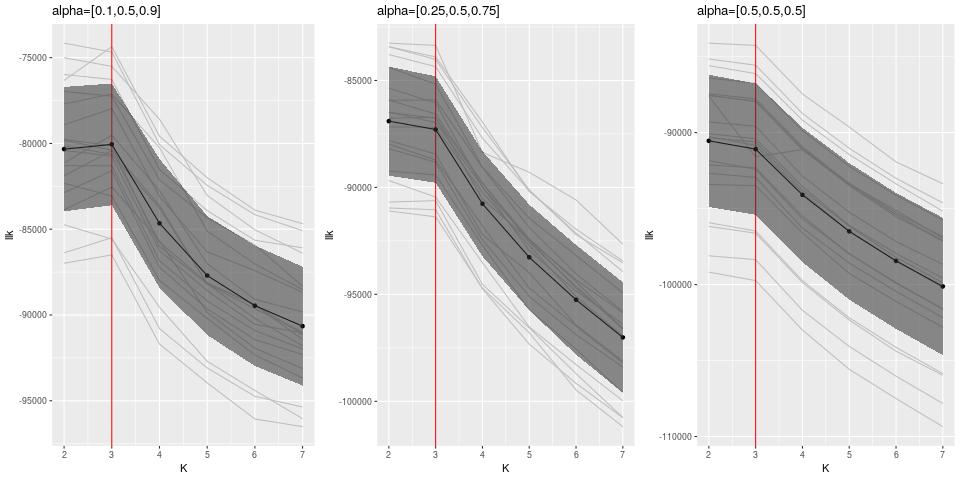}
  \caption{}
  \label{fig:sub.7.6.2}
\end{subfigure}\\
\begin{subfigure}{1\textwidth}
  \centering
  \includegraphics[width=.8\linewidth,height=5cm]{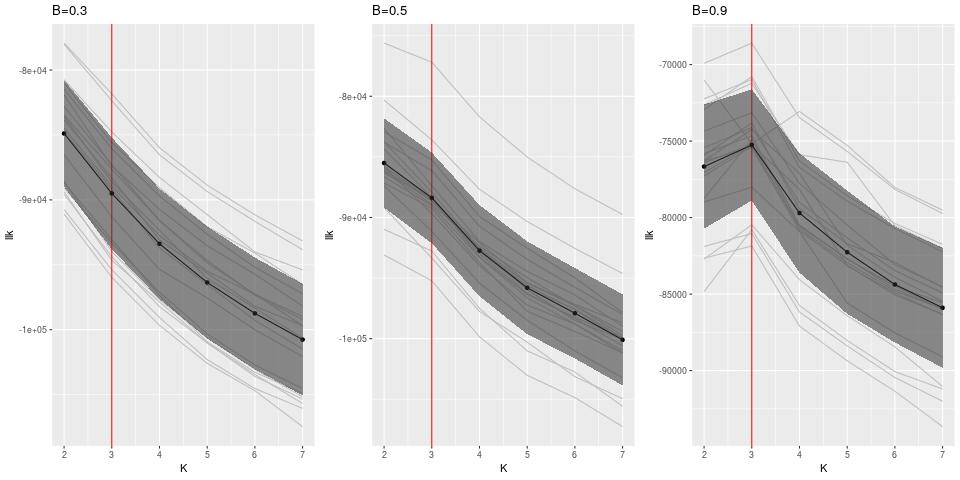}
  \caption{}
  \label{fig:sub.7.6.3}
\end{subfigure}
\caption{The trace plots of the likelihood in different settings. (a) For a varying choice of the underlying truth K, $a=0.8$, $b = \frac{0.2}{(K-1)}$,  $\alpha\sim$Uniform$(0.4,0.8)$; (b) Fix K=3, $a=0.8$, $b=0.1$, $\{\alpha_1,\alpha_2,\alpha_3\}=\{\{0.1,0.5,0.9\},\{0.25,0.5,0.75\},\{0.5,0.5,0.5\}\}$; (c) Fix K=3, $\{\alpha_1,\alpha_2,\alpha_3\}=\{0.25,0.5,0.75\}$ , $\{a,b\}=\{\{0.3,0.35\},\{0.5,0.25\},\{0.9,0.05\}\}$ }
\label{fig:7.6}
\end{figure}

In Section 4.0.3, we briefly discuss a model selection criteria based on the maximal marginal likelihood. Figure~\ref{fig:7.6} shows the likelihood as a function of $K$ in other settings, where the propensity of within/between connections and the power-law parameters are different from the one shown in Section 4.0.3. The K that gives the MAP based on the empirical results is affected not only by the number of clusters in the generative model, but also by the strength of the connectivity and the power-law properties.

\subsection{Likelihood pattern in TalkLife data}

\begin{figure}
\centering
\begin{subfigure}{0.3\textwidth}
  \centering
  \includegraphics[width=.9\linewidth]{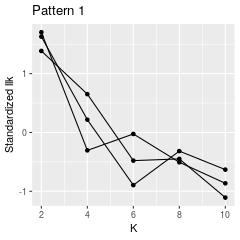}
  \caption{}
  \label{fig:sub.5.2.1.1}
\end{subfigure}%
\begin{subfigure}{0.3\textwidth}
  \centering
  \includegraphics[width=.9\linewidth]{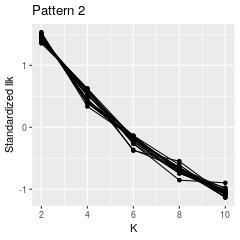}
  \caption{}
  \label{fig:sub.5.2.1.2}
\end{subfigure}
\begin{subfigure}{0.3\textwidth}
  \centering
  \includegraphics[width=.9\linewidth]{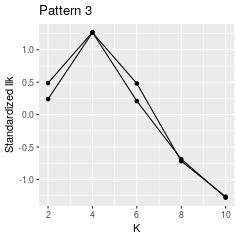}
  \caption{}
  \label{fig:sub.5.2.1.3}
\end{subfigure}
\caption{Different marginal log likelihood patterns observed in TalkLife networks. (a) The likelihood is maximized when K=2, but is not monotonic; (b) The likelihood is maximized when K=2, and is monotonic; (c) The likelihood is maximized when K=4. The log likelihoods in different networks are standardized to the same scale.}
\label{fig:5.2.1}
\end{figure}

The averaged marginal likelihood over Gibbs iterations is shown in Figure~\ref{fig:5.2.1}. In the majority of sub-communities, the averaged marginal likelihood shows a monotonic pattern as function of the number of presumed blocks (Figure~\ref{fig:sub.5.2.1.1}). That is, the likelihood is the largest when $K=2$ and keeps decreasing as $K$ increases. There are several exceptions with the marginal likelihoods being maximized when K is greater than 2 (Figure~\ref{fig:sub.5.2.1.3}). 

\subsection{Case study with model fitting in other network}
\label{sec:case_study}
\begin{figure}
\centering
\begin{subfigure}{0.6\textwidth}
  \centering
  \includegraphics[width=.9\linewidth]{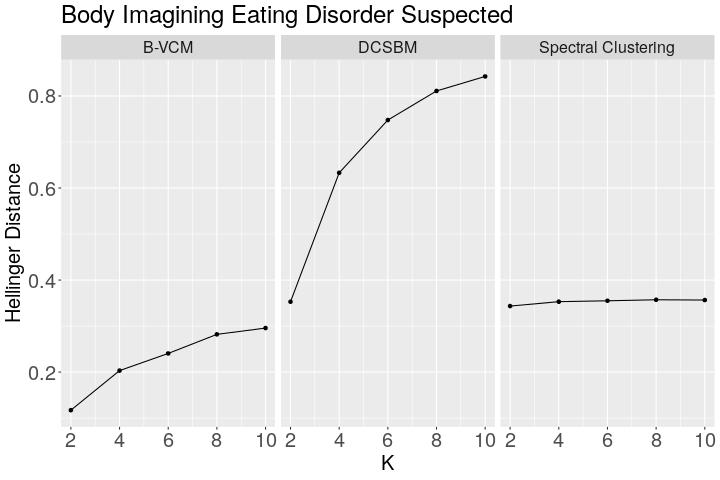}
  \caption{}
  \label{fig:sub.HD_61}
\end{subfigure}%
\begin{subfigure}{0.3\textwidth}
  \centering
  \includegraphics[height=160pt,width=.9\linewidth]{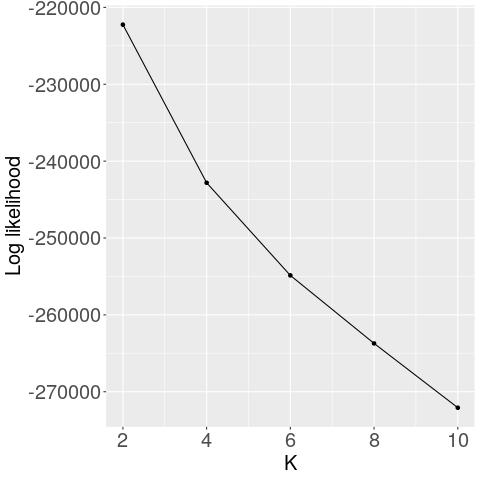}
  \caption{}
  \label{fig:sub.llk_61}
\end{subfigure}

\begin{subfigure}{0.6\textwidth}
  \centering
  \includegraphics[width=.9\linewidth]{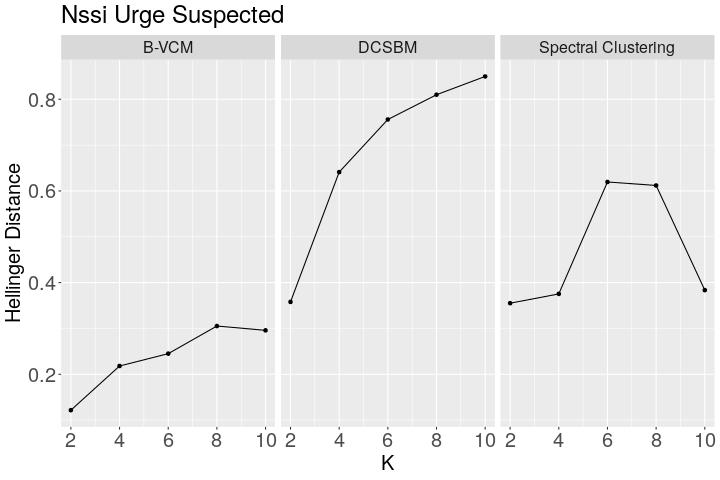}
  \caption{}
  \label{fig:sub.HD_78}
\end{subfigure}%
\begin{subfigure}{0.3\textwidth}
  \centering
  \includegraphics[height=160pt,width=.9\linewidth]{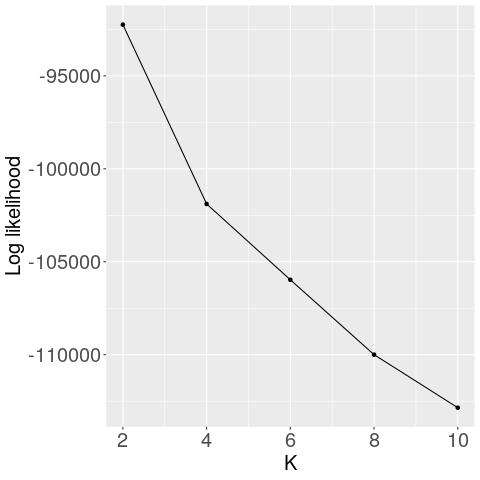}
  \caption{}
  \label{fig:sub.llk_78}
\end{subfigure}

\caption{The Hellinger distances of the block assignment between the first half and the second half of the 2019 data in (a) Body Imagining and Eating Disorder network and (c) Nssi Urge network. The marginal likelihood over different K values in (b) Body Imagining and Eating Disorder network and (d) Nssi Urge network.}
\label{fig:6.5.jpg}
\end{figure}

In Section 5.3, we showed two examples using Alcohol and Substance Abuse network and Behavoiral Symptoms network. Note that not all the networks will give proper results due to the complexity of the underlying community structure. Here, we showed the results from fitting several more networks in Figure~\ref{fig:6.5.jpg}, including Body Imagining Eating Disorder Suspected network and Nssi Urge Suspected network. Our main conclusions remain the same as stated in the main context.  

\subsection{Fitting DC-SBM in the projected network}

\label{sec:dcsbm}

\begin{figure}
\centering
\begin{subfigure}{0.3\textwidth}
  \centering
  \includegraphics[width=.9\linewidth]{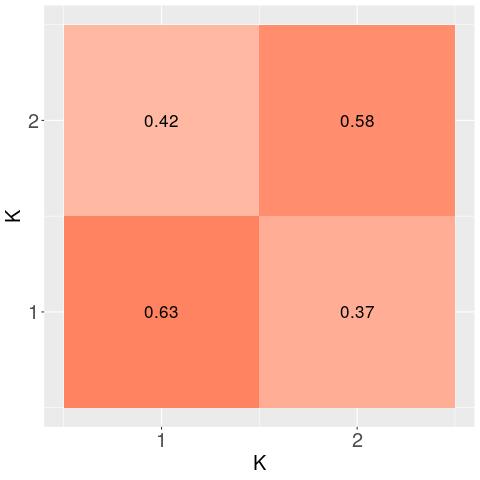}
  \caption{}
  \label{fig:sub.5.3.2.1}
\end{subfigure}%
\begin{subfigure}{0.3\textwidth}
  \centering
  \includegraphics[width=.9\linewidth]{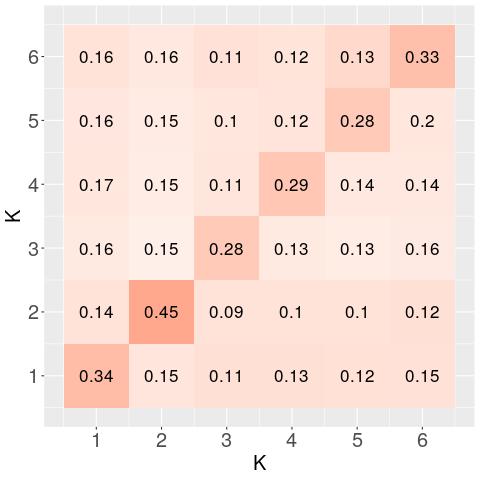}
  \caption{}
  \label{fig:sub.5.3.2.2}
\end{subfigure}
\begin{subfigure}{0.3\textwidth}
  \centering
  \includegraphics[width=.9\linewidth]{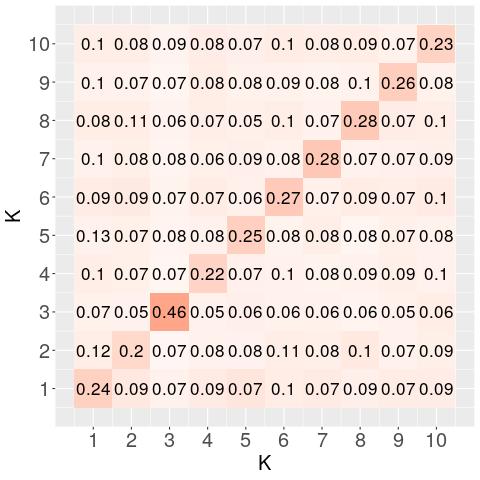}
  \caption{}
  \label{fig:sub.5.3.2.3}
\end{subfigure}
\caption{The inter/intra connectivity of the communities detected by DC-SBM in the projected network, ranging from 0 to 1, indicates the proportion of the interactions that initiated from one block to the other. The number within each cell is the proportion of the interactions initiated from one cluster (y-axis) to another (x-axis), normalized by each row.}
\label{fig:dcsbm-cutoff}
\end{figure}

\begin{figure}[htp]
    \centering
    \includegraphics[width=15cm]{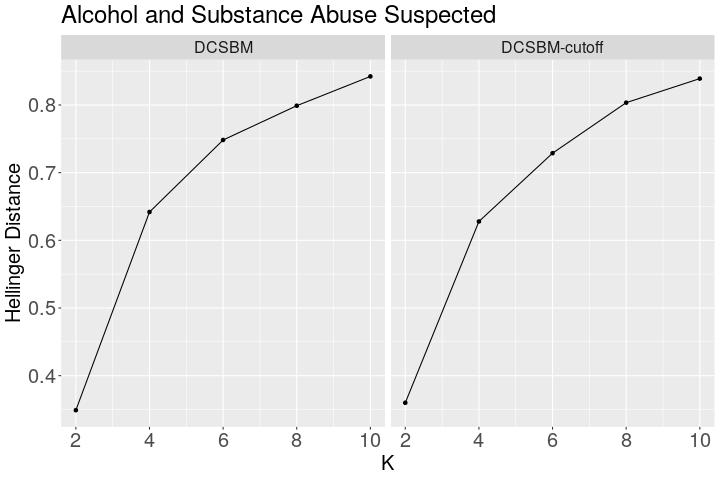}
    \caption{The Hellinger distances of the block assignment between the first half and the second half of the 2019 data in (a) Alcohol and Substance Abuse suspected network and (b) the projected network with the cutoff value being 2.}
    \label{fig:dcsbm-cutoff2}
\end{figure}

We show here the fitting results of the DC-SBM to the project Alcohol and Substance Abuse network. In the projected network, we say there exists an interaction between i and j only if the number of interactions between i and j reaches certain cutoff in the binary graph. We set the cutoff value to be 2. In the original binary graph, there are 30,744 edges. While in the projected network, there exist 4,218 edges, that accounts for only 13\% of the interactions in the original network. Given the amount of the edges being trimmed, we didn't try the cutoff greater than 2. The results are shown in Figure~\ref{fig:dcsbm-cutoff} and Figure~\ref{fig:dcsbm-cutoff2}. The within block connections are more frequent in the projected network than before, but still weaker as compared to the other two methods. The HL distance is similar to what was observed before.

\section{Supplementary Materials}

\subsection{Supplementary Tables}

\begin{table}
\centering
\def\arraystretch{0.6}
\begin{tabular}{|c|}
    \hline
    \textbf{Tag names}\\
    \hline
AgitationOrIrritationSuspected\\
\hline
AlcoholAndSubstanceAbuseSuspected\\
\hline
AnxietyPanicFearSuspected\\
\hline
BehavorialSymptomsSuspected\\
\hline
BodyImageEatingDisordersSuspected\\
\hline
CryingSuspected\\
\hline
DeathOfOtherSuspected\\
\hline
DepressedMoodSuspected\\
\hline
DistortedThinkingSuspected\\
\hline
EmotionalExhaustionSuspected\\
\hline
EmptinessSuspected\\
\hline
FailureSuspected\\
\hline
FamilyIssuesSuspected\\
\hline
FinalTiredFatiguedLowEnergySuspected\\
\hline
HelplessnessHopelessnessSuspected\\
\hline
InpatientOutPatientMedicationSuspected\\
\hline
LonelinessSuspected\\
\hline
MentalHealthTreatmentSuspected\\
\hline
NauseaSuspected\\
\hline
NauseaWithEatingDisorderSuspected\\
\hline
NssiIdeationAndBehaviorSuspected\\
\hline
NssiUrgeSuspected\\
\hline
NumbnessEmptinessSuspected\\
\hline
NumbnessSuspected\\
\hline
SelfHarmRelapseSuspected\\
\hline
SelfHarmRemissionOrRelapseSuspected\\
\hline
SelfHarmRemissionSuspected\\
\hline
SelfHarmSuspectedTakeTwo\\
\hline
SongLyricsSuspected\\
\hline
SuicidalIdeationAndBehaviorSuspected\\
\hline
SuicidalPlanningSuspected\\
\hline
SuicideAttemptSuspected\\
\hline
TiredFatiguedLowEnergySuspected \\
\hline
    \end{tabular}
    \caption{Name of all the Tags in the TalkLife data}
\end{table}

\subsection{Supplementary Figures}
\begin{figure}
\centering
\begin{subfigure}{0.3\textwidth}
  \centering
  \includegraphics[width=.9\linewidth]{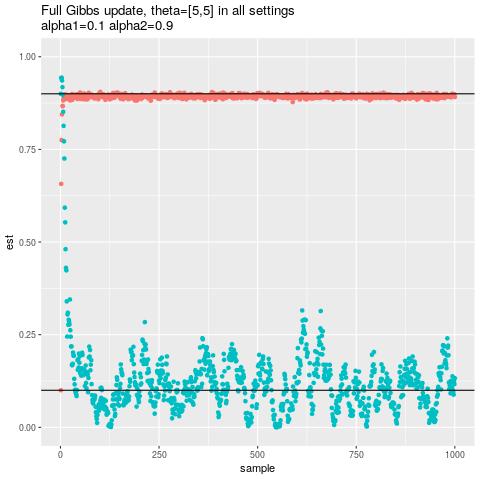}
  \caption{}
  \label{fig:sub.8.1.1}
\end{subfigure}%
\begin{subfigure}{0.3\textwidth}
  \centering
  \includegraphics[width=.9\linewidth]{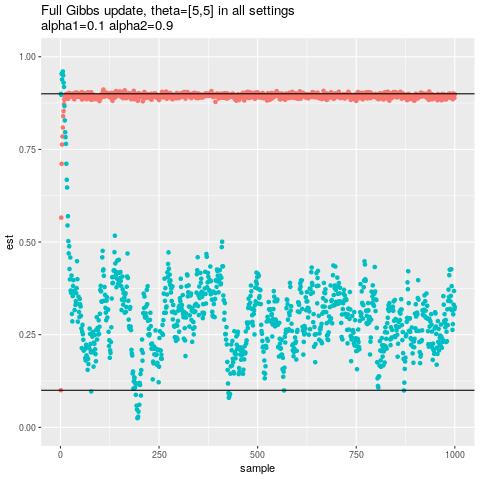}
  \caption{}
  \label{fig:sub.8.1.2}
\end{subfigure}
\begin{subfigure}{0.3\textwidth}
  \centering
  \includegraphics[width=.9\linewidth]{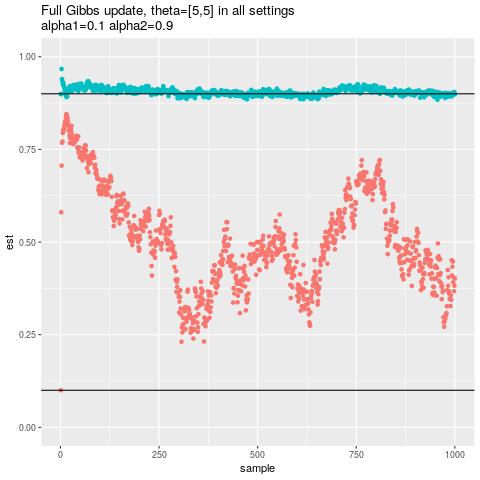}
  \caption{}
  \label{fig:sub.8.1.3}
\end{subfigure}

\begin{subfigure}{0.3\textwidth}
  \centering
  \includegraphics[width=.9\linewidth]{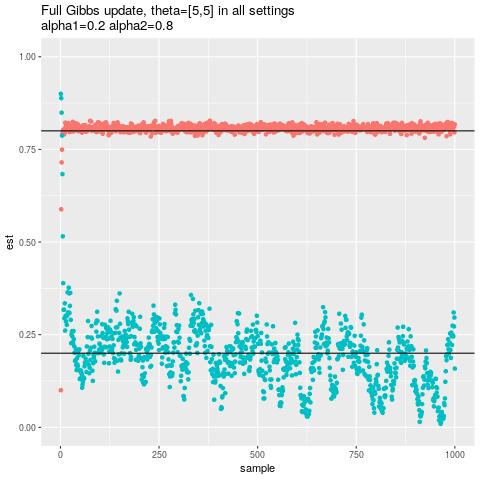}
  \caption{}
  \label{fig:sub.8.1.4}
\end{subfigure}%
\begin{subfigure}{0.3\textwidth}
  \centering
  \includegraphics[width=.9\linewidth]{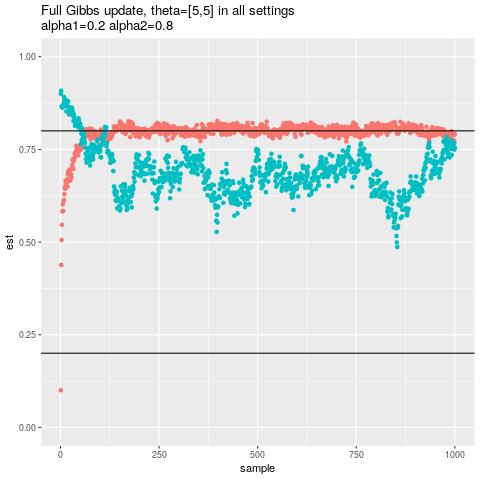}
  \caption{}
  \label{fig:sub.8.1.5}
\end{subfigure}
\begin{subfigure}{0.3\textwidth}
  \centering
  \includegraphics[width=.9\linewidth]{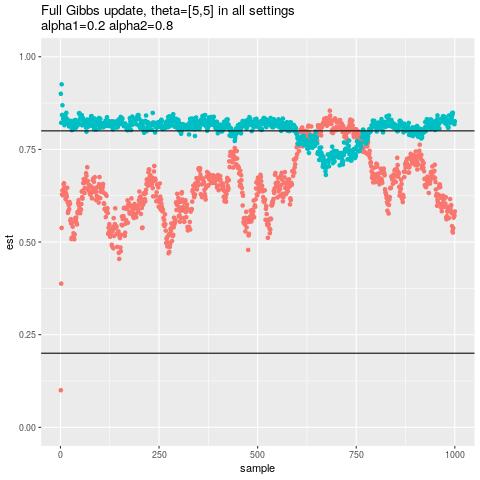}
  \caption{}
  \label{fig:sub.8.1.6}
\end{subfigure}

\begin{subfigure}{0.3\textwidth}
  \centering
  \includegraphics[width=.9\linewidth]{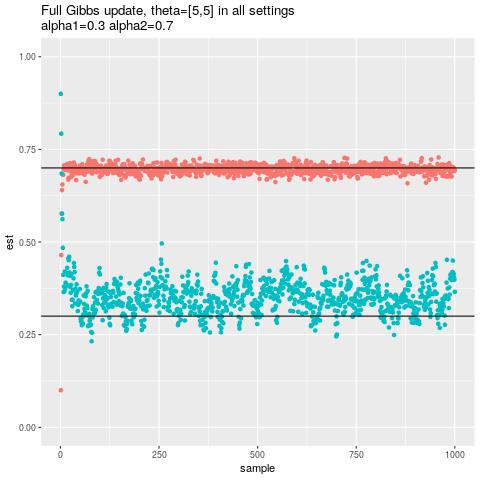}
  \caption{}
  \label{fig:sub.8.1.7}
\end{subfigure}%
\begin{subfigure}{0.3\textwidth}
  \centering
  \includegraphics[width=.9\linewidth]{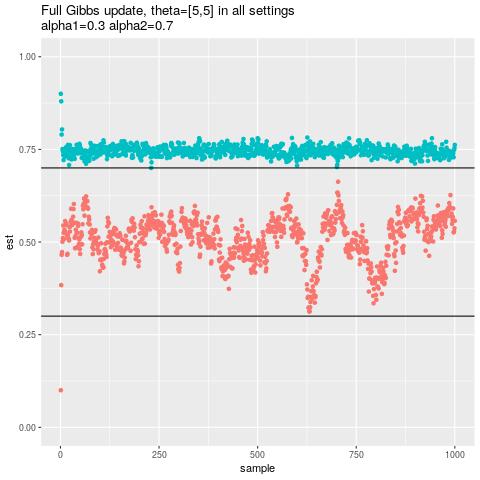}
  \caption{}
  \label{fig:sub.8.1.8}
\end{subfigure}
\begin{subfigure}{0.3\textwidth}
  \centering
  \includegraphics[width=.9\linewidth]{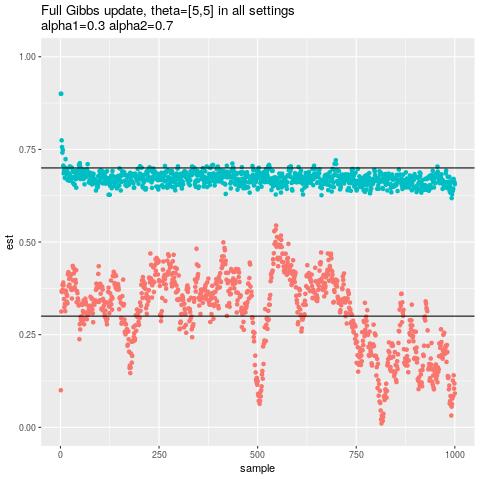}
  \caption{}
  \label{fig:sub.8.1.9}
\end{subfigure}

\begin{subfigure}{0.3\textwidth}
  \centering
  \includegraphics[width=.9\linewidth]{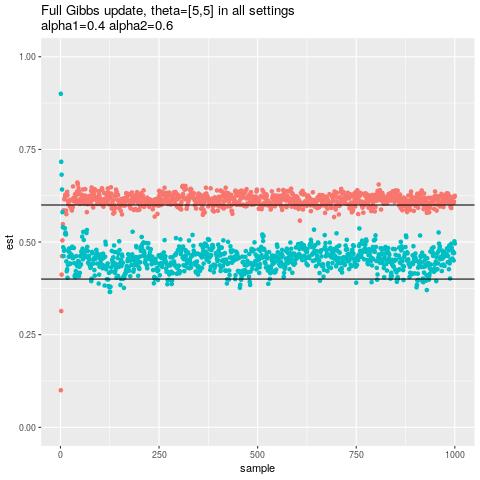}
  \caption{}
  \label{fig:sub.8.1.1rep}
\end{subfigure}%
\begin{subfigure}{0.3\textwidth}
  \centering
  \includegraphics[width=.9\linewidth]{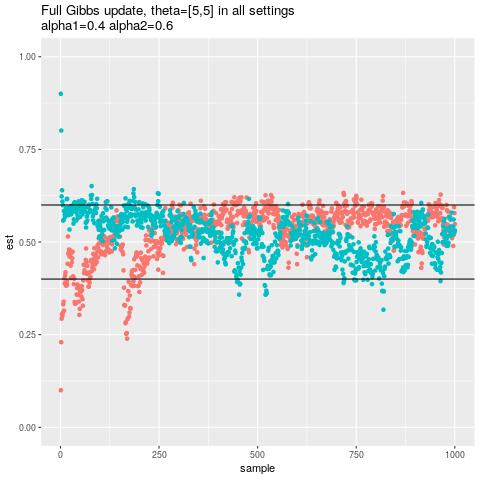}
  \caption{}
  \label{fig:sub.8.1.2rep}
\end{subfigure}
\begin{subfigure}{0.3\textwidth}
  \centering
  \includegraphics[width=.9\linewidth]{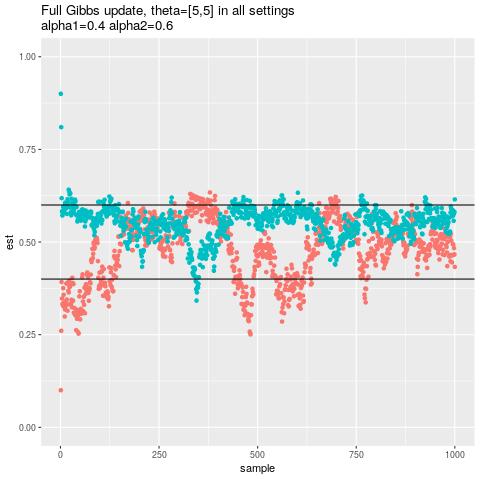}
  \caption{}
  \label{fig:sub.8.1.3rep}
\end{subfigure}

\caption{The trace plots of the Gibbs samplers of $\alpha1$ and $\alpha2$. Each row corresponds to different underlying $\alpha$ values, marked in the solid line; each column corresponds to different $\{a,b\}$, which takes the value from $\{0.1,0.9\}$, $\{0.3,0.7\}$, and $\{0.5,0.5\}$}
\label{fig:8.1}
\end{figure}

\begin{figure}[htp]
    \centering
    \includegraphics[width=15cm]{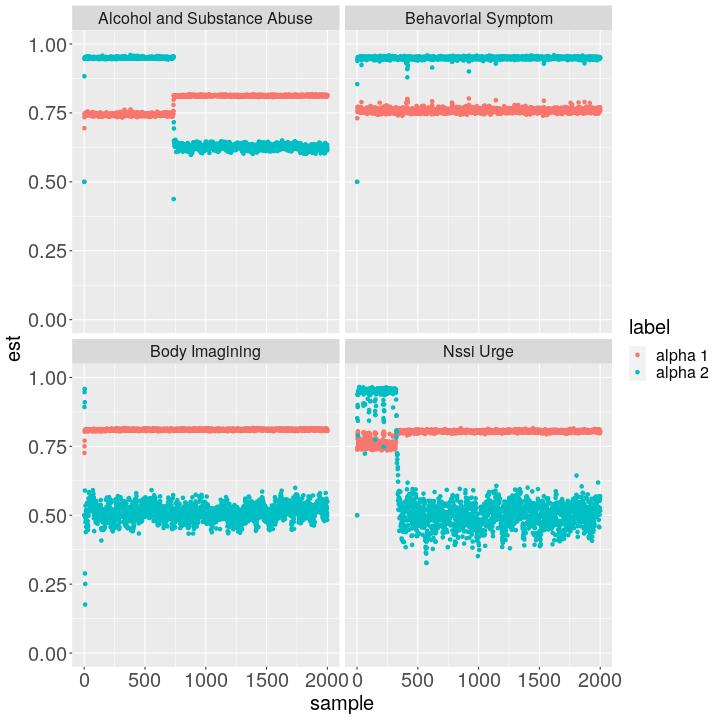}
    \caption{The trace plots of the Gibbs Samplers in different sub networks in TalkLife data, when setting the number of blocks to be 2.}
    \label{fig:traceplot.jpg}
\end{figure}

%\walt{Add extra detail on these figures to explain how they are computed.}

\begin{figure}[htp]
    \centering
    \includegraphics[width=15cm]{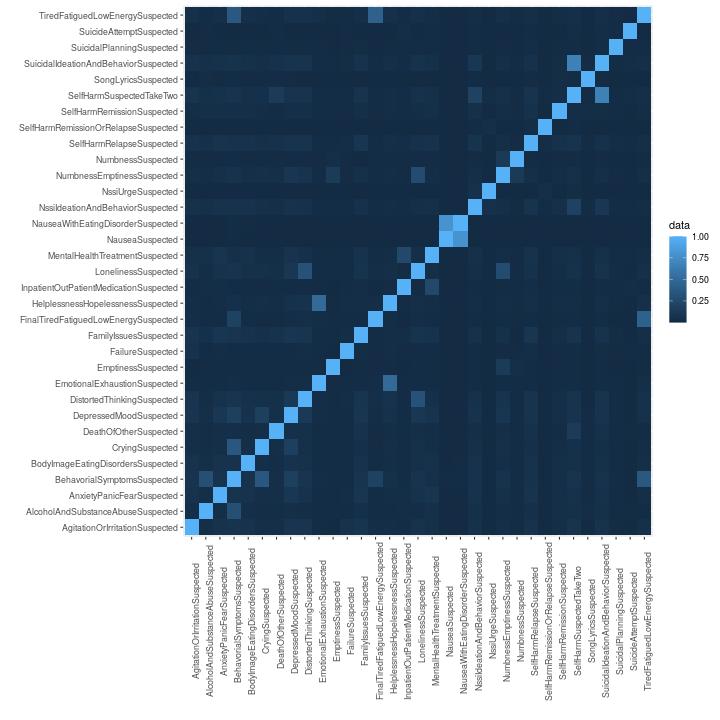}
    \caption{The heat map of the overlapping of posters being categorized as two of the topics. Each spot shown in the figure is calculated by $\frac{\text{Number of posters that have both tags}}{\text{Number of posters that have either of the tags}}$.}
    \label{fig:suppheatmap.jpg}
\end{figure}

\begin{figure}[htp]
    \centering
    \includegraphics[width=15cm]{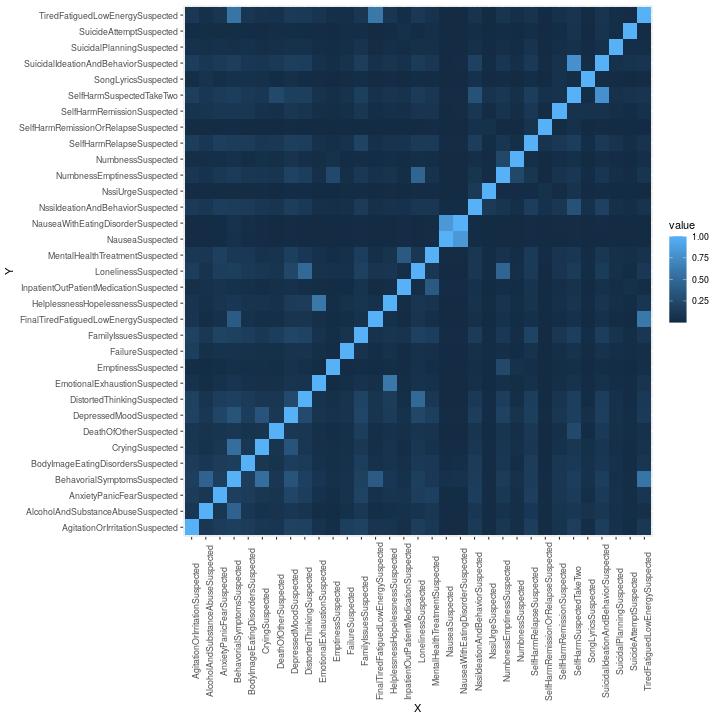}
    \caption{The heat map of the overlapping of the users focusing on different topics. Each spot shown in the figure is calculated by $\frac{\text{Number of users that focus on both topics}}{\text{Number of users that focus on either of the topics}}$.}
    \label{fig:suppheatmap2.jpg}
\end{figure}

\newpage

\section{Code to Replicate Simulation and Case Study Results}
The R code used to generate the simulation experiments and case study results in this paper can be obtained at %\verb"https://github.com/XXXX/XXXX".
\verb"https://github.com/YuhuaZhang1995/B-VCM".

\printbibliography

%% file: z_concep_fig_h.tex
\begin{tikzpicture}
    % \node[inner sep=0pt] (fbpost) at (0,3.75)
    %     {\includegraphics[width=0.75in]{./Mario-Facebook-Wall.png}};
    %  \node[inner sep=0pt] (fbpost) at (0,1.75)
    %  {\includegraphics[width=0.75in]{./Macbeth-Facebook-Wall.png}};i
    %  \node (ell) at (0,0.5) {\LARGE $\vdots$};
    %\node[inner sep=0pt] (fbpost) at (0.5,7.5){};
    % \node[inner sep=0pt] (fbpost) at (2.5,7.5){};

    %\node (ts) at (0.5,4.5) {Time};
    \node (poster) at (0,4.5) {Poster};
    \node (reactors) at (1.75,4.5) {Reactors};
    
    %Interaction 1
    %\node (t1) at (0.5,4) {$t_1$};

    \node (p) at (0,4) {a};
    \node (r1) at (1.75,4) {\{b, c, d\}};
    %Interaction 2
    % \node (t2) at (0.5,3.5) {$t_2$};
    
    \node (p2) at (0,3.5) {e};
    \node (r2) at (1.75,3.5) {\{d, f\}};
    
    \node (p3) at (0,3) {g};
    \node (r3) at (1.75,3) {\{f, h\}};
    
    \node (poster) at (0,2.5) {Block};
    \node (reactors) at (1.75,2.5) {Labels};
    
    \node (p) at (0,2) {1};
    \node (r1) at (1.75,2) {\{1, 1, 1\}};
    %Interaction 2
    % \node (t2) at (0.5,3.5) {$t_2$};
    \node (p) at (0,1.5) {1};
    \node (r1) at (1.75,1.5) {\{1, 2\}};
    
    %Interaction 1
    %\node (t1) at (0.5,4) {$t_1$};
    \node (p) at (0,1) {2};
    \node (r1) at (1.75,1) {\{2, 2\}};

    \node[single arrow, rotate=0, rounded corners=3pt, fill=blue!30,
          draw, align=center, xshift=3.5cm, yshift=2.75cm,
          minimum height=1.75cm, minimum width=1.75cm]{};
    
    \begin{scope}[inner sep=1mm,
                    poster/.style={circle,draw=brown!50,fill=brown!50,thick},
                    transition/.style={rectangle,draw=black!50,fill=black!20,thick},
                    reactor/.style={circle,draw=black!50,fill=black!50,thick}]
                    
    \node (a) [poster] at (5,4) {a};
    \node (b) [poster] at (5.5,2.75) {b};
    \node (c) [poster] at (6.5,3.0) {c};
    \node (d) [poster] at (8,4) {d};
    \node (e) [poster] at (8.5,3) {e};
    \node (f) [reactor] at (7.5,2) {f};
    \node (h) [reactor] at (5,1.75) {h};
    \node (g) [reactor] at (6.25,1.25) {g};
    \end{scope}
    \begin{scope}[blend mode=multiply, fill opacity=0.8, on background layer]
    \filldraw[fill=red!70,opacity=0.7] ($(a)+(-0.5,0)$) 
        to[out=270,in=180] ($(b) - (0,0.5)$) 
        to[out=0,in=230] ($(c) + (0.5,0)$)
        to[out=50,in=270] ($(d) + (0.5,0)$)
        to[out=90,in=0] ($(d)+(0,0.5)$)
        to[out=180,in=0] ($(c)+(0,0.5)$)
        to[out=180,in=0] ($(a) + (0,0.5)$)
        to[out=180,in=90] ($(a)+(-0.5,0)$);
    \filldraw[fill=blue!20,opacity=0.7] ($(d)+(0,0.5)$)
        to[out=0,in=90] ($(e)+(0.5,0)$)
        to[out=270,in=0] ($(f)+(0,-0.5)$)
        to[out=180, in=270] ($(f)+(-0.5,0)$)
        to[out=90,in=270] ($(e)+(-0.5,0)$)
        to[out=90,in=270] ($(d)+(-0.5,0.2)$)
        to[out=90,in=180] ($(d)+(-0,0.5)$);
    \filldraw[fill=yellow!20,opacity=0.7] ($(h)+(-0.5,0)$)
        to[out=270,in=180] ($(g)-(0,0.5)$)
        to[out=0,in=230] ($(g)+(0.5,0)$)
        to[out=50, in=270] ($(f)+(0.5,0)$)
        to[out=90,in=0] ($(f)+(0,0.5)$)
        to[out=180,in=0] ($(g)+(0,0.5)$)
        to[out=180,in=0] ($(h)+(0,0.5)$)
        to[out=180,in=90] ($(h)-(0.5,0)$);
    \end{scope}

    %\node at (0.75,1.5) {$t_1$};
    %\node at (4,1.75) {$t_2$};
    %\node at (3,0.8) {$e_3$};
    %\node at (0.1,0.7) {$e_4$};
  \end{tikzpicture}

  